\documentclass[12pt]{article}
\usepackage{amsmath,amssymb,amsfonts, epsfig}
\usepackage{color}
\usepackage{bm}
\usepackage{epsfig}
\usepackage{psfrag}
\usepackage{latexsym}
\usepackage{indentfirst}
\usepackage{fancyhdr}
\usepackage{dsfont}
\usepackage{amssymb}
\usepackage{amsmath}
\usepackage{amsfonts}
\usepackage{pifont}
\usepackage{cite}
\usepackage{bbold}
\usepackage{color}
\usepackage{colordvi}
\usepackage{fancybox}
\usepackage[footnotesize]{caption2}
\usepackage{graphicx}
\usepackage[center,footnotesize,hang]{subfigure}
\usepackage{bbm}
\usepackage{url}
\usepackage{multirow}
\usepackage{tabularx}
\usepackage{array}
\usepackage{arydshln}
\usepackage{ulem}
\usepackage{enumitem}
\newcommand{\PreserveBackslash}[1]{\let\temp=\\#1\let\\=\temp}
\newcolumntype{C}[1]{>{\PreserveBackslash\centering}p{#1}}
\newcolumntype{R}[1]{>{\PreserveBackslash\raggedleft}p{#1}}
\newcolumntype{L}[1]{>{\PreserveBackslash\raggedright}p{#1}}
\newcommand{\cleqn}{\setcounter{equation}{0}}

\newcommand{\bq}{\begin{eqnarray}}
\newcommand{\nq}{\end{eqnarray}}
\allowdisplaybreaks

\makeatletter
\@addtoreset{equation}{section}
\makeatother

\begin{document}
\title{
\begin{flushright}
\ \\*[-80pt]
\begin{minipage}{0.2\linewidth}
\normalsize
\end{minipage}
\end{flushright}
{\Large \bf
Generalised CP and $\Delta (96)$ Family Symmetry
\\*[20pt]}}

\author{
Gui-Jun~Ding$^{1}$,  \
Stephen~F.~King$^{2}$ \
\\*[20pt]
\centerline{
\begin{minipage}{\linewidth}
\begin{center}
$^1${\it \normalsize
Department of Modern Physics, University of Science and Technology of China,\\
Hefei, Anhui 230026, China}\\
$^2${\it \normalsize
School of Physics and Astronomy,
University of Southampton,
Southampton, SO17 1BJ, U.K.}\\
\end{center}
\end{minipage}}
\\*[50pt]}
\vskip 2 cm
\date{\small
\centerline{ \bf Abstract}
\begin{minipage}{0.9\linewidth}
\medskip
We perform a comprehensive study of the $\Delta (96)$ family symmetry combined with the generalised CP symmetry $H_{\rm{CP}}$. We investigate the lepton mixing parameters which can be obtained from the original symmetry $\Delta (96)\rtimes H_{\rm{CP}}$ breaking to different remnant symmetries in the neutrino and charged lepton sectors, namely $G_{\nu}$ and $G_l$ subgroups in the neutrino and the charged lepton sector respectively, and the remnant CP symmetries from the breaking of $H_{\rm{CP}}$ are $H^{\nu}_{\rm{CP}}$ and $H^{l}_{\rm{CP}}$, respectively, where all cases correspond to a preserved symmetry smaller than the full Klein symmetry, as in the semi-direct approach, leading to predictions which depend on a single undetermined real parameter, which may be fitted to the reactor angle for example. We discuss 26 possible cases, including a global $\chi^2$ determination of the best fit parameters and the correlations between mixing parameters, in each case.
\end{minipage}
}

\begin{titlepage}
\maketitle
\thispagestyle{empty}
\end{titlepage}

\section{Introduction}
\cleqn

Following the measurement of the reactor mixing angle $\theta_{13}$ by the Daya Bay~\cite{An:2012eh}, RENO~\cite{Ahn:2012nd}, and Double Chooz~\cite{Abe:2011fz} reactor neutrino experiments, the three lepton
mixing angles $\theta_{12}$, $\theta_{23}$, $\theta_{13}$ and both
mass-squared differences $\Delta m^2_{sol}$ and $\Delta m^2_{atm}$ have been measured to reasonably good accuracy. However the Dirac CP phase and neutrino mass ordering have not been measured so far. If neutrinos are Majorana particles, there are two more Majorana CP phases which play a role in neutrinoless double-beta decay searches. Determining the neutrino mass ordering and measuring the Dirac and Majorana CP violating phases are primary goals of the next generation neutrino oscillation experiments. The CP violation has been firmly established in the quark sector and therefore it is natural to expect that CP violation occurs in the lepton sector as well. Indeed hints of a nonzero $\delta_{\rm{CP}}$ have begun to show up in global analysis of neutrino oscillation data~\cite{Tortola:2012te,GonzalezGarcia:2012sz,Capozzi:2013csa}.

In recent years, much effort has been devoted to explaining the structure of the lepton mixing angles through the introduction of discrete family symmetries. In this paradigm, one generally assumes a non-abelian discrete flavour group which is broken down to different subgroups in the neutrino and charged lepton sectors. The mismatch between these two subgroups gives rise to particular predictions for the lepton mixing angles. For recent reviews, please see~\cite{Altarelli:2010gt} for the model building and relevant group theory aspects. Inspired by the success of discrete family symmetry, it is conceivable to extend the family symmetry to include a generalised CP symmetry $H_{\rm{CP}}$ ~\cite{Ecker:1981wv,Grimus:1995zi} which will allow the prediction of both CP phases and mixing angles.

The idea of combining a family symmetry with a generalised CP symmetry has begun to be discussed in the literature. For example, the simple $\mu-\tau$ reflection symmetry, which exchanges a muon (tau) neutrino with a tau (muon) antineutrino in the charged lepton diagonal basis, has been discussed and successfully implemented in a number of models where both atmospheric mixing angle $\theta_{23}$ and Dirac CP phase $\delta_{\rm{CP}}$ were determined to be maximal~\cite{Harrison:2002kp,Grimus:2003yn,Farzan:2006vj}. The phenomenological consequences of imposing both an $S_4$ flavour symmetry and a generalised CP symmetry have been investigated in a model-independent way~\cite{Feruglio:2012cw,Ding:2013hpa,Li:2013jya}. All the three lepton mixing angles and CP phases are found to depend on only one free parameter for the symmetry breaking of $S_4\rtimes H_{\rm{CP}}$ to $Z_2\times \rm{CP}$ in the neutrino sector and to some abelian subgroup of $S_4$ in the charged lepton sector. Concrete $S_4$ family models with a generalised CP symmetry have been constructed in Refs.~\cite{Ding:2013hpa,Li:2013jya,Feruglio:2013hia,Luhn:2013lkn} where the spontaneous breaking of the $S_4\rtimes H_{\rm{CP}}$ down to $Z_2\times \rm{CP}$ in the neutrino sector was implemented. A similar generalised analysis has also been considered for $A_4$ family symmetry~\cite{Ding:2013bpa}. Other models with a family symmetry and a generalised CP symmetry can also be found in Refs.~\cite{Krishnan:2012me,Mohapatra:2012tb,Nishi:2013jqa,Ding:2013nsa}. The interplay between flavor symmetries and CP symmetries has been generally discussed in~\cite{Holthausen:2012dk,Chen:2014tpa}.
In addition, there are other theoretical approaches involving both family symmetry and CP violation~\cite{Branco:1983tn,Chen:2009gf,Antusch:2011sx,Girardi:2013sza}.
A generalised CP analysis of $\Delta(6n^2)$ has been performed recently~\cite{King:2014rwa} based on a direct approach with the full Klein symmetry $Z_2\times Z_2$ preserved in the neutrino sector and a $Z_3$ preserved in the charged lepton sector. Here we shall focus on $\Delta (96)$ and relax the requirement of having the full Klein symmetry.

In this paper, then, we study generalised CP symmetry in the context of $\Delta(96)$ where a CP symmetry is assumed to exist at a high energy scale. The generalised CP transformation compatible with an $\Delta(96)$ family symmetry is defined, and a model-independent analysis of the lepton mixing matrix is performed by scanning all the possible remnant subgroups in the neutrino and charged lepton sectors. Relaxing the requirement of having the full Klein symmetry in the neutrino sector given by a subgroup of $\Delta(96)$, as in the semi-direct approach we are led to a large number of possibilities
where the results depend on a single parameter, expressed as an angle which determines the reactor angle. We systematically discuss all such possibilities consistent with existing phenomenological data,
then analyse in detail the resulting predictions for mixing parameters, including a full discussion of correlations between parameters and a $\chi^2$ determination of the best fit point.

The remainder of this paper is organised as follows. In section~\ref{2} we discuss generalised CP with $\Delta(96)$.
In section~\ref{3} we perform a model independent CP analysis of $\Delta (96)$ subgroups, and categorise all the different possibilities for preserved flavour and CP symmetries in the neutrino
and charged lepton sectors. In section~\ref{4} we analyse the lepton mixing predictions arising from 26 different possible cases discussed in the previous section. We perform a $\chi^2$ analysis to determine the best fit to current data.
Section~\ref{5} concludes the paper. The details of the group theory of $\Delta(96)$ are collected in Appendix~\ref{sec:appendix_A}.

\section{Generalised CP with $\Delta(96)$}
\label{2}
\cleqn

It is non-trivial to define a CP transformation consistently in the presence of a family symmetry $G_f$. Generally the so-called consistency condition must be satisfied~\cite{Ecker:1981wv,Grimus:1995zi,Holthausen:2012dk}:
\begin{equation}
\label{eq:consistence}X\rho^{*}(g)X^{-1}=\rho(g^{\prime}),\quad g, g^{\prime} \in G_f\,,
\end{equation}
where $\rho(g)$ denotes the representation matrix for the group element $g$, $X$ is the generalised CP transformation, which maps a field $\varphi$ into
\begin{equation}
\varphi(t,\mathbf{x})\stackrel{CP}{\longrightarrow}X\,\varphi^{*}(t,-\mathbf{x})\,,
\end{equation}
where the obvious action of CP on the spinor indices has been suppressed for the case of $\varphi$ being spinor. Eq.~\eqref{eq:consistence} implies
that the generalised CP transformation $X$ maps the group element $g$ onto $g^{\prime}$ and the family group structure is preserved under this mapping. Because $X$ is unitary and therefore invertible, the generalized CP is an automorphism of the family symmetry group, and all the possible unitary matrices of $X$ forms a representation of the automorphism group. Given a solution $X$ to Eq.~\eqref{eq:consistence}, $\rho(h)X$ with any $h \in G_f$ also satisfies the consistency equation Eq.~\eqref{eq:consistence},
\begin{equation}
\big(\rho(h)X\big)\rho^{*}(g)\big(\rho(h)X\big)^{-1}=\rho(h)\big[X\rho^{*}(g)X^{-1}\big]\rho^{-1}(h)=\rho(hg^{\prime}h^{-1})\,,
\end{equation}
which indicates that the CP transformation $\rho(h)X$ maps the element $g$ into $hg^{\prime}h^{-1}$, and the corresponding automorphism is the composition of the automorphism of $X$ followed by an inner automorphism $\mathrm{conj}(h):g^{\prime}\rightarrow hg^{\prime}h^{-1}$. Therefore, when we investigate the groups of generalised CP transformations consistent with a family symmetry, it is sufficient to only consider the outer automorphism of $G_f$ with the inner automorphism modded out, since the inner automorphism doesn't impose any constraint.

For our family symmetry of interest $G_f=\Delta(96)$ in the present work, only the identity element commutes with all other elements and hence the inner automorphism group is isomorphic to $\Delta(96)$.
The group theory of $\Delta (96)$ is discussed in Appendix~\ref{sec:appendix_A}.
There is only one non-trivial outer automorphism (up to inner automorphism) $\mathfrak{u}$ with
\begin{equation}
\label{eq:outer_STU}S\stackrel{\mathfrak{u}}{\longrightarrow}S,\quad T\stackrel{\mathfrak{u}}{\longrightarrow}T^2,\quad U\stackrel{\mathfrak{u}}{\longrightarrow}U\,.
\end{equation}
Therefore the structure of the automorphism group of $\Delta(96)$ can be summarized as
\begin{eqnarray}
\begin{array}{ll}
\mathrm{Z}(\Delta(96))\cong Z_1,& \quad \mathrm{Aut}(\Delta(96))\cong\Delta(96)\rtimes Z_2,\\
\mathrm{Inn}(\Delta(96))\cong\Delta(96),& \quad \mathrm{Out}(\Delta(96))\cong Z_2=\left\{id,\mathfrak{u}\right\}\,,
\end{array}
\end{eqnarray}
where $\mathrm{Z}(\Delta(96))$, $\mathrm{Aut}(\Delta(96))$, $\mathrm{Inn}(\Delta(96))$ and $\mathrm{Out}(\Delta(96))$ denote the center, automorphism group, inner automorphism group and outer automorphism group of $\Delta(96)$ respectively. Under the action of $\mathfrak{u}$, another set of $\Delta(96)$ generators $a$, $b$, $c$ and $d$ defined in Eq.~\eqref{eq:generators} is mapped into
\begin{equation}
\label{eq:outer_abcd}a\stackrel{\mathfrak{u}}{\longrightarrow}a^2c^2d^2,\quad b\stackrel{\mathfrak{u}}{\longrightarrow}abc^2,\quad c\stackrel{\mathfrak{u}}{\longrightarrow}c^3d^3,\quad d\stackrel{\mathfrak{u}}{\longrightarrow}d\,.
\end{equation}
Applying this mapping, we see that the three-dimensional representations transform as
\begin{equation}
\label{eq:representation_outer}\mathbf{3}\leftrightarrow\mathbf{\overline{3}},\quad \mathbf{3^{\prime}}\leftrightarrow\mathbf{\overline{3}^{\prime}}\,.
\end{equation}
The other representations are not changed. This transformation is a symmetry of the character table shown in Table~\ref{tab:character_table}, if we exchange the conjugacy classes $3C_4\leftrightarrow3C^{\prime}_4$ and $12C_8\leftrightarrow12C^{\prime}_8$. As a result, the non-trivial CP  transformation of $\Delta(96)$ has to be a representation of $\mathfrak{u}$ in the sense of Eq.~\eqref{eq:consistence}, i.e.
\begin{equation}
X(\mathfrak{u})\rho^{*}(g)X^{-1}(\mathfrak{u})=\rho(\mathfrak{u}(g))\,.
\end{equation}
Notice that it is sufficient to only impose the consistency equation on the group's generators for discrete family symmetry:
\begin{eqnarray}
\nonumber&&X(\mathfrak{u})\rho^{*}(S)X^{-1}(\mathfrak{u})=\rho(\mathfrak{u}(S))=\rho(S),\\
\nonumber&&X(\mathfrak{u})\rho^{*}(T)X^{-1}(\mathfrak{u})=\rho(\mathfrak{u}(T))=\rho(T^2),\\
\label{eq:consistence_outer}&&X(\mathfrak{u})\rho^{*}(U)X^{-1}(\mathfrak{u})=\rho(\mathfrak{u}(U))=\rho(U)\,,
\end{eqnarray}
For the two-dimensional representation $\mathbf{2}$, we have
\begin{equation}
\rho^{*}_{\mathbf{2}}(S)=\rho_{\mathbf{2}}(S),\quad \rho^{*}_{\mathbf{2}}(T)=\rho_{\mathbf{2}}(T^2),\quad  \rho^{*}_{\mathbf{2}}(U)=\rho_{\mathbf{2}}(U)\,.
\end{equation}
Therefore the corresponding generalized CP transformation is of the form
\begin{equation}
\label{eq:cp_two_STU}X(\mathfrak{u})=X_{\mathbf{2}}(\mathfrak{u})\equiv\left(\begin{array}{cc}
1  &  0 \\
0  &  1
\end{array}
\right)\,,
\end{equation}
which represents the outer automorphism $\mathfrak{u}$ via $X_{\mathbf{2}}(\mathfrak{u})\rho^{*}_{\mathbf{2}}(g)X^{-1}_{\mathbf{2}}(\mathfrak{u})=\rho_{\mathbf{2}}(\mathfrak{u}(g))$. In the same way, for the three-dimensional representations $\mathbf{3}$, $\mathbf{3^{\prime}}$, $\mathbf{\overline{3}}$, $\mathbf{\overline{3}^{\prime}}$, $\mathbf{\widetilde{3}}$ and $\mathbf{\widetilde{3}^{\prime}}$, the CP transformation satisfying the consistency equation Eq.\eqref{eq:consistence_outer} is determined to be
\begin{equation}
\label{eq:cp_three1_STU}X_{\mathbf{3}}(\mathfrak{u})=X_{\mathbf{3^{\prime}}}(\mathfrak{u})=X_{\mathbf{\overline{3}}}(\mathfrak{u})
=X_{\mathbf{\overline{3}^{\prime}}}(\mathfrak{u})=X_{\mathbf{\widetilde{3}}}(\mathfrak{u})=X_{\mathbf{\widetilde{3}^{\prime}}}(\mathfrak{u})=
\begin{pmatrix}
 1 & 0 & 0 \\
 0 & 1 & 0 \\
 0 & 0 & 1
\end{pmatrix}\equiv\mathds{1}_{3\times3}\,.
\end{equation}
For the six-dimensional representation $\mathbf{6}$, the associated CP transformation $X_{\mathbf{6}}(\mathfrak{u})$ is
\begin{equation}
\label{eq:cp_six_STU}X_{\mathbf{6}}(\mathfrak{u})=\mathds{1}_{6\times6}\,.
\end{equation}
with again $X_{\mathbf{6}}(\mathfrak{u})\rho^{*}_{\mathbf{6}}(S)X^{-1}_{\mathbf{6}}(\mathfrak{u})=\rho_{\mathbf{6}}(S)$, $X_{\mathbf{6}}(\mathfrak{u})\rho^{*}_{\mathbf{6}}(T)X^{-1}_{\mathbf{6}}(\mathfrak{u})=\rho_{\mathbf{6}}(T^2)$ and $X_{\mathbf{6}}(\mathfrak{u})\rho^{*}_{\mathbf{6}}(U)X^{-1}_{\mathbf{6}}(\mathfrak{u})=\rho_{\mathbf{6}}(U)$.
The set of consistence equations in Eq.~\eqref{eq:consistence_outer} are trivially satisfied for the one dimensional representations $\mathbf{1}$ and $\mathbf{1^{\prime}}$, and we simply take
\begin{equation}
\label{eq:cp_one_STU}X_{\mathbf{1}}(\mathfrak{u})=X_{\mathbf{1^{\prime}}}(\mathfrak{u})=1\,.
\end{equation}
Therefore we conclude that the working basis listed in Table~\ref{tab:representation} is the so-called ``CP basis''. Including the inner automorphism (the family symmetry transforamtion), the most general CP transformation $H_{CP}$ consistent with $\Delta(96)$ family symmetry is given by
\begin{equation}
\rho_{\mathbf{r}}(h)X_{\mathbf{r}}(\mathfrak{u})=\rho_{\mathbf{r}}(h),\qquad h\in \Delta(96)\,,
\end{equation}
where $h$ can be any of the 96 group elements of $\Delta(96)$, and $\rho_{\mathbf{r}}(h)$ denotes the representation matrix of $h$ in the irreducible representation $\mathbf{r}$. Hence the generalised CP transformation consistent with the $\Delta(96)$ family symmetry is of the same form as the family group transformation in the chosen basis.

\section{Model independent CP analysis of $\Delta (96)$ subgroups}
\label{3}
\cleqn

\subsection{Subgroups of $\Delta (96)$}
In the notation of Appendix~\ref{sec:appendix_A}, one finds that
$\Delta(96)$ has fifteen $Z_2$ subgroups, sixteen $Z_3$ subgroups,  seven $K_4$ subgroups, twelve $Z_4$ subgroups and six $Z_8$ subgroups, which in terms of the generators $a$, $b$, $c$ and $d$,
can be expressed as follows:
\begin{itemize}[leftmargin=1.5em]
\item{$Z_2$ subgroups}
\begin{equation}
\nonumber\begin{array}{llll}
Z^{(1)}_2=\left\{1,c^2\right\},\quad\quad & Z^{(2)}_2=\left\{1,d^2\right\},\quad\quad & Z^{(3)}_2=\left\{1,c^2d^2\right\}, \quad\quad  & Z^{(4)}_2=\left\{1,ab\right\},\quad\quad \\
Z^{(5)}_2=\left\{1,abc\right\},\quad\quad & Z^{(6)}_2=\left\{1,abc^2\right\},\quad\quad & Z^{(7)}_2=\left\{1,abc^3\right\},\quad\quad & Z^{(8)}_2=\left\{1,a^2b\right\},\\
Z^{(9)}_2=\left\{1,a^2bd\right\},\quad\quad & Z^{(10)}_2=\left\{1,a^2bd^2\right\},\quad\quad & Z^{(11)}_2=\left\{1,a^2bd^3\right\},\quad\quad &  Z^{(12)}_2=\left\{1,b\right\},\\
Z^{(13)}_2=\left\{1,bcd\right\},\quad\quad &  Z^{(14)}_2=\left\{1,bc^2d^2\right\},\quad\quad & Z^{(15)}_2=\left\{1,bc^3d^3\right\},\quad\quad &
\end{array}\,.
\end{equation}
The first three $Z_2$ subgroups $Z^{(1)}_2$, $Z^{(2)}_2$ and $Z^{(3)}_2$ are related with each under the group conjugation, and the same holds to be true for the remaining $Z_2$ subgroups $Z^{(4)}_2\ldots Z^{(15)}_2$.
\item{$Z_3$ subgroups}
\begin{equation}
\nonumber\begin{array}{lll}
Z^{(1)}_3=\{1,a,a^2\},\quad\quad & Z^{(2)}_3=\{1,ac,a^2cd\},\quad\quad &  Z^{(3)}_3=\{1,ac^2,a^2c^2d^2\}, \\
Z^{(4)}_3=\{1,ac^3,a^2c^3d^3\}, \quad\quad & Z^{(5)}_3=\{1,ad,a^2c^3\}, \quad\quad & Z^{(6)}_3=\{1,ad^2,a^2c^2\}, \\
Z^{(7)}_3=\{1,ad^3,a^2c\}, \quad\quad & Z^{(8)}_3=\{1,acd,a^2d\}, \quad\quad &  Z^{(9)}_3=\{1,acd^2,a^2c^3d\}, \\
Z^{(10)}_3=\{1,acd^3,a^2c^2d\}, \quad\quad & Z^{(11)}_3=\{1,ac^2d,a^2cd^2\}, \quad\quad & Z^{(12)}_3=\{1,ac^2d^2,a^2d^2\}, \\
Z^{(13)}_3=\{1,ac^2d^3,a^2c^3d^2\}, \quad\quad & Z^{(14)}_3=\{1,ac^3d,a^2c^2d^3\}, \quad\quad & Z^{(15)}_3=\{1,ac^3d^2,a^2cd^3\}, \\ Z^{(16)}_3=\{1,ac^3d^3,a^2d^3\} &  &
\end{array}\,,
\end{equation}
which can be written compactly as
\begin{equation}
Z^{(x,y)}_3=\left\{1, ac^xd^y, a^2c^{x-y}d^x\right\},\quad x,y=0,1,2,3\,.
\end{equation}
All the above $Z_3$ subgroups are found to be conjugate to each other.
\item{$K_4$ subgroups}
\begin{equation}
\nonumber\begin{array}{lll}
K^{(1)}_4=\{1,c^2,d^2,c^2d^2\},\quad\quad & K^{(2)}_4=\{1,ab,c^2,abc^2\},\quad\quad & K^{(3)}_4=\{1,abc,c^2,abc^3\},\\
K^{(4)}_4=\{1,a^2b,d^2,a^2bd^2\},\quad\quad & K^{(5)}_4=\{1,a^2bd,d^2,a^2bd^3\},\quad\quad & K^{(6)}_4=\{1,b,c^2d^2,bc^2d^2\},\\
K^{(7)}_4=\{1,bcd,c^2d^2,bc^3d^3\}\,.
\end{array}
\end{equation}
Note that $K^{(1)}_4$ is a normal subgroup of $\Delta(96)$, and the remaining $K_4$ subgroups are conjugate to each other.
\item{$Z_4$ subgroups}
\begin{equation}
\nonumber\begin{array}{lll}
Z^{(1)}_4=\{1,cd^2,c^2,c^3d^2\}, \quad & Z^{(2)}_4=\{1,cd^3,c^2d^2,c^3d\}, \quad & Z^{(3)}_4=\{1,c^2d^3,d^2,c^2d\},\\
Z^{(4)}_4=\{1,c,c^2,c^3\}, \quad & Z^{(5)}_4=\{1,d,d^2,d^3\},\quad & Z^{(6)}_4=\{1,cd,c^2d^2,c^3d^3\},\\
Z^{(7)}_4=\{1,abd^2,c^2,abc^2d^2\}, \quad  & Z^{(8)}_4=\{1,abcd^2,c^2,abc^3d^2\}, \quad & Z^{(9)}_4=\{1,a^2bc^2,d^2,a^2bc^2d^2\},\\
Z^{(10)}_4=\{1,a^2bc^2d,d^2,a^2bc^2d^3\},\quad & Z^{(11)}_4=\{1,bc^2,c^2d^2,bd^2\}, \quad & Z^{(12)}_4=\{1,bc^3d,c^2d^2,bcd^3\}
\end{array}\,.
\end{equation}
The twelve $Z_4$ subgroups fall into three categories applying similarity transformations belonging to $\Delta(96)$: the first contains $Z^{1}_4$, $Z^{(2)}_4$ and $Z^{(3)}_4$, the second one $Z^{(4)}_4$, $Z^{(5)}_4$ and $Z^{(6)}_4$ and the third the others $Z^{(7)}_4\ldots Z^{(12)}_4$. The generating elements of the $Z_4$ subgroups $Z^{(1)}_4$, $Z^{(2)}_4$ and $Z^{(3)}_4$ have two degenerate eigenvalues, the lepton mixing matrix can not be determined uniquely if the flavor symmetry $\Delta(96)$ in broken to $Z^{(1)}_4$, $Z^{(2)}_4$ or $Z^{(3)}_4$ in the charged lepton sector. As a result, we don't consider these cases in the present work.
\item{$Z_8$ subgroups}
\begin{equation}
\nonumber\begin{array}{l}
Z^{(1)}_8=\{1,abd,cd^2,abcd^3,c^2,abc^2d,c^3d^2,abc^3d^3\},\\
Z^{(2)}_8=\{1,abcd,cd^2,abc^2d^3,c^2,abc^3d,c^3d^2,abd^3\},\\
Z^{(3)}_8=\{1,a^2bc^3,c^2d^3,a^2bcd^3,d^2,a^2bc^3d^2,c^2d,a^2bcd\},\\
Z^{(4)}_8=\{1,a^2bc^3d,c^2d^3,a^2bc,d^2,a^2bc^3d^3,c^2d,a^2bcd^2\},\\
Z^{(5)}_8=\{1,bc,cd^3,bc^2d^3,c^2d^2,bc^3d^2,c^3d,bd\},\\
Z^{(6)}_8=\{1,bc^2d,cd^3,bc^3,c^2d^2,bd^3,c^3d,bcd^2\}\,.
\end{array}
\end{equation}
All the six $Z_8$ subgroups are conjugate to each other.
\end{itemize}

\subsection{Leptonic Mixing from Remnant Symmetries}

Lepton mixing can be derived from a flavor symmetry $G_f$ and the generalised CP symmetry $H_{\rm{CP}}$ breaking to remnant symmetries in the charged lepton and neutrino sectors respectively. In concrete models, this is usually achieved via a spontaneous breaking using some scalar fields charged under this symmetry into different subgroups of the full symmetry group. The charge assignments are chosen such that there are different residual symmetries in the charged lepton and neutrino sectors. The misalignment between the two residual symmetries leads to particular predictions for the PMNS matrix. In this method, only the structure of full symmetry group and its remnant symmetries are assumed and we do not need to consider the breaking mechanism, i.e. how the required vacuum alignment achieving the remnant symmetries is dynamically realized. In the following, we assume that the family symmetry $G_f$ is spontaneously broken to the $G_{\nu}$ and $G_l$ subgroups in the neutrino and the charged lepton sector respectively, and the remnant CP symmetries from the breaking of $H_{CP}$ are $H^{\nu}_{CP}$ and $H^{l}_{CP}$,
respectively. The mismatch between the remnant symmetry groups $G_{\nu}\rtimes H^{\nu}_{CP}$ and $G_l\rtimes H^{l}_{CP}$ gives rise to particular values for both mixing angles and CP phases. As usual, the three generations of the left-handed (LH) lepton
doublets are unified into a three-dimensional representation
$\mathbf{3}$ of $G_f$. The same results would be obtained if the lepton doublets were assigned to $\mathbf{3^{\prime}}$ of $\Delta(96)$, since the representation $\mathbf{3^{\prime}}$ differs from $\mathbf{3}$ only in the overall sign of the generator $U$. Furthermore, if the LH lepton doublets are embedded into the $\Delta(96)$ triplets $\mathbf{\overline{3}}$ or $\mathbf{\overline{3}^{\prime}}$, the following predictions for the lepton mass matrices and the diagonalization matrices would become their complex conjugate. The invariance under the residual family
symmetries $G_{\nu}$ and $G_l$ implies that the neutrino mass matrix
$m_{\nu}$ and the charged lepton mass matrix $m_{l}$ satisfy
\begin{eqnarray}
\nonumber&&\rho^{T}_{\mathbf{3}}(g_{\nu_i})m_{\nu}\rho_{\mathbf{3}}(g_{\nu_
i})=m_{\nu},\quad g_{\nu_i}\in G_{\nu},\\
\label{eq:inv_family}&&\rho^{\dagger}_{\mathbf{3}}(g_{l_i})m^{\dagger}_{l}m_{l}\rho_{\mathbf{3}}(g_{l_i})=m^{\dagger}_{l}m_{l},\quad
g_{l_i}\in G_l\,,
\end{eqnarray}
where the charged lepton mass matrix $m_l$ is given in the so-called right-left convention, $l^{c}m_{l}l$, and $\rho_{\mathbf{3}}(g)$ denotes the representation matrix of the element $g$ in the irreducible representation $\mathbf{3}$. Furthermore, both mass matrices $m_{\nu}$ and $m_{l}$ are constrained by the residual CP symmetry as follows:
\begin{eqnarray}
\nonumber&&X^{T}_{\nu\mathbf{3}}m_{\nu}X_{\nu\mathbf{3}}=m^{*}_{\nu}, \qquad
\quad X_{\nu\mathbf{3}}\in H^{\nu}_{CP},\\
\label{eq:inv_CP}&&X^{\dagger}_{l\mathbf{3}}m^{\dagger}_{l}m_{l}X_{l\mathbf{3}}=(m^{\dagger}_{l}m_{l})^{*},\quad
X_{l\mathbf{3}}\in H^{l}_{CP}\,.
\end{eqnarray}
Since the theory still preserves both remnant family symmetry and remnant CP symmetries after symmetry breaking, they have to be compatible with each other, and the corresponding consistency equation should be fulfilled
\begin{eqnarray}
\nonumber&&X_{\nu\mathbf{r}}\rho^{*}_{\mathbf{r}}(g_{\nu_i})X^{-1}_{\nu\mathbf{r}}=\rho_{\mathbf{r}}(g_{\nu_j}),\qquad
g_{\nu_i},g_{\nu_j}\in G_{\nu},\\
\label{eq:consistency_remnant}&&X_{l\mathbf{r}}\rho^{*}_{\mathbf{r}}(g_{l_i})X^{-1}_{l\mathbf{r}}=\rho_{\mathbf{r}}(g_{l_j}),\qquad
g_{l_i},g_{l_j}\in G_{l}\,,
\end{eqnarray}
where $X_{\nu\mathbf{r}}$ and $X_{l\mathbf{r}}$ are the elements of $H^{\nu}_{CP}$ and $H^{l}_{CP}$, respectively. Given a set of solutions $X_{\nu\mathbf{r}}$ and $X_{l\mathbf{r}}$, we can
straightforwardly see that $\rho_{\mathbf{r}}(g_{\nu_i})X_{\nu\mathbf{r}}$
and $\rho_{\mathbf{r}}(g_{l_i})X_{l\mathbf{r}}$ are also solutions to the above consistency equations in Eq.~\eqref{eq:consistency_remnant}. The invariance conditions of Eqs.~(\ref{eq:inv_family})-(\ref{eq:inv_CP}) allow us to reconstruct the lepton mass matrices $m_{\nu}$ and $m^{\dagger}_lm_l$, and ultimately we can determine the lepton mixing matrix $U_{PMNS}$. Furthermore, if the residual family symmetries are another pair of subgroup $G^{\prime}_{\nu}$ and $G^{\prime}_{l}$ which are conjugate to $G_{\nu}$ and $G_{l}$ under the action of the group element $h\in G_f$, i.e.
\begin{equation}
 G^{\prime}_{\nu}=hG_{\nu}h^{-1},\qquad G^{\prime}_{l}=hG_{l}h^{-1}\,.
\end{equation}
Then the consistent residual CP symmetries $H^{\nu'}_{CP}$ and $H^{l'}_{CP}$ are related to $H^{\nu}_{CP}$ and $H^{l}_{CP}$ by
\begin{equation}
\label{eq:CP_conju}H^{\nu^{\prime}}_{CP}=\rho_{\mathbf{r}}(h)H^{\nu}_{CP}\rho^{T}_{\mathbf{r}}(h),\qquad
H^{l^{\prime}}_{CP}=\rho_{\mathbf{r}}(h)H^{l}_{CP}\rho^{T}_{\mathbf{r}}(h)\,,
\end{equation}
and the resulting neutrino and charged lepton mass matrices are of the form
\begin{equation}
\label{eq:mass_matr_conju}m^{\prime}_{\nu}=\rho^{*}_{\mathbf{3}}(h)m_{\nu}\rho^{\dagger}_{\mathbf{3}}(h),\qquad
m^{\prime\dagger}_{l}m^{\prime}_{l}=\rho_{\mathbf{3}}(h)m^{\dagger}_{l}m_{l}\rho^{\dagger}_{\mathbf{3}}(h).
\end{equation}
Hence the remnant subgroups $G^{\prime}_{\nu}$ and $G^{\prime}_{l}$ lead to the same predictions for the lepton mixing matrix $U_{PMNS}$ as $G_{\nu}$, $G_{l}$ case~\cite{Ding:2013bpa}.

Having completed a general discussion of the implementation of a generalised CP symmetry with a family symmetry, we now concentrate on the case of interest in which the family symmetry $G_f=\Delta(96)$ and a generalised CP symmetry $H_{\rm{CP}}$ consistent with $\Delta(96)$ is imposed. Thus, the theory respects the full symmetry $\Delta(96)\rtimes H_{\rm{CP}}$. In the following, we shall perform a model independent study of the constraints that these symmetries impose on the neutrino mass matrix, the charged lepton mass matrix and the PMNS matrix by scanning all the possible remnant symmetries $G^{\nu}_{\rm{CP}}\cong G_{\nu}\rtimes H^{\nu}_{\rm{CP}}$ and $G^{l}_{\rm{CP}}\cong G_{l}\rtimes H^{l}_{\rm{CP}}$. Note that all the possible lepton mixing patterns derived from $\Delta(96)$ family symmetry breaking has been completed by one of us in Ref.~\cite{Ding:2012xx}, where the generalised CP symmetry is not imposed. Other related work on $\Delta(96)$ flavor symmetry can be found in Refs.~\cite{Toorop:2011jn,King:2012in}. It is sufficient to consider only a small number of representative cases which leads to different results for mixing angles and CP phases, since different choices of $G_{\nu}$ and $G_{l}$ related by group conjugation generate the same result. We further restrict ourselves to the case of Majorana neutrinos, which implies that the remnant family symmetry $G_{\nu}$ can only be $K_4\cong Z_2\times Z_2$ or $Z_2$ subgroups. For the case of $G_{\nu}=K_4$, the lepton flavor mixing is completely fixed as shown in Ref.~\cite{Ding:2012xx}, and seven mass independent textures including the well-known tri-bimaximal, bimaximal and Toorop-Feruglio-Hagedorn (TFH) mixing patterns can be produced. In the following, we shall concentrate on the case of $G_{\nu}=Z_2$ and generalised CP symmetry is imposed. By considering all the possible family symmetry breaking, we find only six viable cases listed below.
\begin{itemize}[leftmargin=2em]
\item{$G_{l}=Z^{(2)}_{3}$,  $G_{\nu}=Z^{(2)}_{2}$}

\item{$G_{l}=Z^{(2)}_{3}$,  $G_{\nu}=Z^{(9)}_{2}$}

\item{$G_{l}=Z^{(2)}_{3}$,  $G_{\nu}=Z^{(10)}_{2}$}

\item{$G_{l}=K^{(3)}_{4}$,  $G_{\nu}=Z^{(9)}_{2}$}

\item{$G_{l}=Z^{(7)}_{4}$,  $G_{\nu}=Z^{(9)}_{2}$}

\item{$G_{l}=Z^{(1)}_{8}$,  $G_{\nu}=Z^{(9)}_{2}$}\,.

\end{itemize}
We begin this study with an analysis of the charged lepton sector.

\subsection{Charged lepton sector}

\subsubsection{$G_{l}=Z^{(2)}_{3}=\left\{1, ac, a^2cd\right\}$}
Now the full symmetry $\Delta(96)\rtimes H_{CP}$ is broken down to $G^{l}_{CP}\cong Z^{(2)}_{3}\rtimes H^{l}_{CP}$ in the charged lepton sector. The element $X_{l\mathbf{r}}$ of $H^{l}_{CP}$ should satisfy the consistency equation
\begin{equation}
X_{l\mathbf{r}}\rho^{*}_{\mathbf{r}}(ac)X^{-1}_{l\mathbf{r}}=\rho_{\mathbf{r}}(g^{\prime}),\quad g^{\prime}=ac, a^2cd\,.
\end{equation}
It is found that the remnant CP transformation $H^{l}_{CP}$ can be
\begin{equation}
H^{l}_{CP}=\left\{\rho_{\mathbf{r}}(1), \rho_{\mathbf{r}}(ac), \rho_{\mathbf{r}}(a^2cd), \rho_{\mathbf{r}}(a^2b), \rho_{\mathbf{r}}(bcd), \rho_{\mathbf{r}}(abc) \right\}\,.
\end{equation}
The charged lepton mass matrix $m_{l}$ must respect both the residual family symmetry $Z^{(2)}_{3}$ and the generalised CP symmetry $H^{l}_{CP}$, i.e.
\begin{subequations}
\begin{eqnarray}
\label{eq:charged_lepton_Z3_flavor}&& \rho^{\dagger}_{\mathbf{3}}(ac)m^{\dagger}_{l}m_{l}\rho_{\mathbf{3}}(ac)=m^{\dagger}_{l}m_{l}\,\\
\label{eq:charged_lepton_Z3_CP}&& X^{\dagger}_{l\mathbf{3}}m^{\dagger}_{l}m_{l}X_{l\mathbf{3}}=\left(m^{\dagger}_{l}m_{l}\right)^{*}\,.
\end{eqnarray}
\end{subequations}
Eq.~\eqref{eq:charged_lepton_Z3_flavor} implies that the $m^{\dagger}_{l}m_{l}$ is diagonal, i.e.,
\begin{equation}
m^{\dagger}_{l}m_{l}=\text{diag}(m^2_{e},m^{2}_{\mu},m^2_{\tau})\,,
\end{equation}
up to permutation of diagonal entries. For $X_{l\mathbf{r}}=\rho_{\mathbf{r}}(1), \rho_{\mathbf{r}}(ac), \rho_{\mathbf{r}}(a^2cd)$, the invariance condition of Eq.~\eqref{eq:charged_lepton_Z3_CP} is automatically satisfied, and no additional constraints are imposed. For the remaining values of $X_{l\mathbf{r}}=\rho_{\mathbf{r}}(a^2b), \rho_{\mathbf{r}}(bcd), \rho_{\mathbf{r}}(abc)$, the residual CP invariant condition of Eq.~\eqref{eq:charged_lepton_Z3_CP} implies $m_{e}=m_{\mu}$. Hence, this case is not viable phenomenologically.

\subsubsection{$G_{l}=K^{(3)}_4=\left\{1, abc, c^2, abc^3\right\}$}

The hermitian combination $m^{\dagger}_lm_{l}$ is constrained by the remnant family symmetry $K^{(3)}_4$ as
\begin{eqnarray}
\nonumber&&\rho^{\dagger}_{\mathbf{3}}(abc)m^{\dagger}_{l}m_{l}\rho_{\mathbf{3}}(abc)=m^{\dagger}_{l}m_{l}\,,\\
&&\rho^{\dagger}_{\mathbf{3}}(c^2)m^{\dagger}_{l}m_{l}\rho_{\mathbf{3}}(c^2)=m^{\dagger}_{l}m_{l}\,.
\end{eqnarray}
Then the most general charged lepton mass matrix satisfying these equations is of the form
\begin{equation}
\label{eq:charged_lepton_K4}m^{\dagger}_{l}m_{l}=\begin{pmatrix}
R_{11}    &  (1+i\sqrt{3})R_{12}     &    (1-i\sqrt{3}) R_{13}  \\
(1-i\sqrt{3})R_{12}     &   R_{11}   &   (1+i\sqrt{3}) R_{13}  \\
(1+i\sqrt{3}) R_{13}    &   (1-i\sqrt{3})R_{13}   &   R_{11}-2R_{12}+2R_{13}
\end{pmatrix}\,,
\end{equation}
where $R_{11}$, $R_{12}$ and $R_{13}$ are real parameters. It is diagonalized by the unitary matrix
\begin{equation}
\label{eq:Ul_K43}U_{l}=\begin{pmatrix}
\frac{1}{\sqrt{2}}e^{\pi i/3}   &   \frac{1}{\sqrt{3}}e^{2\pi i/3}   & -\frac{1}{\sqrt{6}} e^{2\pi i/3}  \\
\frac{1}{\sqrt{2}}     &    -\frac{1}{\sqrt{3}} e^{\pi i/3}  &   \frac{1}{\sqrt{6}} e^{\pi i/3}  \\
0    &   \frac{1}{\sqrt{3}}     &   \sqrt{\frac{2}{3}}
\end{pmatrix}\,,
\end{equation}
with
\begin{equation}
U^{\dagger}_{l}m^{\dagger}_{l}m_{l}U_{l}=\text{diag}\left(R_{11}+2R_{12}, R_{11}-2R_{12}-2R_{13}, R_{11}-2R_{12}+4R_{13}\right)\,.
\end{equation}
Note that the unitary matrix $U_{l}$ is determined up permutations of columns and phases of its column vectors, because the the charged lepton masses can not be predicted in the present approach. The same comment applies to the following cases with different remnant symmetry in the charged lepton sector. The mass matrix $m^{\dagger}_{l}m_{l}$ of Eq.~\eqref{eq:charged_lepton_K4} also respects the residual CP symmetry $H^{l}_{CP}$, which is determined by the so-called consistency equation:
\begin{equation}
X_{l\mathbf{r}}\rho^{*}_{\mathbf{r}}(abc)X^{-1}_{l\mathbf{r}}=\rho_{\mathbf{r}}(g^{\prime}_1),\qquad
X_{l\mathbf{r}}\rho^{*}_{\mathbf{r}}(c^2)X^{-1}_{l\mathbf{r}}=\rho_{\mathbf{r}}(g^{\prime}_2),\qquad g^{\prime}_1, g^{\prime}_2\in K^{(3)}_4\,.
\end{equation}
By considering all possible values for $g^{\prime}_1$ and $g^{\prime}_2$, we find that only 16 of the 96 non-trivial CP transformations are acceptable,
\begin{eqnarray}
\nonumber&&H^{l}_{CP}=\Big\{\rho_{\mathbf{r}}(a^2cd), \rho_{\mathbf{r}}(a^2b), \rho_{\mathbf{r}}(a^2c^3d^3), \rho_{\mathbf{r}}(a^2bc^2d^2), \rho_{\mathbf{r}}(a^2c^2), \rho_{\mathbf{r}}(a^2bcd^3), \rho_{\mathbf{r}}(a^2d^2), \rho_{\mathbf{r}}(a^2bc^3d), \\
&&~~\rho_{\mathbf{r}}(a^2), \rho_{\mathbf{r}}(a^2bc^3d^3), \rho_{\mathbf{r}}(a^2c^2d^2), \rho_{\mathbf{r}}(a^2bcd), \rho_{\mathbf{r}}(a^2bd^2), \rho_{\mathbf{r}}(a^2cd^3), \rho_{\mathbf{r}}(a^2bc^2), \rho_{\mathbf{r}}(a^2c^3d)\Big\}\,.
\end{eqnarray}
The invariance under the action of $H^{l}_{CP}$ yields
\begin{equation}
\label{eq:charged_lepton_K4_CP}X^{\dagger}_{l\mathbf{3}}m^{\dagger}_{l}m_{l}X_{l\mathbf{3}}=\left(m^{\dagger}_{l}m_{l}\right)^{*}\,,
\end{equation}
which further constrains the charged lepton mass matrix $m^{\dagger}_{l}m_{l}$ of Eq.~\eqref{eq:charged_lepton_K4} for different preserved CP transformations. We find that the 16 elements of $H^{l}_{CP}$ can be divided  into two classes. For the case of $X_{l\mathbf{r}}=\rho_{\mathbf{r}}(a^2cd)$, $\rho_{\mathbf{r}}(a^2b)$, $\rho_{\mathbf{r}}(a^2c^3d^3)$, $\rho_{\mathbf{r}}(a^2bc^2d^2)$, $\rho_{\mathbf{r}}(a^2c^2)$, $\rho_{\mathbf{r}}(a^2bcd^3)$, $\rho_{\mathbf{r}}(a^2d^2)$, $\rho_{\mathbf{r}}(a^2bc^3d)$, the remnant CP invariance condition of Eq.~\eqref{eq:charged_lepton_K4_CP} is satisfied, and no new constraint is generated. For the remaining case of $X_{l\mathbf{r}}=\rho_{\mathbf{r}}(a^2)$, $\rho_{\mathbf{r}}(a^2bc^3d^3)$, $\rho_{\mathbf{r}}(a^2c^2d^2)$, $\rho_{\mathbf{r}}(a^2bcd)$, $\rho_{\mathbf{r}}(a^2bd^2)$, $\rho_{\mathbf{r}}(a^2cd^3)$, $\rho_{\mathbf{r}}(a^2bc^2)$, $\rho_{\mathbf{r}}(a^2c^3d)$, the residual CP invariant condition of Eq.~\eqref{eq:charged_lepton_K4_CP} leads to the constraint $R_{12}=R_{13}$. As a result, the charged lepton masses are predicted to be partially degenerate with $m_{e}=m_{\tau}$. Hence this case is not viable phenomenologically.

\subsubsection{$G_{l}=Z^{(7)}_4=\left\{1, abd^2, c^2, abc^2d^2\right\}$}

The underlying symmetry $\Delta(96)\rtimes G_{CP}$ is broken down to $G^{l}_{CP}\cong Z^{(7)}_4\rtimes H^{l}_{CP}$ in this case, and the element $X_{l\mathbf{r}}$ of $H^{l}_{CP}$ fulfill
\begin{equation}
X_{l\mathbf{r}}\rho^{*}_{\mathbf{r}}(abd^2)X^{-1}_{l\mathbf{r}}=\rho_{\mathbf{r}}(g^{\prime}),\qquad \text{with}~~ g^{\prime}=abd^2, abc^2d^2\,.
\end{equation}
It is easy to check that the remnant CP symmetry $H^{l}_{CP}$ can take the value
\begin{eqnarray}
\nonumber&&H^{l}_{CP}=\Big\{\rho_{\mathbf{r}}(a^2bd^2), \rho_{\mathbf{r}}(a^2), \rho_{\mathbf{r}}(a^2bc^2), \rho_{\mathbf{r}}(a^2c^2d^2), \rho_{\mathbf{r}}(a^2c^3d), \rho_{\mathbf{r}}(a^2bcd), \rho_{\mathbf{r}}(a^2cd^3), \rho_{\mathbf{r}}(a^2bc^3d^3),\\
&&~~\rho_{\mathbf{r}}(a^2cd), \rho_{\mathbf{r}}(a^2bc^3d), \rho_{\mathbf{r}}(a^2c^3d^3), \rho_{\mathbf{r}}(a^2bcd^3), \rho_{\mathbf{r}}(a^2b), \rho_{\mathbf{r}}(a^2d^2), \rho_{\mathbf{r}}(a^2bc^2d^2), \rho_{\mathbf{r}}(a^2c^2)\Big\}\,.
\end{eqnarray}
The charged lepton mass matrix $m_{l}$ respects both the residual family symmetry $Z^{(7)}_4$ and the generalised CP symmetry $H^{l}_{CP}$, i.e.
\begin{subequations}
\begin{eqnarray}
\label{eq:charged_lepton_Z4_flavor}&&\rho^{\dagger}_{\mathbf{3}}(abd^2)m^{\dagger}_{l}m_{l}\rho_{\mathbf{3}}(abd^2)=m^{\dagger}_{l}m_{l}\,,\\
\label{eq:charged_lepton_Z4_CP}&&X^{\dagger}_{l\mathbf{3}}m^{\dagger}_{l}m_{l}X_{l\mathbf{3}}=\left(m^{\dagger}_{l}m_{l}\right)^{*}\,,
\end{eqnarray}
\end{subequations}
where Eq.~\eqref{eq:charged_lepton_Z4_flavor} is the invariance condition under $Z^{(7)}_4$, and it constrains the charged lepton mass matrix to take the following form
\begin{equation}
\label{eq:charged_lepton_Z4_mass_matrix}m^{\dagger}_{l}m_{l}=\begin{pmatrix}
R_{11}    &   (1+i\sqrt{3})R_{12}    &    (1-i\sqrt{3})R_{13}  \\
(1-i\sqrt{3})R_{12}   &  R_{11}    &   (1+i\sqrt{3})R_{13}  \\
(1+i\sqrt{3})R_{13}   &  (1-i\sqrt{3})R_{13}   &   R_{11}-2R_{12}+2R_{13}
\end{pmatrix}\,,
\end{equation}
where $R_{11}$, $R_{12}$ and $R_{13}$ are real. The unitary matrix $U_{l}$ which diagonalizes $m^{\dagger}_{l}m_{l}$ is then of the form
\begin{equation}
U_{l}=\begin{pmatrix}
\frac{1}{\sqrt{2}}e^{\pi i/3}   &   \frac{1}{\sqrt{3}} e^{2\pi i/3}   &   -\frac{1}{\sqrt{6}} e^{2\pi i/3}  \\
\frac{1}{\sqrt{2}}    &     -\frac{1}{\sqrt{3}} e^{\pi i/3}   &   \frac{1}{\sqrt{6}}e^{\pi i/3}  \\
0    &     \frac{1}{\sqrt{3}}    &     \sqrt{\frac{2}{3}}
\end{pmatrix}\,,
\end{equation}
with
\begin{equation}
U^{\dagger}_{l}m^{\dagger}_{l}m_{l}U_{l}=\text{diag}\left(R_{11}+2R_{12},R_{11}-2R_{12}-2R_{13}, R_{11}-2R_{12}+4R_{13}\right)\,.
\end{equation}
The charged lepton mass matrix is further constrained by the residual CP symmetry as Eq.~\eqref{eq:charged_lepton_Z4_CP}. For the values of $X_{l\mathbf{r}}=\rho_{\mathbf{r}}(a^2cd)$, $\rho_{\mathbf{r}}(a^2bc^3d)$, $\rho_{\mathbf{r}}(a^2c^3d^3)$, $\rho_{\mathbf{r}}(a^2bcd^3)$, $\rho_{\mathbf{r}}(a^2b)$, $\rho_{\mathbf{r}}(a^2d^2)$, $\rho_{\mathbf{r}}(a^2bc^2d^2)$, $\rho_{\mathbf{r}}(a^2c^2)$, it is easy to check that $m^{\dagger}_{l}m_{l}$ of Eq.~\eqref{eq:charged_lepton_Z4_mass_matrix} respects the CP invariant condition of Eq.~\eqref{eq:charged_lepton_Z4_CP}, and no new constraints are introduced. For the case of $X_{l\mathbf{r}}=\rho_{\mathbf{r}}(a^2bd^2)$, $\rho_{\mathbf{r}}(a^2)$, $\rho_{\mathbf{r}}(a^2bc^2)$, $\rho_{\mathbf{r}}(a^2c^2d^2)$, $\rho_{\mathbf{r}}(a^2c^3d)$, $\rho_{\mathbf{r}}(a^2bcd)$, $\rho_{\mathbf{r}}(a^2cd^3)$, $\rho_{\mathbf{r}}(a^2bc^3d^3)$, the equality $R_{12}=R_{13}$ is required to be fulfilled. As a consequence, the degeneracy $m_{e}=m_{\tau}$ arises, and therefore this case is not viable. Comparing the charged lepton mass matrix $m^{\dagger}_{l}m_{l}$ predicted in Eq.~\eqref{eq:charged_lepton_K4} and Eq.~\eqref{eq:charged_lepton_Z4_mass_matrix}, we see that the remnant family symmetries $G_{l}=K^{(3)}_4$ and $G_{l}=Z^{(7)}_4$ lead to the same constraints on the charged lepton mass, and hence the diagonalization matrix $U_{l}$ and the charged lepton masses are predicted to be of the same forms in both cases.

\subsubsection{$G_{l}=Z^{(1)}_{8}=\left\{1, abd, cd^2, abcd^3, c^2, abc^2d, c^3d^2, abc^3d^3\right\}$}

The remnant family symmetry $G_{l}=Z^{(1)}_{8}$ imposes the following constraint on the charged lepton mass matrix:
\begin{equation}
\rho^{\dagger}_{\mathbf{3}}(abd)m^{\dagger}_{l}m_{l}\rho_{\mathbf{3}}(abd)=m^{\dagger}_{l}m_{l}\,.
\end{equation}
The the mass matrix $m^{\dagger}_{l}m_{l}$ is determined to be of the form
\begin{equation}
m^{\dagger}_{l}m_{l}=
\begin{pmatrix}
\begin{smallmatrix}
R_{11}    &   (1+i\sqrt{3})R_{12}    &   (1-i\sqrt{3})R_{13}  \\
(1-i\sqrt{3})R_{12}    &  R_{11}-2(1-\sqrt{3})(R_{12}-R_{13})  &  -(1-\sqrt{3})(1+i\sqrt{3})R_{12}+(2-\sqrt{3})(1+i\sqrt{3})R_{13}  \\
(1+i\sqrt{3})R_{13}    &   -(1-\sqrt{3})(1-i\sqrt{3})R_{12}+(2-\sqrt{3})(1-i\sqrt{3})R_{13}   &  R_{11}-2(2-\sqrt{3})(R_{12}-R_{13})
\end{smallmatrix}
\end{pmatrix}\,,
\end{equation}
where $R_{11}$, $R_{12}$ and $R_{13}$ are real parameters. It is diagonalized by the following unitary transformation $U_l$
\begin{eqnarray}
\nonumber U_{l}&=&\frac{1}{2\sqrt{3}}\begin{pmatrix}
\sqrt{4+\sqrt{2}-\sqrt{6}}\;e^{2\pi i/3}   &   2e^{2\pi i/3}     &    -\sqrt{4-\sqrt{2}+\sqrt{6}}\;e^{2\pi i/3} \\
\sqrt{4+\sqrt{2}+\sqrt{6}}\;e^{\pi i/3}   &   -2e^{\pi i/3}  & \sqrt{4-\sqrt{2}-\sqrt{6}}\;e^{\pi i/3}  \\
\sqrt{4-2\sqrt{2}}     &    2      &     \sqrt{4+2\sqrt{2}}
\end{pmatrix}\\
\label{eq:Ul_Z81}&=&\frac{1}{\sqrt{3}}
\begin{pmatrix}
 \sqrt{2}\;e^{\frac{2i\pi}{3}}\sin\frac{5\pi}{24} & ~e^{\frac{2i\pi}{3}}~ & -\sqrt{2}\;e^{\frac{2i\pi}{3}}\cos\frac{5\pi}{24} \\
 \sqrt{2}\; e^{\frac{i\pi}{3}}\cos\frac{\pi}{24} &
   ~-e^{\frac{i\pi}{3}}~ & \sqrt{2}\; e^{\frac{i\pi}{3}}\sin\frac{\pi}{24} \\
 \sqrt{2}\;\sin\frac{\pi}{8} & ~1~ & \sqrt{2}\;\cos\frac{\pi}{8}
\end{pmatrix}\,,
\end{eqnarray}
with
\begin{equation}
U^{\dagger}_{l}m^{\dagger}_{l}m_{l}U_{l}=\text{daig}\left(m^2_{e}, m^2_{\mu}, m^2_{\tau}\right)\,,
\end{equation}
where
\begin{eqnarray}
\nonumber&& m^2_{e}=R_{11}+(-2+3\sqrt{2}+2\sqrt{3}-\sqrt{6})R_{12}+(4-3\sqrt{2}-2\sqrt{3}+\sqrt{6})R_{13}\\
\nonumber&& m^2_{\mu}=R_{11}-2R_{12}-2R_{13}\\
&&m^2_{\tau}=R_{11}+(-2-3\sqrt{2}+2\sqrt{3}+\sqrt{6})R_{12}+(4+3\sqrt{2}-2\sqrt{3}-\sqrt{6})R_{13}\,.
\end{eqnarray}
The mass matrix $m^{\dagger}_{l}m_{l}$ also respects the CP symmetry $H^{l}_{CP}$ which should be compatible with the remnant family symmetry $G_{l}=Z^{(1)}_{8}$, i.e.,
\begin{equation}
X_{l\mathbf{r}}\rho^{*}_{\mathbf{r}}(abd)X^{-1}_{l\mathbf{r}}=\rho_{\mathbf{r}}(g^{\prime}),\qquad g^{\prime}\in Z^{(1)}_{8}\,.
\end{equation}
One can straightforwardly obtain that there are 16 possible choices for $X_{l\mathbf{r}}$,
\begin{eqnarray}
\nonumber&&H^{l}_{CP}=\Big\{\rho_{\mathbf{r}}(a^2bc^2d), \rho_{\mathbf{r}}(a^2c^2), \rho_{\mathbf{r}}(a^2bc^3), \rho_{\mathbf{r}}(a^2c^3d^3), \rho_{\mathbf{r}}(a^2bd^3), \rho_{\mathbf{r}}(a^2d^2), \rho_{\mathbf{r}}(a^2bcd^2), \rho_{\mathbf{r}}(a^2cd),\\
&&~~\rho_{\mathbf{r}}(a^2bc^2d^3), \rho_{\mathbf{r}}(a^2c^2d^2), \rho_{\mathbf{r}}(a^2bc^3d^2), \rho_{\mathbf{r}}(a^2c^3d), \rho_{\mathbf{r}}(a^2bd),\rho_{\mathbf{r}}(a^2), \rho_{\mathbf{r}}(a^2bc), \rho_{\mathbf{r}}(a^2cd^3)\Big\}\,.
\end{eqnarray}
The invariance under the action of the remnant CP symmetry $H^{l}_{CP}$ implies that
\begin{equation}
X^{\dagger}_{l\mathbf{3}}m^{\dagger}_{l}m_{l}X_{l\mathbf{3}}=\left(m^{\dagger}_{l}m_{l}\right)^{*}\,.
\end{equation}
We find no additional constraint for $X_{l\mathbf{r}}=\rho_{\mathbf{r}}(a^2bc^2d)$, $\rho_{\mathbf{r}}(a^2c^2)$, $\rho_{\mathbf{r}}(a^2bc^3)$, $\rho_{\mathbf{r}}(a^2c^3d^3)$, $\rho_{\mathbf{r}}(a^2bd^3)$, $\rho_{\mathbf{r}}(a^2d^2)$, $\rho_{\mathbf{r}}(a^2bcd^2)$, $\rho_{\mathbf{r}}(a^2cd)$, while $X_{l\mathbf{r}}=\rho_{\mathbf{r}}(a^2bc^2d^3)$, $\rho_{\mathbf{r}}(a^2c^2d^2)$, $\rho_{\mathbf{r}}(a^2bc^3d^2)$, $\rho_{\mathbf{r}}(a^2c^3d)$, $\rho_{\mathbf{r}}(a^2bd)$, $\rho_{\mathbf{r}}(a^2)$, $\rho_{\mathbf{r}}(a^2bc)$, $\rho_{\mathbf{r}}(a^2cd^3)$ leads to $R_{12}=R_{13}$ such that the mass degeneracy $m_{e}=m_{\tau}$ follows.

\subsection{Neutrino sector}

\subsubsection{$G_{\nu}=Z^{(2)}_2=\left\{1, d^2\right\}$}

The symmetry $\Delta(96)\rtimes H_{CP}$ is spontaneously broken to $G^{\nu}_{CP}=Z^{(2)}_2\times H^{\nu}_{CP}$ in the neutrino sector. The residual CP symmetry $H^{\nu}_{CP}$ should be consistent with the residual family symmetry $G_{\nu}=Z^{(2)}_2$, and therefore its element $X_{\nu\mathbf{r}}$ has to fulfill the consistency equation
\begin{equation}
X_{\nu\mathbf{r}}\rho^{*}_{\mathbf{r}}(d^2)X^{-1}_{\nu\mathbf{r}}=\rho_{\mathbf{r}}(d^2)\,.
\end{equation}
We find that 32 generalised CP transformations are acceptable,
\begin{equation}
\label{eq:residual_CP_Z22}H^{\nu}_{CP}=\left\{\rho_{\mathbf{r}}(c^{m}d^{n}), \rho_{\mathbf{r}}(a^2bc^{m}d^{n})|m, n=0,1,2,3\right\}\,.
\end{equation}
We can construct the light neutrino mass matrix $m_{\nu}$ from its invariance under both the remnant family symmetry $Z^{(2)}_2$ and the remnant CP symmetry $H^{\nu}_{CP}$ as follows:
\begin{subequations}
\begin{eqnarray}
\label{eq:neutrino_z22_flavor}&&\rho^{T}_{\mathbf{3}}(d^2)m_{\nu}\rho_{\mathbf{3}}(d^2)=m_{\nu}\,,\\
\label{eq:neutrino_z22_CP}&&X^{T}_{\nu\mathbf{3}}m_{\nu}X_{\nu\mathbf{3}}=m^{*}_{\nu}\,.
\end{eqnarray}
\end{subequations}
The most general neutrino mass matrix satisfying Eq.~\eqref{eq:neutrino_z22_flavor} is of the form
\begin{equation}
\label{eq:neutrino_matrix_Z22}m_{\nu}=\alpha\begin{pmatrix}
2   & -1   &  -1   \\
-1  &  2   &  -1  \\
-1  &  -1  &  2
\end{pmatrix}+\beta\begin{pmatrix}
2  &   1   &  0  \\
1  &   0   &  2   \\
0  &   2   &  1
\end{pmatrix}+\gamma\begin{pmatrix}
0  &  1   &   2  \\
1  &  2   &   0 \\
2  &  0   &   1
\end{pmatrix}+\delta\begin{pmatrix}
0  &  1  &  0  \\
1  &  0  &  0 \\
0  &  0  &  1
\end{pmatrix}\,,
\end{equation}
where $\alpha$, $\beta$, $\gamma$ and $\delta$ are complex parameters, and they are further constrained by the remnant CP symmetry as shown in Eq.~\eqref{eq:neutrino_z22_CP}. In order to diagonalize this neutrino mass matrix, it is useful to first perform a THF transformation $U_{THF}$ and yield
\begin{equation}
m^{\prime}_{\nu}=U^{T}_{TFH}m_{\nu}U_{TFH}=\begin{pmatrix}
3\alpha+\sqrt{3}\;(\beta-\gamma)   &    0    &    \delta  \\
0     &   3\beta+3\gamma+\delta     &    0   \\
\delta   &    0    &     3\alpha-\sqrt{3}\;(\beta-\gamma)
\end{pmatrix}\,,
\end{equation}
where
\begin{equation}
U_{TFH}=\frac{1}{6}\begin{pmatrix}
-3-\sqrt{3}   &   ~2\sqrt{3}~      &    3-\sqrt{3}  \\
3-\sqrt{3}   &   ~2\sqrt{3}~       &  -3-\sqrt{3}  \\
2\sqrt{3}     &  ~2\sqrt{3}~       &   2\sqrt{3}
\end{pmatrix}\,.
\end{equation}
$m^{\prime}_{\nu}$ can be further diagonalized by a unitary matrix $U^{\prime}_{\nu}$ as
\begin{equation}
U^{\prime T}_{\nu}m^{\prime}_{\nu} U^{\prime}_{\nu}=\text{diag}(m_1,m_2,m_3)\,.
\end{equation}
As we shall show in the following, $U^{\prime}_{\nu}$ can be written into the form
\begin{equation}
\label{eq:unup_z22}U^{\prime}_{\nu}=\mathds{U}R(\theta)P\,,  
\end{equation}
where $\mathds{U}$ is a constant unitary matrix such that $\mathds{U}^{T}m^{\prime}_{\nu}\mathds{U}$ becomes a real matrix, and $R(\theta)$ is a rotation matrix in the (1,3) sector with
\begin{equation}
R(\theta)=\begin{pmatrix}
\cos\theta   &    0   &  \sin\theta  \\
0   &   1   &   0   \\
-\sin\theta   &   0   &   \cos\theta
\end{pmatrix}\,.
\end{equation}
Finally, the unitary matrix $P$ is diagonal with entries $\pm1$ and $\pm i$ which encode the CP parity of the neutrino states and renders the light neutrino masses $m_{1,2,3}$ positive. Hence the neutrino mass matrix $m_{\nu}$ in Eq.~\eqref{eq:neutrino_matrix_Z22} is diagonalized by the the unitary matrix $U_{\nu}$ as
\begin{equation}
U^{T}_{\nu}m_{\nu}U_{\nu}=\text{diag}(m_1,m_2,m_3)\,,
\end{equation}
with
\begin{equation}
U_{\nu}=U_{TFH}\mathds{U}R(\theta)P\,.
\end{equation}
Notice that the neutrino diagonalization matrix $U_{\nu}$ is fixed up to permutations of the columns, since neutrino masses are unconstrained in the present framework. Now we turn to investigate the implication of the remnant CP invariant condition of Eq.~\eqref{eq:neutrino_z22_CP}. The predictions for the neutrino diagonalization matrix $U_{\nu}$ and the light neutrino masses would be presented for different residual CP transformations in Eq.~\eqref{eq:residual_CP_Z22}.
\begin{itemize}[leftmargin=*]

\item{$X_{\nu\mathbf{r}}=\rho_{\mathbf{r}}(1), \rho_{\mathbf{r}}(d^2)$}

In this case, the residual CP invariant requirement of Eq.~\eqref{eq:neutrino_z22_CP} leads to the constraint
\begin{equation}
\text{Im}\alpha=\text{Im}\beta=\text{Im}\gamma=\text{Im}\delta=0\,,
\end{equation}
where $``\text{Im}"$ denotes the imaginary part, and hence all the four parameters $\alpha$, $\beta$, $\gamma$ and $\delta$ are real. The unitary transformation $\mathds{U}$ is a unit matrix, i.e.
\begin{equation}
\mathds{U}=\mathds{1}_{3\times3}\,.
\end{equation}
Therefore the neutrino diagonalization matrix $U_{\nu}$ is
\begin{equation}
U_{\nu}=\frac{1}{\sqrt{3}}
\begin{pmatrix}
-\sqrt{2}\cos\left(\frac{\pi}{12}-\theta\right) &  ~1~ &   \sqrt{2}\sin\left(\frac{\pi}{12}-\theta\right) \\
\sqrt{2}\sin\left(\frac{\pi}{12}+\theta\right)  &  ~1~ &   -\sqrt{2}\cos\left(\frac{\pi}{12}+\theta\right) \\
\sqrt{2}\cos\left(\frac{\pi}{4}+\theta\right)   &  ~1~ &  \sqrt{2}\sin\left(\frac{\pi}{4}+\theta\right)
\end{pmatrix}P\,,
\end{equation}
where the rotation angle $\theta$ is determined to be
\begin{equation}
\tan2\theta=\frac{\text{Re}\delta}{\sqrt{3}\left(\text{Re}\gamma-\text{Re}\beta\right)}\,.
\end{equation}
Finally the light neutrino masses are
\begin{eqnarray}
\nonumber&&m_1=\left|3\text{Re}\alpha+\text{sign}\left(\left(\text{Re}\beta-\text{Re}\gamma\right)\cos2\theta\right)\sqrt{3\left(\text{Re}\beta-\text{Re}\gamma\right)^2+\left(\text{Re}\delta\right)^2}\right|\,,\\
\nonumber&&m_2=\left|3\text{Re}\beta+3\text{Re}\gamma+\text{Re}\delta\right|\,,\\
&&m_3=\left|3\text{Re}\alpha-\text{sign}\left(\left(\text{Re}\beta-\text{Re}\gamma\right)\cos2\theta\right)\sqrt{3\left(\text{Re}\beta-\text{Re}\gamma\right)^2+\left(\text{Re}\delta\right)^2}\right|\,.
\end{eqnarray}

\item{$X_{\nu\mathbf{r}}=\rho_{\mathbf{r}}(c^2d), \rho_{\mathbf{r}}(c^2d^3)$}

In this case, the parameters $\alpha$, $\beta$, $\gamma$ and $\delta$ are constrained to satisfy
\begin{equation}
\text{Re}\alpha=\text{Re}\delta=0, \quad \text{Re}\gamma=\text{Re}\beta, \quad \text{Im}\delta=-3\left(\text{Im}\beta+\text{Im}\gamma\right)\,,
\end{equation}
where $``\text{Re}"$ denotes the real part. We find that the unitary transformation is given by
\begin{equation}
\mathds{U}=\text{diag}(e^{i\pi/4}, 1, e^{i\pi/4})\,.
\end{equation}
The resulting neutrino diagonalization $U_{\nu}$ matrix reads
\begin{equation}
U_{\nu}=\frac{1}{\sqrt{3}}
\begin{pmatrix}
-\sqrt{2}\,e^{i\pi/4}\cos\left(\frac{\pi}{12}-\theta\right) & ~~1~~ &   \sqrt{2}\,e^{i\pi/4}\sin\left(\frac{\pi}{12}-\theta\right) \\
\sqrt{2}\,e^{i\pi/4}\sin\left(\frac{\pi}{12}+\theta\right)  & ~~1~~ &   -\sqrt{2}\,e^{i\pi/4}\cos\left(\frac{\pi}{12}+\theta\right) \\
\sqrt{2}\,e^{i\pi/4}\cos\left(\frac{\pi}{4}+\theta\right)   & ~~1~~ &  \sqrt{2}\,e^{i\pi/4}\sin\left(\frac{\pi}{4}+\theta\right)
\end{pmatrix}\,,
\end{equation}
where the trivial phase matrix $P$ has been omitted here, and it would be neglected as well in the following cases. The angle $\theta$ is
\begin{equation}
\tan2\theta=\frac{\sqrt{3}\,\left(\text{Im}\beta+\text{Im}\gamma\right)}{\text{Im}\beta-\text{Im}\gamma}\,.
\end{equation}
The light neutrino mass are predicted to be
\begin{eqnarray}
\nonumber&&\hskip-0.35in m_1=\left|3\text{Im}\alpha+\text{sign}\left(\left(\text{Im}\beta-\text{Im}\gamma\right)\cos2\theta\right)2\sqrt{3\left(\text{Im}\alpha\right)^2+3\text{Im}\beta\text{Im}\gamma+3\left(\text{Im}\gamma\right)^2}\right|\,,\\
\nonumber&&\hskip-0.35in  m_2=6\left|\text{Re}\beta\right|\,,\\
&&\hskip-0.35in m_3=\left|3\text{Im}\alpha-\text{sign}\left(\left(\text{Im}\beta-\text{Im}\gamma\right)\cos2\theta\right)2\sqrt{3\left(\text{Im}\alpha\right)^2+3\text{Im}\beta\text{Im}\gamma+3\left(\text{Im}\gamma\right)^2}\right|\,.
\end{eqnarray}

\item{$X_{\nu\mathbf{r}}=\rho_{\mathbf{r}}(a^2b), \rho_{\mathbf{r}}(a^2bd^2)$}

This residual CP symmetry implies that
\begin{equation}
\text{Re}\gamma=\text{Re}\beta,\quad \text{Im}\alpha=\text{Im}\delta=0,\quad \text{Im}\gamma=-\text{Im}\beta\,.
\end{equation}
The unitary transformation $\mathds{U}$ takes the form
\begin{equation}
\mathds{U}=\frac{1}{\sqrt{2}}
\begin{pmatrix}
 1 & ~0~ & i \\
 0 & ~\sqrt{2}~~ & 0 \\
 -1 & ~0~ & i
\end{pmatrix}\,.
\end{equation}
The light neutrino mass matrix is diagonalized by
\begin{equation}
U_{\nu}=\frac{1}{\sqrt{6}}
\begin{pmatrix}
-\sqrt{3}\cos\theta +i\sin\theta &   ~~\sqrt{2}~~  & -\sqrt{3}\sin\theta-i\cos\theta \\
\sqrt{3}\cos\theta+i\sin\theta   &   ~~\sqrt{2}~~  & \sqrt{3}\sin\theta-i\cos\theta \\
-2i\sin\theta  & ~~\sqrt{2}~~ & 2i\cos\theta
\end{pmatrix}\,,
\end{equation}
with
\begin{equation}
\tan2\theta=\frac{2\text{Im}\beta}{\sqrt{3}\;\text{Re}\alpha}\,.
\end{equation}
The light neutrino masses are given by
\begin{eqnarray}
\nonumber&&m_1=\left|\text{Re}\delta-\text{sign}\left(\text{Re}\alpha\cos2\theta\right)\sqrt{9\left(\text{Re}\alpha\right)^2+12\left(\text{Im}\beta\right)^2}\right|\,,\\
\nonumber&&m_1=\left|6\text{Re}\beta+\text{Re}\delta\right|\,,\\
&&m_3=\left|\text{Re}\delta+\text{sign}\left(\text{Re}\alpha\cos2\theta\right)\sqrt{9\left(\text{Re}\alpha\right)^2+12\left(\text{Im}\beta\right)^2}\right|\,.
\end{eqnarray}

\item{$X_{\nu\mathbf{r}}=\rho_{\mathbf{r}}(a^2bd), \rho_{\mathbf{r}}(a^2bd^3)$}

The invariance of the neutrino mass matrix under the residual CP transformation leads to
\begin{equation}
\text{Re}\delta=\text{Im}\alpha=0,\quad \text{Im}\gamma=\text{Im}\beta,\quad \text{Im}\delta=-6\text{Im}\beta\,.
\end{equation}
The unitary transformation $\mathds{U}$ is of the form
\begin{equation}
\mathds{U}=\text{diag}\left(1,1,i\right)\,.
\end{equation}
The corresponding neutrino diagonalization matrix is given by
\begin{equation}
U_{\nu}=\frac{1}{2\sqrt{3}}
\begin{pmatrix}
-\sqrt{3}\;e^{i\theta}-e^{-i\theta}  & ~~2~~ & i  \left(\sqrt{3}\;e^{i\theta}-e^{-i\theta}\right) \\
\sqrt{3}\;e^{i\theta}-e^{-i\theta}   & ~~2~~ & -i \left(\sqrt{3}\;e^{i\theta}+e^{-i\theta}\right) \\
 2 e^{-i \theta } & ~~2~~ & 2 i e^{-i \theta }
\end{pmatrix}\,,
\end{equation}
where the rotation angle $\theta$ is
\begin{equation}
\tan2\theta=-\frac{2\text{Im}\beta}{\text{Re}\alpha}\,.
\end{equation}
The light neutrino masses in this cases are
\begin{eqnarray}
\nonumber&&m_1=\left|\sqrt{3}\left(\text{Re}\beta-\text{Re}\gamma\right)+3\text{sign}\left(\text{Re}\alpha\cos2\theta\right)\sqrt{\left(\text{Re}\alpha\right)^2+4\left(\text{Im}\beta\right)^2}\right|\,,\\
\nonumber&&m_2=3\left|\text{Re}\beta+\text{Re}\gamma\right|\,,\\
&&m_3=\left|\sqrt{3}\left(\text{Re}\beta-\text{Re}\gamma\right)-3\text{sign}\left(\text{Re}\alpha\cos2\theta\right)\sqrt{\left(\text{Re}\alpha\right)^2+4\left(\text{Im}\beta\right)^2}\right|\,.
\end{eqnarray}

\item{$X_{\nu\mathbf{r}}=\rho_{\mathbf{r}}(a^2bc), \rho_{\mathbf{r}}(a^2bcd^2)$}

The parameter $\alpha$, $\beta$, $\gamma$ and $\delta$ are constrained to satisfy
\begin{eqnarray}
\nonumber&\text{Re}\delta=-3\left(\text{Re}\beta+\text{Re}\gamma\right),\quad \text{Im}\alpha=\text{Re}\alpha\,,\\
&\text{Im}\gamma=\text{Im}\beta-\sqrt{3}\;\left(\text{Re}\beta+\text{Re}\gamma\right),\quad \text{Im}\delta=-\sqrt{3}\;\left(\text{Re}\beta-\text{Re}\gamma\right)\,.
\end{eqnarray}
The unitary transformation $\mathds{U}$ takes the form
\begin{equation}
\mathds{U}=\begin{pmatrix}
e^{\frac{7i\pi}{8}} \cos\frac{\pi}{8} & 0 & e^{\frac{3i\pi}{8}}\sin\frac{\pi }{8} \\
 0 & e^{\frac{i\pi}{4}} & 0 \\
-e^{\frac{7i\pi}{8}} \sin\frac{\pi}{8} & 0 & e^{\frac{3i\pi}{8}}\cos\frac{\pi }{8}
\end{pmatrix}\,.
\end{equation}
Therefore the neutrino diagonalization matrix $U_{\nu}$ is
{\small
\begin{equation}
U_{\nu}=\frac{1}{\sqrt{3}}
\begin{pmatrix}
e^{\frac{i\pi}{8}}\cos\left(\frac{\pi}{24}-\theta\right)-e^{\frac{5i\pi}{8}}\cos\left(\frac{\pi}{24}+\theta\right) &  ~e^{\frac{i\pi}{4}}~ & -e^{\frac{i\pi}{8}}\sin\left(\frac{\pi}{24}-\theta\right)-e^{\frac{5i\pi}{8}}\sin\left(\frac{\pi}{24} +\theta\right) \\
-e^{\frac{i\pi}{8}}\sin\left(\frac{5\pi}{24}-\theta\right)+e^{\frac{5i\pi}{8}}\sin\left(\frac{5\pi}{24}+\theta\right) &  ~e^{\frac{i\pi}{4}}~ & -e^{\frac{i\pi}{8}}\cos\left(\frac{5\pi}{24}-\theta\right)-e^{\frac{5i\pi}{8}}\cos\left(\frac{5\pi}{24}+\theta\right)\\
-e^{\frac{i\pi}{8}}\sin\left(\frac{\pi}{8}+\theta\right)+e^{\frac{5i\pi}{8}}\sin\left(\frac{\pi}{8}-\theta\right) &   ~e^{\frac{i\pi}{4}}~ & e^{\frac{i\pi}{8}}\cos\left(\frac{\pi}{8}+\theta\right)+e^{\frac{5i\pi}{8}}\cos\left(\frac{\pi}{8}-\theta\right)
\end{pmatrix}\,,
\end{equation}
}
with
\begin{equation}
\tan2\theta=\frac{\left(\sqrt{3}-1\right)\text{Re}\beta+\left(\sqrt{3}+1\right)\text{Re}\gamma}{-\sqrt{6}\;\text{Re}\alpha}\,.
\end{equation}
The light neutrino masses are determined to be
{\footnotesize\begin{eqnarray*}
\nonumber&&\hskip-0.2in m_1=\left|(3+\sqrt{3})\text{Re}\beta+(3-\sqrt{3})\text{Re}\gamma+\text{sign}\left(\text{Re}\alpha\cos2\theta\right)\sqrt{18(\text{Re}\alpha)^2+\left((3-\sqrt{3})\text{Re}\beta+(3+\sqrt{3})\text{Re}\gamma\right)^2}\right|\,,\\
\nonumber&&\hskip-0.2in m_2=\left|2\sqrt{3}\left(2\text{Re}\beta+\text{Re}\gamma\right)-6\text{Im}\beta\right|\,,\\
\nonumber&&\hskip-0.2in
m_3=\left|(3+\sqrt{3})\text{Re}\beta+(3-\sqrt{3})\text{Re}\gamma-\text{sign}\left(\text{Re}\alpha\cos2\theta\right)\sqrt{18(\text{Re}\alpha)^2+\left((3-\sqrt{3})\text{Re}\beta+(3+\sqrt{3})\text{Re}\gamma\right)^2}\right|\,.
\end{eqnarray*}}

\item{$X_{\nu\mathbf{r}}=\rho_{\mathbf{r}}(a^2bcd), \rho_{\mathbf{r}}(a^2bcd^3)$}

The invariance under the residual CP symmetry implies that
\begin{eqnarray}
\nonumber&\text{Re}\delta=-3\left(\text{Re}\beta+\text{Re}\gamma\right),\quad \text{Im}\alpha=\text{Re}\alpha\,,\\
&\text{Im}\gamma=\text{Im}\beta+\sqrt{3}\left(\text{Re}\beta+\text{Re}\gamma\right),\quad \text{Im}\delta=\sqrt{3}\left(\text{Re}\beta-\text{Re}\gamma\right)\,.
\end{eqnarray}
The unitary transformation $\mathds{U}$ is found to be
\begin{equation}
\mathds{U}=\begin{pmatrix}
e^{\frac{7i\pi}{8}}\sin\frac{\pi}{8} & 0 &  e^{\frac{3i\pi}{8}}\cos\frac{\pi}{8} \\
0 & e^{\frac{i\pi}{4}} & 0 \\
-e^{\frac{7i\pi}{8}}\cos\frac{\pi}{8} & 0 &  e^{\frac{3i\pi}{8}}\sin\frac{\pi}{8}
\end{pmatrix}\,.
\end{equation}
The resulting neutrino diagonalization matrix reads
{\small\begin{equation}
U_{\nu}=\frac{1}{\sqrt{3}}
\begin{pmatrix}
e^{\frac{i\pi}{8}}\sin\left(\frac{5\pi}{24}+\theta\right)-e^{\frac{5i\pi}{8}}\sin\left(\frac{5\pi}{24}-\theta\right) &  ~e^{\frac{i\pi}{4}}~ & -e^{\frac{i\pi}{8}}\cos\left(\frac{5\pi}{24}+\theta\right)-e^{\frac{5i\pi}{8}}\cos\left(\frac{5\pi}{24}-\theta\right)\\
-e^{\frac{i\pi}{8}}\cos\left(\frac{\pi}{24}+\theta\right)+e^{\frac{5i\pi}{8}}\cos\left(\frac{\pi}{24}-\theta\right)  &  ~e^{\frac{i\pi}{4}}~ & -e^{\frac{i\pi}{8}}\sin\left(\frac{\pi}{24}+\theta\right)-e^{\frac{5i\pi}{8}}\sin\left(\frac{\pi}{24}-\theta\right)\\
e^{\frac{i\pi}{8}}\sin\left(\frac{\pi}{8}-\theta\right)-e^{\frac{5i\pi}{8}}\sin\left(\frac{\pi}{8}+\theta\right)     &  ~e^{\frac{i\pi}{4}}~ & e^{\frac{i\pi}{8}}\cos\left(\frac{\pi}{8}-\theta\right)+e^{\frac{5i\pi}{8}}\cos\left(\frac{\pi}{8}+\theta\right)
\end{pmatrix}\,,
\end{equation}
}
with
\begin{equation}
\tan2\theta=\frac{(\sqrt{3}+1)\text{Re}\beta+(\sqrt{3}-1)\text{Re}\gamma}{\sqrt{6}\;\text{Re}\alpha}\,.
\end{equation}
The light neutrino masses are given by
{\footnotesize\begin{eqnarray*}
\nonumber&&\hskip-0.2in m_1=\left|(3-\sqrt{3})\text{Re}\beta+(3+\sqrt{3})\text{Re}\gamma+\text{sign}\left(\text{Re}\alpha\cos2\theta\right)\sqrt{18(\text{Re}\alpha)^2+\left((3+\sqrt{3})\text{Re}\beta+(3-\sqrt{3})\text{Re}\gamma\right)^2}\right|\,,\\
\nonumber&&\hskip-0.2in m_2=\left|2\sqrt{3}\left(2\text{Re}\beta+\text{Re}\gamma\right)+6\text{Im}\beta\right|\,,\\
\nonumber&&\hskip-0.2in m_3=\left|(3-\sqrt{3})\text{Re}\beta+(3+\sqrt{3})\text{Re}\gamma-\text{sign}\left(\text{Re}\alpha\cos2\theta\right)\sqrt{18(\text{Re}\alpha)^2+\left((3+\sqrt{3})\text{Re}\beta+(3-\sqrt{3})\text{Re}\gamma\right)^2}\right|\,.
\end{eqnarray*}
}

\item{$X_{\nu\mathbf{r}}=\rho_{\mathbf{r}}(a^2bc^2), \rho_{\mathbf{r}}(a^2bc^2d^2)$}

In this case, the parameters $\alpha$, $\beta$, $\gamma$ and $\delta$ are constrained to satisfy
\begin{equation}
\text{Re}\alpha=\text{Im}\delta=0,\quad \text{Re}\gamma=\text{Re}\beta,\quad \text{Im}\gamma=-\text{Im}\beta\,.
\end{equation}
The unitary transformation $\mathds{U}$ is given by
\begin{equation}
\mathds{U}=\begin{pmatrix}
 e^{\frac{i\pi}{4}} & 0 & 0 \\
 0 & 1 & 0 \\
 0 & 0 & e^{\frac{3 i\pi}{4}}
\end{pmatrix}\,.
\end{equation}
Hence the light neutrino mass matrix is diagonalized by
\begin{equation}
U_{\nu}=\frac{1}{2\sqrt{3}}
\begin{pmatrix}
-e^{\frac{i\pi}{4}}\left(\sqrt{3}\;e^{i\theta}+e^{-i\theta}\right) & ~2~ &  e^{\frac{3i\pi}{4}}\left(\sqrt{3}\;e^{i\theta}-e^{-i\theta}\right)\\
e^{\frac{i\pi}{4}}\left(\sqrt{3}\;e^{i\theta}-e^{-i\theta}\right)  & ~~2~~ &   -e^{\frac{3i\pi}{4}}\left(\sqrt{3}\;e^{i\theta}+e^{-i\theta}\right)
   \\
2e^{i\left(\frac{\pi}{4}-\theta\right)} & ~2~ & 2 e^{i\left(\frac{3\pi}{4}-\theta\right)}
\end{pmatrix}\,,
\end{equation}
where
\begin{equation}
\tan2\theta=-\frac{\text{Re}\delta}{3\text{Im}\alpha}\,.
\end{equation}
The light neutrino masses are
\begin{eqnarray}
\nonumber&&m_1=\left|2\sqrt{3}\;\text{Im}\beta+\text{sign}(\text{Im}\alpha\cos2\theta)\sqrt{9\left(\text{Im}\alpha\right)^2+\left(\text{Re}\delta\right)^2}\right|\,,\\
\nonumber&&m_2=\left|6\text{Re}\beta+\text{Re}\delta\right|\,,\\
&&m_3=\left|2\sqrt{3}\;\text{Im}\beta-\text{sign}(\text{Im}\alpha\cos2\theta)\sqrt{9\left(\text{Im}\alpha\right)^2+\left(\text{Re}\delta\right)^2}\right|\,.
\end{eqnarray}

\item{$X_{\nu\mathbf{r}}=\rho_{\mathbf{r}}(a^2bc^2d), \rho_{\mathbf{r}}(a^2bc^2d^3)$}

We find that the following relations should be satisfied in this case,
\begin{equation}
\text{Re}\alpha=\text{Re}\delta=0, \quad \text{Im}\gamma=\text{Im}\beta,\quad \text{Im}\delta=-6\text{Im}\beta\,.
\end{equation}
The unitary transformation $\mathds{U}$ is given by
\begin{equation}
\mathds{U}=\frac{1}{\sqrt{2}}
\begin{pmatrix}
 e^{-\frac{i\pi}{4}} & 0 & e^{\frac{i\pi}{4}} \\
 0 & ~\sqrt{2}~~ & 0 \\
 -e^{-\frac{i\pi}{4}} & 0 & e^{\frac{i\pi}{4}}
\end{pmatrix}\,.
\end{equation}
Hence the neutrino diagonalization matrix $U_{\nu}$ is
\begin{equation}
U_{\nu}=\frac{1}{\sqrt{6}}
\begin{pmatrix}
 e^{\frac{i\pi}{4}} \left(\sin\theta+i\sqrt{3} \cos\theta\right) & ~\sqrt{2}~ & ~-e^{\frac{i\pi}{4}}\left(\cos\theta-i \sqrt{3}\sin\theta\right) \\
 e^{\frac{i\pi}{4}} \left(\sin\theta-i\sqrt{3}\cos\theta\right)  & ~\sqrt{2}~ &   ~-e^{\frac{i\pi}{4}}\left(\cos\theta+i\sqrt{3}\sin\theta\right) \\
 -2 e^{\frac{i\pi}{4}} \sin\theta & ~\sqrt{2}~ &
   2 e^{\frac{i\pi}{4}} \cos\theta
\end{pmatrix}\,,
\end{equation}
with
\begin{equation}
\tan2\theta=\frac{\text{Re}\gamma-\text{Re}\beta}{\sqrt{3}\;\text{Im}\alpha}\,.
\end{equation}
The light neutrino masses are
\begin{eqnarray}
\nonumber&&m_1=\left|6\text{Im}\beta+\text{sign}\left(\text{Im}\alpha\cos2\theta\right)\sqrt{3\left(\text{Re}\beta-\text{Re}\gamma\right)^2+9\left(\text{Im}\alpha\right)^2}\right|\,,\\
\nonumber&&m_2=3\left|\text{Re}\beta+\text{Re}\gamma\right|\,,\\
&&m_3=\left|6\text{Im}\beta-\text{sign}\left(\text{Im}\alpha\cos2\theta\right)\sqrt{3\left(\text{Re}\beta-\text{Re}\gamma\right)^2+9\left(\text{Im}\alpha\right)^2}\right|\,.
\end{eqnarray}

\item{$X_{\nu\mathbf{r}}=\rho_{\mathbf{r}}(a^2bc^3), \rho_{\mathbf{r}}(a^2bc^3d^2)$}

The remnant CP symmetry leads to
\begin{eqnarray}
\nonumber& \text{Re}\delta=-3\left(\text{Re}\beta+\text{Re}\gamma\right),\quad  \text{Im}\alpha=-\text{Re}\alpha\,\\
& \text{Im}\gamma=\text{Im}\beta-\sqrt{3}\left(\text{Re}\beta+\text{Re}\gamma\right),\quad \text{Im}\delta=-\sqrt{3}\left(\text{Re}\beta-\text{Re}\gamma\right)\,.
\end{eqnarray}
The unitary transformation $\mathds{U}$ is
\begin{equation}
\mathds{U}=
\begin{pmatrix}
e^{\frac{5i\pi}{8}}\sin\frac{\pi}{8}  & 0 & e^{\frac{i\pi}{8}}\cos\frac{\pi}{8} \\
 0 & e^{\frac{i\pi}{4}} & 0 \\
-e^{\frac{5i\pi}{8}}\cos\frac{\pi}{8} & 0 & e^{\frac{i\pi}{8}}\sin\frac{\pi}{8}
\end{pmatrix}\,.
\end{equation}
Hence the light neutrino mass matrix is diagonalized by
{\small
\begin{equation}
U_{\nu}=\frac{1}{\sqrt{3}}
\begin{pmatrix}
e^{-\frac{i\pi}{8}}\sin\left(\frac{5\pi}{24}+\theta\right)-e^{\frac{3i\pi}{8}}\sin\left(\frac{5\pi}{24}-\theta\right) &  ~e^{\frac{i\pi}{4}}~ & -e^{-\frac{i\pi}{8}}\cos\left(\frac{5\pi}{24}+\theta\right)-e^{\frac{3i\pi}{8}}\cos\left(\frac{5\pi}{24}-\theta\right)\\

-e^{-\frac{i\pi}{8}}\cos\left(\frac{\pi}{24}+\theta\right)+e^{\frac{3i\pi}{8}}\cos\left(\frac{\pi}{24}-\theta\right) &   ~e^{\frac{i\pi}{4}}~ & -e^{-\frac{i\pi}{8}}\sin\left(\frac{\pi}{24}+\theta\right)-e^{\frac{3i\pi}{8}}\sin\left(\frac{\pi}{24}-\theta\right)\\

e^{-\frac{i\pi}{8}}\sin\left(\frac{\pi}{8}-\theta\right)-e^{\frac{3i\pi}{8}}\sin\left(\frac{\pi}{8}+\theta\right) &   ~e^{\frac{i\pi}{4}}~ & e^{-\frac{i\pi}{8}}\cos\left(\frac{\pi}{8}-\theta\right)+e^{\frac{3i\pi}{8}}\cos\left(\frac{\pi}{8}+\theta\right)
\end{pmatrix}\,,
\end{equation}}
where
\begin{equation}
\tan2\theta=\frac{(1+\sqrt{3})\text{Re}\beta+(\sqrt{3}-1)\gamma}{-\sqrt{6}\;\text{Re}\alpha}\,.
\end{equation}
The light neutrino masses are determined to be
{\footnotesize\begin{eqnarray*}
\nonumber&&\hskip-0.2in m_1=\left|(3-\sqrt{3})\text{Re}\beta+(3+\sqrt{3})\text{Re}\gamma+\text{sign}\left(\text{Re}\alpha\cos2\theta\right)\sqrt{18(\text{Re}\alpha)^2+\left((3+\sqrt{3})\text{Re}\beta+(3-\sqrt{3})\text{Re}\gamma\right)^2}\right|\,,\\
\nonumber&&\hskip-0.2in m_2=\left|2\sqrt{3}\left(2\text{Re}\beta+\text{Re}\gamma\right)-6\text{Im}\beta\right|\,,\\
\nonumber&&\hskip-0.2in m_3=\left|(3-\sqrt{3})\text{Re}\beta+(3+\sqrt{3})\text{Re}\gamma-\text{sign}\left(\text{Re}\alpha\cos2\theta\right)\sqrt{18(\text{Re}\alpha)^2+\left((3+\sqrt{3})\text{Re}\beta+(3-\sqrt{3})\text{Re}\gamma\right)^2}\right|\,.
\end{eqnarray*}}

\item{$X_{\nu\mathbf{r}}=\rho_{\mathbf{r}}(a^2bc^3d), \rho_{\mathbf{r}}(a^2bc^3d^3)$}

The invariance of the neutrino mass matrix under the remnant CP transformations leads to
\begin{eqnarray}
\nonumber& \text{Re}\delta=-3\left(\text{Re}\beta+\text{Re}\gamma\right),\quad \text{Im}\alpha=-\text{Re}\alpha\,,\\
&\text{Im}\gamma=\text{Im}\beta+\sqrt{3}\left(\text{Re}\beta+\text{Re}\gamma\right),\quad \text{Im}\delta=\sqrt{3}\left(\text{Re}\beta-\text{Re}\gamma\right)\,.
\end{eqnarray}
The unitary transformation $\mathds{U}$ is
\begin{equation}
\mathds{U}=
\begin{pmatrix}
e^{\frac{5i\pi}{8}}\cos\frac{\pi}{8} & 0 &   e^{\frac{i\pi}{8}}\sin\frac{\pi}{8} \\
0 & e^{\frac{i\pi}{4}} & 0 \\
-e^{\frac{5i\pi}{8}}\sin\frac{\pi}{8} & 0 &   e^{\frac{i\pi}{8}}\cos\frac{\pi}{8}
\end{pmatrix}\,.
\end{equation}
The neutrino diagonalization matrix is of the form
{\small\begin{equation}
U_{\nu}=\frac{1}{\sqrt{3}}
\begin{pmatrix}
e^{-\frac{i\pi}{8}}\cos\left(\frac{\pi}{24}-\theta\right)-e^{\frac{3i\pi}{8}}\cos\left(\frac{\pi}{24}+\theta\right) &   e^{\frac{i\pi}{4}} & -e^{-\frac{i\pi}{8}}\sin\left(\frac{\pi}{24}-\theta\right)-e^{\frac{3i\pi}{8}}\sin\left(\frac{\pi}{24}+\theta\right)\\

-e^{-\frac{i\pi}{8}}\sin\left(\frac{5\pi}{24}-\theta\right)+e^{\frac{3i\pi}{8}}\sin\left(\frac{5\pi}{24}+\theta\right) & e^{\frac{i\pi}{4}} &  -e^{-\frac{i\pi}{8}}\cos\left(\frac{5\pi}{24}-\theta\right)-e^{\frac{3i\pi}{8}}\cos\left(\frac{5\pi}{24}+\theta\right)\\

-e^{-\frac{i\pi}{8}}\sin\left(\frac{\pi}{8}+\theta\right)+e^{\frac{3i\pi}{8}}\sin\left(\frac{\pi}{8}-\theta\right) &  e^{\frac{i\pi}{4}} &   e^{-\frac{i\pi}{8}}\cos\left(\frac{\pi}{8}+\theta\right)+e^{\frac{3i\pi}{8}}\cos\left(\frac{\pi}{8}-\theta\right)
\end{pmatrix}\,,
\end{equation}}
with
\begin{equation}
\tan2\theta=\frac{\left(\sqrt{3}-1\right)\text{Re}\beta+\left(1+\sqrt{3}\right)\text{Re}\gamma}{\sqrt{6}\;\text{Re}\alpha}\,.
\end{equation}
In the end, the light neutrino masses are
{\footnotesize\begin{eqnarray*}
\nonumber&&\hskip-0.2in m_1=\left|(3+\sqrt{3})\text{Re}\beta+(3-\sqrt{3})\text{Re}\gamma+\text{sign}\left(\text{Re}\alpha\cos2\theta\right)\sqrt{18(\text{Re}\alpha)^2+\left((3-\sqrt{3})\text{Re}\beta+(3+\sqrt{3})\text{Re}\gamma\right)^2}\right|\,,\\
\nonumber&&\hskip-0.2in m_2=\left|2\sqrt{3}\left(2\text{Re}\beta+\text{Re}\gamma\right)+6\text{Im}\beta\right|\,,\\
\nonumber&&\hskip-0.2in m_3=\left|(3+\sqrt{3})\text{Re}\beta+(3-\sqrt{3})\text{Re}\gamma-\text{sign}\left(\text{Re}\alpha\cos2\theta\right)\sqrt{18(\text{Re}\alpha)^2+\left((3-\sqrt{3})\text{Re}\beta+(3+\sqrt{3})\text{Re}\gamma\right)^2}\right|\,.
\end{eqnarray*}}
For the remaining twelve remnant CP transformations $X_{\nu\mathbf{r}}=\rho_{\mathbf{r}}(d)$, $\rho_{\mathbf{r}}(d^3)$, $\rho_{\mathbf{r}}(c)$, $\rho_{\mathbf{r}}(cd^2)$, $\rho_{\mathbf{r}}(cd)$, $\rho_{\mathbf{r}}(cd^3)$, $\rho_{\mathbf{r}}(c^2)$, $\rho_{\mathbf{r}}(c^2d^2)$, $\rho_{\mathbf{r}}(c^3)$, $\rho_{\mathbf{r}}(c^3d^2)$, $\rho_{\mathbf{r}}(c^3d)$ and $\rho_{\mathbf{r}}(c^3d^3)$, the light neutrino masses are predicted to be partially degenerate, i.e., two of the light neutrinos are of the same masses, therefore these cases are not phenomenologically viable.

\end{itemize}

\subsubsection{\label{subsubsec:Gnu_z29}$G_{\nu}=Z^{(9)}_2=\left\{1, a^2bd\right\}$}

In this scenario, the residual CP symmetry $H^{\nu}_{CP}$, which should be consistent with the residual family symmetry $G_{\nu}=Z^{(9)}_2$, and is determined by the consistency equation
\begin{equation}
X_{\nu\mathbf{r}}\rho^{*}_{\mathbf{r}}(a^2bd)X^{-1}_{\nu\mathbf{r}}=\rho_{\mathbf{r}}(a^2bd)\,.
\end{equation}
One can easily check that there are 8 possible choices for $X_{\nu\mathbf{r}}$, i.e.,
\begin{equation}
\label{eq:residual_CP_Z29} H^{\nu}_{CP}=\left\{\rho_{\mathbf{r}}(1), \rho_{\mathbf{r}}(d^2), \rho_{\mathbf{r}}(c^2d), \rho_{\mathbf{r}}(c^2d^3), \rho_{\mathbf{r}}(a^2bd), \rho_{\mathbf{r}}(a^2bc^2), \rho_{\mathbf{r}}(a^2bd^3), \rho_{\mathbf{r}}(a^2bc^2d^2)\right\}\,.
\end{equation}
The light neutrino mass matrix $m_{\nu}$ is constrained by the remnant family and remnant CP symmetries as
\begin{subequations}
\begin{eqnarray}
\label{eq:neutrino_z29_flavor}&&\rho^{T}_{\mathbf{3}}(a^2bd)m_{\nu}\rho_{\mathbf{3}}(a^2bd)=m_{\nu}\,,\\
\label{eq:neutrino_z29_CP}&&X^{T}_{\nu\mathbf{3}}m_{\nu}X_{\nu\mathbf{3}}=m^{*}_{\nu}\,.
\end{eqnarray}
\end{subequations}
The most general neutrino mass matrix, which is invariant under the residual family symmetry $G_{\nu}=Z^{(9)}_2$ and satisfies Eq.~\eqref{eq:neutrino_z22_flavor}, takes the following form
{\small
\begin{equation}
\label{eq:neutrino_matrix_Z29}m_{\nu}=\alpha\begin{pmatrix}
2   & -1   &  -1   \\
-1  &  2   &  -1  \\
-1  &  -1  &  2
\end{pmatrix}+\beta\begin{pmatrix}
1  &   1   &  1  \\
1  &   1   &  1   \\
1  &   1   &  1
\end{pmatrix}+\gamma\begin{pmatrix}
1  &  0   &   -1  \\
0  &  -1   &  1 \\
-1  &  1   &   0
\end{pmatrix}+\delta\begin{pmatrix}
2\left(2+\sqrt{3}\right)  &  0  &  -1  \\
0   &  2  &  -2-\sqrt{3} \\
-1  & -2-\sqrt{3}  &  0
\end{pmatrix}\,,
\end{equation}}
where $\alpha$, $\beta$, $\gamma$ and $\delta$ are complex parameters, and they are further constrained by the remnant CP symmetry as shown in Eq.~\eqref{eq:neutrino_z29_CP}. After perform the THF transformation $U_{THF}$, we have
\begin{eqnarray}
\nonumber\hskip-0.3in m^{\prime}_{\nu}&=&U^{T}_{TFH}m_{\nu}U_{TFH}\\
\hskip-0.3in &=&
\begin{pmatrix}
3\alpha+\sqrt{3}\;\gamma+\left(3+\sqrt{3}\right)\delta  & ~~-\left(3+\sqrt{3}\right)\delta~  & 0 \\

-\left(3+\sqrt{3}\right)\delta  & ~~3\beta~  & 0 \\

 0 & ~~0~ & 3\alpha -\sqrt{3}\; \gamma+\left(3+\sqrt{3}\right) \delta
\end{pmatrix}\,.
\end{eqnarray}
$m^{\prime}_{\nu}$ can be further diagonalized by a (1,2) rotation $U^{\prime}_{\nu}$,
\begin{equation}
U^{\prime T}_{\nu}m^{\prime}_{\nu} U^{\prime}_{\nu}=\text{diag}(m_1,m_2,m_3)\,.
\end{equation}
Hence the light neutrino mass matrix is diagonalized by the unitary matrix $U_{\nu}$ with
\begin{equation}
U_{\nu}=U_{TFH}U^{\prime}_{\nu}\,.
\end{equation}
Analogous to the previous case, the neutrino diagonalization matrix $U_{\nu}$ is fixed up to permutations of the columns. In the following, we shall discuss the constraints of the different remnant CP transformation shown in Eq.~\eqref{eq:residual_CP_Z29} one by one.

\begin{itemize}[leftmargin=*]

\item{$X_{\nu\mathbf{r}}=\rho_{\mathbf{r}}(1), \rho_{\mathbf{r}}(a^2bd)$}

In this case, the neutrino mass matrix is constrained to be real such that we have
\begin{equation}
\text{Im}\alpha=\text{Im}\beta=\text{Im}\gamma=\text{Im}\delta=0\,.
\end{equation}
Then the unitary matrix $U^{\prime}_{\nu}$ becomes a real rotation matrix with
\begin{equation}
U^{\prime}_{\nu}=
\begin{pmatrix}
\cos\theta &   ~\sin\theta~ & 0 \\
-\sin\theta &   ~\cos\theta~  & 0 \\
 0 & ~0~ & 1
\end{pmatrix}\,,
\end{equation}
where the diagonal phase matrix $P$, which renders the light neutrino mass positive, has been omitted here and hereinafter. The rotation angle $\theta$ is given by
\begin{equation}
\tan2\theta=\frac{2(1+\sqrt{3})\text{Re}\delta}{\sqrt{3}\left(\text{Re}\alpha-\text{Re}\beta\right)+\text{Re}\gamma+(1+\sqrt{3})\text{Re}\delta}\,.
\end{equation}
As a result, the neutrino mass matrix is diagonalized by
{\small\begin{equation}
U_{\nu}=\frac{1}{2\sqrt{3}}
\begin{pmatrix}
 -\left(1+\sqrt{3}\right)\cos\theta-2\sin\theta   &    ~-\left(1+\sqrt{3}\right)\sin\theta+2\cos\theta~ &
   \sqrt{3}-1 \\

\left(\sqrt{3}-1\right)\cos\theta-2\sin\theta & ~\left(\sqrt{3}-1\right)\sin\theta+2\cos\theta~ &   -1-\sqrt{3} \\

2\sqrt{2}\cos\left(\frac{\pi}{4}+\theta\right) &   ~2\sqrt{2}\cos\left(\frac{\pi}{4}-\theta\right)~ & 2
\end{pmatrix}\,.
\end{equation}}
The light neutrino masses are determined to be
\begin{eqnarray}
\nonumber&&m_1=\frac{1}{2}\left|a_{11}+a_{22}-\text{sign}\left(\left(a_{22}-a_{11}\right)\cos2\theta\right)\sqrt{\left(a_{22}-a_{11}\right)^2+4a^2_{12}}\right|\,,\\
\nonumber&&m_2=\frac{1}{2}\left|a_{11}+a_{22}+\text{sign}\left(\left(a_{22}-a_{11}\right)\cos2\theta\right)\sqrt{\left(a_{22}-a_{11}\right)^2+4a^2_{12}}\right|\,,\\
&&m_3=\left|3\text{Re}\alpha-\sqrt{3}\;\text{Re}\gamma+(3+\sqrt{3})\text{Re}\delta\right|\,,
\end{eqnarray}
where
\begin{equation}
a_{11}=3\text{Re}\alpha+\sqrt{3}\;\text{Re}\gamma+(3+\sqrt{3})\text{Re}\delta,\quad a_{22}=3\text{Re}\beta,\quad a_{12}=-(3+\sqrt{3})\text{Re}\delta\,.
\end{equation}

\item{$X_{\nu\mathbf{r}}=\rho_{\mathbf{r}}(d^2), \rho_{\mathbf{r}}(a^2bd^3)$}

The invariance of the neutrino mass matrix under residual CP symmetry leads to
\begin{equation}
\text{Re}\delta=\text{Im}\beta=\text{Im}\gamma=0,\quad \text{Im}\delta=\frac{1}{2}\left(-3+\sqrt{3}\right)\text{Im}\alpha\,.
\end{equation}
The unitary transformation $U^{\prime}_{\nu}$ diagonalizing the neutrino mass matrix $m^{\prime}_{\nu}$ is of the form
\begin{equation}
U^{\prime}_{\nu}=
\begin{pmatrix}
\cos\theta    & ~\sin\theta~   & 0 \\
-i\sin\theta  & ~i\cos\theta~ & 0 \\
0 & ~0~ & 1
\end{pmatrix}\,.
\end{equation}
Therefore the neutrino diagonalization matrix $U_{\nu}$ is
{\small\begin{equation}
U_{\nu}=\frac{1}{2\sqrt{3}}
\begin{pmatrix}
-\left(1+\sqrt{3}\right)\cos\theta-2i\sin\theta  &  ~-\left(1+\sqrt{3}\right)\sin\theta+2i\cos\theta~ &   \sqrt{3}-1 \\

\left(\sqrt{3}-1\right)\cos\theta-2i\sin\theta  &  ~\left(\sqrt{3}-1\right)\sin\theta+2i\cos\theta~  &  -1-\sqrt{3}\\
 2 e^{-i \theta } & ~2ie^{-i\theta}~ & 2
\end{pmatrix}\,.
\end{equation}}
The light neutrino masses are
\begin{eqnarray}
\nonumber&&m_1=\frac{1}{2}\left|a_{11}+a_{22}-\text{sign}\left(\left(a_{22}-a_{11}\right)\cos2\theta\right)\sqrt{\left(a_{22}-a_{11}\right)^2+4a^2_{12}}\right|\,,\\
\nonumber&&m_2=\frac{1}{2}\left|a_{11}+a_{22}+\text{sign}\left(\left(a_{22}-a_{11}\right)\cos2\theta\right)\sqrt{\left(a_{22}-a_{11}\right)^2+4a^2_{12}}\right|\,,\\
&&m_3=\left|3\text{Re}\alpha-\sqrt{3}\;\text{Re}\gamma\right|\,,
\end{eqnarray}
where
\begin{equation}
a_{11}=3\text{Re}\alpha+\sqrt{3}\;\text{Re}\gamma,\quad a_{22}=-3\text{Re}\beta,\quad a_{12}=-3\text{Im}\alpha\,.
\end{equation}

\item{$X_{\nu\mathbf{r}}=\rho_{\mathbf{r}}(c^2d), \rho_{\mathbf{r}}(a^2bc^2d^2)$}

In this case, the parameters $\alpha$, $\beta$, $\gamma$ and $\delta$ are constrained by the remnant CP symmetry to satisfy
\begin{equation}
\text{Re}\gamma=\text{Im}\beta=0,\quad \text{Re}\delta=\frac{1}{2}\left(-3+\sqrt{3}\right)\text{Re}\alpha,\quad \text{Im}\delta=\frac{3\text{Re}\alpha}{3+\sqrt{3}}\,.
\end{equation}
The unitary matrix $U^{\prime}_{\nu}$ takes the form
\begin{equation}
U^{\prime}_{\nu}=
\begin{pmatrix}
e^{\frac{i\pi}{4}}\cos\theta &  ~~e^{\frac{i\pi}{4}}\sin\theta~ & 0 \\
-\sin\theta  &  ~\cos\theta~  & 0 \\
 0 & ~0~ & e^{\frac{i\pi}{4}}
\end{pmatrix}\,.
\end{equation}
Then the neutrino mass matrix is diagonalized by
{\small\begin{equation}
U_{\nu}=\frac{1}{2\sqrt{3}}
\begin{pmatrix}
-\left(1+\sqrt{3}\right)e^{\frac{i\pi}{4}}\cos\theta-2 \sin\theta &   ~-\left(1+\sqrt{3}\right)e^{\frac{i\pi}{4}}\sin\theta+2\cos\theta~ & \left(\sqrt{3}-1\right)e^{\frac{i\pi}{4}}  \\

\left(\sqrt{3}-1\right) e^{\frac{i\pi}{4}}\cos\theta-2\sin\theta &  ~\left(\sqrt{3}-1\right)e^{\frac{i\pi}{4}}\sin\theta+2\cos\theta~  & -\left(1+\sqrt{3}\right)e^{\frac{i\pi}{4}}   \\

2e^{\frac{i\pi}{4}}\cos\theta-2\sin\theta   &   ~2e^{\frac{i\pi}{4}}\sin\theta+2\cos\theta~  &  2e^{\frac{i\pi}{4}}
\end{pmatrix}\,,
\end{equation}}
where
\begin{equation}
\tan2\theta=\frac{2\sqrt{6}\;\text{Re}\alpha}{\sqrt{3}\;\left(\text{Re}\alpha+\text{Re}\beta+\text{Im}\alpha\right)+\text{Im}\gamma}\,.
\end{equation}
The light neutrino masses are determined to be
\begin{eqnarray}
\nonumber&&m_1=\frac{1}{2}\left|a_{11}+a_{22}-\text{sign}\left(\left(a_{22}-a_{11}\right)\cos2\theta\right)\sqrt{\left(a_{22}-a_{11}\right)^2+4a^2_{12}}\right|\,,\\
\nonumber&&m_2=\frac{1}{2}\left|a_{11}+a_{22}+\text{sign}\left(\left(a_{22}-a_{11}\right)\cos2\theta\right)\sqrt{\left(a_{22}-a_{11}\right)^2+4a^2_{12}}\right|\,,\\
&&m_3=\left|3\text{Re}\alpha+3\text{Im}\alpha-\sqrt{3}\;\text{Im}\gamma\right|\,,
\end{eqnarray}
where
\begin{equation}
a_{11}=-3\text{Re}\alpha-3\text{Im}\alpha-\sqrt{3}\;\text{Im}\gamma,\quad a_{22}=3\text{Re}\beta,\quad a_{12}=3\sqrt{2}\;\text{Re}\alpha\,.
\end{equation}

\item{$X_{\nu\mathbf{r}}=\rho_{\mathbf{r}}(c^2d^3), \rho_{\mathbf{r}}(a^2bc^2)$}

This remnant CP symmetry implies that
\begin{equation}
\text{Re}\gamma=\text{Im}\beta=0,\quad \text{Re}\delta=\text{Im}\delta=\frac{1}{2}\left(-3+\sqrt{3}\right)\text{Re}\alpha\,.
\end{equation}
The unitary transformation $U^{\prime}_{\nu}$ is given by
\begin{equation}
U^{\prime}_{\nu}=
\begin{pmatrix}
e^{-\frac{i\pi}{4}}\cos\theta   &   ~~e^{-\frac{i\pi}{4}}\sin\theta~ & 0 \\
-\sin\theta &   ~~\cos\theta~ & 0 \\
0 & ~~0~ & e^{\frac{i\pi}{4}}
\end{pmatrix}\,.
\end{equation}
Hence the neutrino diagonalization matrix $U_{\nu}$ is of the form
{\small\begin{equation}
U_{\nu}=\frac{1}{2\sqrt{3}}
\begin{pmatrix}
-\left(1+\sqrt{3}\right)e^{-\frac{i\pi}{4}}\cos\theta-2\sin\theta & ~-\left(1+\sqrt{3}\right)e^{-\frac{i\pi}{4}}\sin\theta+2\cos\theta~ &
\left(\sqrt{3}-1\right) e^{\frac{i\pi}{4}}   \\

\left(\sqrt{3}-1\right)e^{-\frac{i\pi}{4}}\cos\theta-2\sin\theta & ~\left(\sqrt{3}-1\right)e^{-\frac{i\pi}{4}}\sin\theta+2\cos\theta~ &   -\left(1+\sqrt{3}\right)e^{\frac{i\pi}{4}}   \\

2e^{-\frac{i\pi}{4}}\cos\theta-2\sin\theta &  ~2e^{-\frac{i\pi}{4}}\sin\theta+2\cos\theta~ & 2 e^{\frac{i\pi}{4}}
\end{pmatrix}\,,
\end{equation}}
with
\begin{equation}
\tan2\theta=\frac{2\sqrt{6}\;\text{Re}\alpha}{\sqrt{3}\left(\text{Re}\alpha+\text{Re}\beta-\text{Im}\alpha\right)-\text{Im}\gamma}\,.
\end{equation}
Finally, the light neutrino masses are given by
\begin{eqnarray}
\nonumber&&m_1=\frac{1}{2}\left|a_{11}+a_{22}-\text{sign}\left(\left(a_{22}-a_{11}\right)\cos2\theta\right)\sqrt{\left(a_{22}-a_{11}\right)^2+4a^2_{12}}\right|\,,\\
\nonumber&&m_2=\frac{1}{2}\left|a_{11}+a_{22}+\text{sign}\left(\left(a_{22}-a_{11}\right)\cos2\theta\right)\sqrt{\left(a_{22}-a_{11}\right)^2+4a^2_{12}}\right|\,,\\
&&m_3=\left|3\text{Re}\alpha-3\text{Im}\alpha+\sqrt{3}\;\text{Im}\gamma\right|\,,
\end{eqnarray}
where
\begin{equation}
a_{11}=-3\text{Re}\alpha+3\text{Im}\alpha+\sqrt{3}\;\text{Im}\gamma,\quad a_{22}=3\text{Re}\beta,\quad a_{12}=3\sqrt{2}\;\text{Re}\alpha\,.
\end{equation}

\end{itemize}

\subsubsection{$G_{\nu}=Z^{(10)}_2=\left\{1, a^2bd^2\right\}$}

The residual CP symmetry $H^{\nu}_{CP}$ consistent with the remnant family symmetry $Z^{(9)}_2$, should satisfy the consistency equation:
\begin{equation}
X_{\nu\mathbf{r}}\rho^{*}_{\mathbf{r}}(a^2bd^2)X^{-1}_{\nu\mathbf{r}}=\rho_{\mathbf{r}}(a^2bd^2)\,.
\end{equation}
We find that only 8 of the 96 generalized CP transformations are acceptable,
\begin{equation}
\label{eq:residual_CP_Z210} H^{\nu}_{CP}=\left\{\rho_{\mathbf{r}}(1), \rho_{\mathbf{r}}(d^2), \rho_{\mathbf{r}}(c^2d), \rho_{\mathbf{r}}(c^2d^3), \rho_{\mathbf{r}}(a^2b), \rho_{\mathbf{r}}(a^2bd^2), \rho_{\mathbf{r}}(a^2bc^2d), \rho_{\mathbf{r}}(a^2bc^2d^3)\right\}\,.
\end{equation}
The light neutrino mass matrix $m_{\nu}$ is subject to the following constraints:
\begin{subequations}
\begin{eqnarray}
\label{eq:neutrino_z210_flavor}&&\rho^{T}_{\mathbf{3}}(a^2bd^2)m_{\nu}\rho_{\mathbf{3}}(a^2bd^2)=m_{\nu}\,,\\
\label{eq:neutrino_z210_CP}&&X^{T}_{\nu\mathbf{3}}m_{\nu}X_{\nu\mathbf{3}}=m^{*}_{\nu}\,,
\end{eqnarray}
\end{subequations}
where Eq.~\eqref{eq:neutrino_z210_flavor} is the invariance condition of the neutrino mass matrix under the residual family symmetry $G_{\nu}=Z^{(10)}_2$, and it implies that the neutrino mass matrix is of the form
\begin{equation}
\label{eq:neutrino_matrix_Z210}m_{\nu}=\alpha\begin{pmatrix}
2   & -1   &  -1   \\
-1  &  2   &  -1  \\
-1  &  -1  &  2
\end{pmatrix}+\beta\begin{pmatrix}
1  &   1   &  1  \\
1  &   1   &  1   \\
1  &   1   &  1
\end{pmatrix}+\gamma\begin{pmatrix}
0  &  1   &  0  \\
1  &  0   &  0 \\
0  &   0  &  1
\end{pmatrix}+\delta\begin{pmatrix}
-2  &  0  &  -1  \\
0   &  2  &  1 \\
-1  & 1  &  0
\end{pmatrix}\,,
\end{equation}
where $\alpha$, $\beta$, $\gamma$ and $\delta$ are complex parameters, and they are further constrained by the remnant CP symmetry shown in Eq.~\eqref{eq:neutrino_z210_CP}. In order to diagonalize the neutrino mass matrix of Eq.~\eqref{eq:neutrino_matrix_Z210}, we first perform the following unitary transformation
\begin{equation}
\nonumber m^{\prime}_{\nu}=U^{T}_{TBP}m_{\nu}U_{TBP}=
\begin{pmatrix}
3\alpha+\gamma  & 0 & ~0 \\
 0 & 3\beta+\gamma  & ~\sqrt{6}\;\delta \\
 0 & \sqrt{6}\;\delta  & ~3\alpha-\gamma
\end{pmatrix}\,,
\end{equation}
where
\begin{equation}
U_{TBP}=\begin{pmatrix}
-\frac{1}{\sqrt{6}} & ~\frac{1}{\sqrt{3}}~  & -\frac{1}{\sqrt{2}} \\
-\frac{1}{\sqrt{6}} & ~\frac{1}{\sqrt{3}}~  & \frac{1}{\sqrt{2}} \\
\sqrt{\frac{2}{3}}  & ~\frac{1}{\sqrt{3}}~  & 0
\end{pmatrix}\,,
\end{equation}
which can be obtained by the permutating the first and the third rows of the tri-bimaximal mixing matrix. Furthermore, $m^{\prime}_{\nu}$ can be diagonalized by another unitary matrix $U^{\prime}_{\nu}$,
\begin{equation}
U^{\prime T}_{\nu}m^{\prime}_{\nu} U^{\prime}_{\nu}=\text{diag}(m_1,m_2,m_3)\,.
\end{equation}
Therefore the unitary transformation $U_{\nu}$ diagonalizing the neutrino mass matrix $m_{\nu}$ in Eq.~\eqref{eq:neutrino_matrix_Z210} is of the form
\begin{equation}
U_{\nu}=U_{TBP}U^{\prime}_{\nu}.
\end{equation}
Here we would like to emphasize again that the neutrino diagonalization matrix $U_{\nu}$ is fixed up to permutations of the columns. In the following, we shall investigate the implications of the remnant CP invariant condition of Eq.~\eqref{eq:neutrino_z210_CP}. The eight possible $X_{\nu\mathbf{r}}$ lead to four different phenomenological predictions.

\begin{itemize}[leftmargin=*]

\item{$X_{\nu\mathbf{r}}=\rho_{\mathbf{r}}(1), \rho_{\mathbf{r}}(a^2bd^2)$}

In this case, the neutrino mass matrix is constrained to be real. As a result, we have
\begin{equation}
\text{Im}\alpha=\text{Im}\beta=\text{Im}\gamma=\text{Im}\delta=0\,.
\end{equation}
Hence $m^{\prime}_{\nu}$ becomes a real symmetry matrix and can
be diagonalized by a rotation matrix in the (2,3) sector with
\begin{equation}
U^{\prime}_{\nu}=\begin{pmatrix}
 1 & ~0~ & 0 \\
 0 & ~\cos\theta~ & \sin\theta \\
 0 & ~-\sin\theta~ & \cos\theta
\end{pmatrix}\,,
\end{equation}
where
\begin{equation}
\tan2\theta=\frac{2\sqrt{6}\;\text{Re}\delta}{3\text{Re}\alpha-3\text{Re}\beta-2\text{Re}\gamma}\,.
\end{equation}
As a consequence, the neutrino diagonalization matrix $U_{\nu}$ is
\begin{equation}
U_{\nu}=\frac{1}{\sqrt{6}}
\begin{pmatrix}
-1 & ~~\sqrt{2}\;\cos\theta+\sqrt{3}\;\sin\theta~~ & \sqrt{2}\;\sin\theta-\sqrt{3}\;\cos\theta \\
-1 & ~~\sqrt{2}\;\cos\theta-\sqrt{3}\;\sin\theta~~ & \sqrt{2}\;\sin\theta+\sqrt{3}\;\cos\theta \\
2 & ~~\sqrt{2}\;\cos\theta~~ & \sqrt{2}\;\sin\theta
\end{pmatrix}\,.
\end{equation}
The light neutrino masses are determined to be
{\footnotesize\begin{eqnarray*}
\nonumber&& m_1=\left|3\text{Re}\alpha+\text{Re}\gamma\right|\,,\\
\nonumber&& m_2=\frac{1}{2}\left|3\text{Re}\alpha+3\text{Re}\beta-\text{sign}\left(\left(3\text{Re}\alpha-3\text{Re}\beta-2\text{Re}\gamma\right)\cos2\theta\right)\sqrt{\left(3\text{Re}\alpha-3\text{Re}\beta-2\text{Re}\gamma\right)^2+24\left(\text{Re}\delta\right)^2}\right|\,,\\
\nonumber&& m_3=\frac{1}{2}\left|3\text{Re}\alpha+3\text{Re}\beta+\text{sign}\left(\left(3\text{Re}\alpha-3\text{Re}\beta-2\text{Re}\gamma\right)\cos2\theta\right)\sqrt{\left(3\text{Re}\alpha-3\text{Re}\beta-2\text{Re}\gamma\right)^2+24\left(\text{Re}\delta\right)^2}\right|\,.
\end{eqnarray*}}

\item{$X_{\nu\mathbf{r}}=\rho_{\mathbf{r}}(d^2), \rho_{\mathbf{r}}(a^2b)$}

The invariance of the neutrino mass matrix under the action of residual CP transformation leads to
\begin{equation}
\text{Re}\delta=\text{Im}\alpha=\text{Im}\beta=\text{Im}\gamma=0\,.
\end{equation}
The unitary transformation $U^{\prime}_{\nu}$ is
\begin{equation}
U^{\prime}_{\nu}=\begin{pmatrix}
 1 & ~0~ & 0 \\
 0 & ~\cos\theta~   & \sin\theta \\
 0 & ~-i\sin\theta~ & i\cos\theta
\end{pmatrix}\,,
\end{equation}
where
\begin{equation}
\tan2\theta=\frac{2\sqrt{6}\;\text{Im}\delta}{3\left(\text{Re}\alpha+\text{Re}\beta\right)}\,.
\end{equation}
Therefore the neutrino mass matrix is diagonalized by the following unitary matrix
\begin{equation}
U_{\nu}=\frac{1}{\sqrt{6}}
\begin{pmatrix}
-1  & ~~\sqrt{2}\;\cos\theta+i\sqrt{3}\;\sin\theta~~ &    \sqrt{2}\;\sin\theta-i\sqrt{3}\;\cos\theta \\
 -1 & ~~\sqrt{2}\;\cos\theta-i\sqrt{3}\;\sin\theta~~ & \sqrt{2}\;\sin\theta+i\sqrt{3}\;\cos\theta \\
2 & ~~\sqrt{2}\;\cos\theta~~ & \sqrt{2}\;\sin\theta
\end{pmatrix}\,.
\end{equation}
The light neutrino masses are
{\footnotesize\begin{eqnarray}
\nonumber&&\hskip-0.4in  m_1=\left|3\text{Re}\alpha+\text{Re}\gamma\right|\,,\\
\nonumber&&\hskip-0.4in m_2=\frac{1}{2}\left|-3\text{Re}\alpha+3\text{Re}\beta+2\text{Re}\gamma+\text{sign}\left(\left(\text{Re}\alpha+\text{Re}\beta\right)\cos2\theta\right)\sqrt{9\left(\text{Re}\alpha+\text{Re}\beta\right)^2+24\left(\text{Im}\delta\right)^2}\right|\,,\\
&&\hskip-0.4in m_3=\frac{1}{2}\left|-3\text{Re}\alpha+3\text{Re}\beta+2\text{Re}\gamma-\text{sign}\left(\left(\text{Re}\alpha+\text{Re}\beta\right)\cos2\theta\right)\sqrt{9\left(\text{Re}\alpha+\text{Re}\beta\right)^2+24\left(\text{Im}\delta\right)^2}\right|\,.
\end{eqnarray}}

\item{$X_{\nu\mathbf{r}}=\rho_{\mathbf{r}}(c^2d), \rho_{\mathbf{r}}(a^2bc^2d^3)$}

In this case, the parameters $\alpha$, $\beta$, $\gamma$ and $\delta$ are constrained to satisfy
\begin{equation}
\text{Re}\alpha=\text{Re}\gamma=0,\quad \text{Im}\gamma=-3\text{Im}\beta,\quad \text{Im}\delta=-\text{Re}\delta\,.
\end{equation}
The unitary matrix $U^{\prime}_{\nu}$ takes the form
\begin{equation}
U^{\prime}_{\nu}=\begin{pmatrix}
e^{\frac{i\pi}{4}} & ~0~ & 0 \\
0 & ~\cos\theta~~      & \sin\theta \\
0 & ~-e^{\frac{i\pi}{4}}\sin\theta~~   & e^{\frac{i\pi}{4}}\cos\theta
\end{pmatrix}\,,
\end{equation}
with
\begin{equation}
\tan2\theta=-\frac{4\text{Re}\delta}{\sqrt{3}\;\left(\text{Im}\alpha+\text{Re}\beta+\text{Im}\beta\right)}\,.
\end{equation}
Therefore the resulting neutrino diagonalization matrix is
\begin{equation}
U_{\nu}=\frac{1}{\sqrt{6}}
\begin{pmatrix}
-e^{\frac{i\pi}{4}}   & ~~\sqrt{2}\;\cos\theta+\sqrt{3}\;e^{\frac{i\pi}{4}}\sin\theta~~  & \sqrt{2}\;\sin\theta-\sqrt{3}\;e^{\frac{i\pi}{4}}\cos\theta \\

-e^{\frac{i\pi}{4}} &  ~~\sqrt{2}\;\cos\theta-\sqrt{3}\;e^{\frac{i\pi}{4}}\sin\theta~~  & \sqrt{2}\;\sin\theta+\sqrt{3}\;e^{\frac{i\pi}{4}}\cos\theta  \\

2e^{\frac{i\pi}{4}} & ~~\sqrt{2}\;\cos\theta~~   & \sqrt{2}\;\sin\theta
\end{pmatrix}\,.
\end{equation}
The light neutrino masses are
{\footnotesize\begin{eqnarray}
\nonumber&&m_1=3\left|\text{Im}\alpha-\text{Im}\beta\right|\,,\\
\nonumber&&m_2=\frac{1}{2}\left|3\left(\text{Im}\alpha-\text{Re}\beta+\text{Im}\beta\right)-\text{sign}\left(\left(\text{Im}\alpha+\text{Re}\beta+\text{Im}\beta\right)\cos2\theta\right)\sqrt{9\left(\text{Im}\alpha+\text{Re}\beta+\text{Im}\beta\right)^2+48\left(\text{Re}\delta\right)^2}\right|\,,\\
\nonumber&&m_3=\frac{1}{2}\left|3\left(\text{Im}\alpha-\text{Re}\beta+\text{Im}\beta\right)+\text{sign}\left(\left(\text{Im}\alpha+\text{Re}\beta+\text{Im}\beta\right)\cos2\theta\right)\sqrt{9\left(\text{Im}\alpha+\text{Re}\beta+\text{Im}\beta\right)^2+48\left(\text{Re}\delta\right)^2}\right|\,.
\end{eqnarray}}

\item{$X_{\nu\mathbf{r}}=\rho_{\mathbf{r}}(c^2d^3), \rho_{\mathbf{r}}(a^2bc^2d)$}

In this case, the residual CP symmetry constrains the parameters as
\begin{equation}
\text{Re}\alpha=\text{Re}\gamma=0,\quad \text{Im}\gamma=-3\text{Im}\beta,\quad \text{Im}\delta=\text{Re}\delta\,.
\end{equation}
The unitary transformation $U^{\prime}_{\nu}$ is given by
\begin{equation}
U^{\prime}_{\nu}=\begin{pmatrix}
e^{\frac{i\pi}{4}} & ~~0~~ & 0 \\
0 & ~~\cos\theta~~    & \sin\theta \\
0 & ~~-e^{-\frac{i\pi}{4}}\sin\theta~~ & e^{-\frac{i\pi}{4}}\cos\theta
\end{pmatrix}\,,
\end{equation}
where
\begin{equation}
\tan2\theta=\frac{4\text{Re}\delta}{\sqrt{3}\;\left(\text{Im}\alpha-\text{Re}\beta+\text{Im}\beta\right)}\,.
\end{equation}
Hence the neutrino mass matrix is diagonalized by
\begin{equation}
U_{\nu}=\frac{1}{\sqrt{6}}
\begin{pmatrix}
-e^{\frac{i\pi}{4}} & ~~\sqrt{2}\cos\theta+\sqrt{3}e^{-\frac{i\pi}{4}}\sin\theta~~ & \sqrt{2}\sin\theta-\sqrt{3}e^{-\frac{i\pi}{4}}\cos\theta \\

-e^{\frac{i\pi}{4}} & ~~\sqrt{2}\cos\theta-\sqrt{3}e^{-\frac{i\pi}{4}}\sin\theta~~ & \sqrt{2}\sin\theta+\sqrt{3}e^{-\frac{i\pi}{4}}\cos\theta \\

2 e^{\frac{i\pi}{4}} &  ~~\sqrt{2}\cos\theta~~   & \sqrt{2}\sin\theta
\end{pmatrix}\,.
\end{equation}
Finally, the light neutrino masses are
{\footnotesize\begin{eqnarray*}
\nonumber&& m_1=3\left|\text{Im}\alpha-\text{Im}\beta\right|\,,\\
\nonumber&&m_2=\frac{1}{2}\left|3\left(\text{Im}\alpha+\text{Re}\beta+\text{Im}\beta\right)-\text{sign}\left(\left(\text{Im}\alpha-\text{Re}\beta+\text{Im}\beta\right)\cos2\theta\right)\sqrt{9\left(\text{Im}\alpha-\text{Re}\beta+\text{Im}\beta\right)^2+48\left(\text{Re}\delta\right)^2}\right|\,,\\
\nonumber&&m_3=\frac{1}{2}\left|3\left(\text{Im}\alpha+\text{Re}\beta+\text{Im}\beta\right)+\text{sign}\left(\left(\text{Im}\alpha-\text{Re}\beta+\text{Im}\beta\right)\cos2\theta\right)\sqrt{9\left(\text{Im}\alpha-\text{Re}\beta+\text{Im}\beta\right)^2+48\left(\text{Re}\delta\right)^2}\right|\,.
\end{eqnarray*}}

\end{itemize}

\section{Lepton mixing predictions}
\label{4}
\cleqn

In this section, we perform a comprehensive analysis of all possible lepton mixing matrices obtainable from the implementation of a $\Delta(96)$ family symmetry and its corresponding generalised CP symmetry by considering all possible residual symmetries $G^{\nu}_{\rm{CP}}$ and $G^{l}_{\rm{CP}}$ discussed in previous sections. In all the cases, both leptonic mixing angles and CP phases (including both Dirac and Majorana CP phases) are found to depend on only one parameter $\theta$. As a consequence, the lepton mixing parameters are strongly correlated with each other in this context, and obviously it is highly nontrivial to be able to fit all the observed lepton mixing angles with the sole parameter $\theta$.
As a measure of to which extent the resulting lepton mixing angles
can be close to the accurately measured values of $\theta_{12}$, $\theta_{23}$ and $\theta_{13}$, we use the $\chi^2$ function defined in the conventional way. Since the octant of the atmospheric mixing angle has not been determined so far and $\sin^2\theta_{23}$ has two best fit values $\sin^2\theta_{23}=0.413$ and $\sin^2\theta_{23}=0.594$~\cite{GonzalezGarcia:2012sz}, we define two different $\chi^2$ functions. The smaller the minimum of the $\chi^2$ function is, the better the corresponding PMNS matrix can explain the data. Notice that, without a particular model, the lepton mixing matrix is only determined up to permutations of rows and columns, since neither charged lepton nor neutrino masses are constrained in the present framework. Therefore all the possible permutation of rows and columns are taken into account for each symmetry breaking pattern and the corresponding global minimum of the $\chi^2$ functions is calculated, and subsequently we choose the best one.

The analytical formulas for the mixing parameters and the best fitting results are summarized in Tables~\ref{tab:caseI_II}-\ref{tab:caseXIX_XX}. Because the sign of the Jarlskog invariant $J_{\text{CP}}$ depends on the ordering of rows and
columns, while the sign of $\sin\alpha\;(\tan\alpha)$ and $\sin\beta\; (\tan\beta)$ depends on the CP parity of the neutrino states
which is encoded in the matrix $P$, please see Eq.~\eqref{eq:unup_z22}, all these quantities are presented in terms of absolute values. In addition, if we assign the LH lepton doublets to be the triplet $\mathbf{\overline{3}}$ instead of $\mathbf{3}$, the sign of the CP phases $\delta_{CP}$, $\alpha$ and $\beta$ would be changed. In the following, we shall present the resulting PMNS matrix and its predictions for the lepton mixing parameters for each possible symmetry breaking chains. It is remarkable that the best arrangements of the PMNS matrix for the first octant of $\theta_{23}$ and the second octant of $\theta_{23}$ turn out to be related by the exchange of the second and the third rows. Hence only the form of the PMNS matrix for the first octant of $\theta_{23}$ would be shown in the following if not mentioned explicitly.

\subsection{$G_{l}=Z^{(2)}_{3}$, $G_{\nu}=Z^{(2)}_{2}$}

In this case, the charged lepton mass matrix is diagonal, therefore the lepton mixing is completely determined by the neutrino sector, and the lepton mixing matrix coincides the neutrino diagonalization matrix $U_{\nu}$ up to permutations of rows and columns.

\begin{figure}[t!]\hskip-0.2in
\hskip-5mm\includegraphics[width=1.1\textwidth]{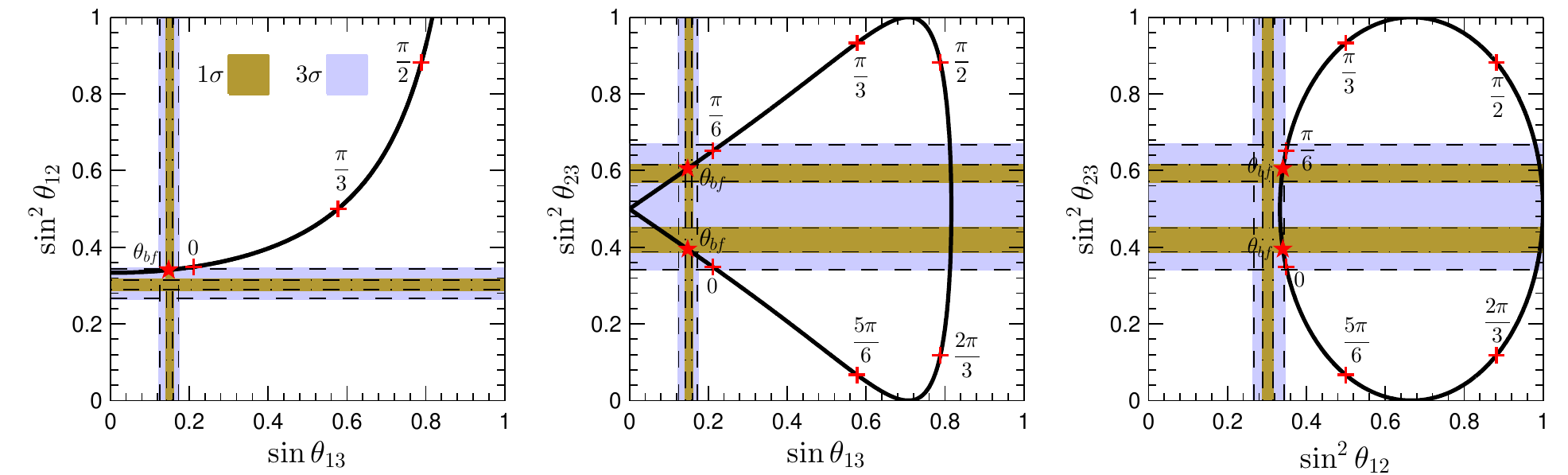} ~~~\caption{\label{fig:caseI_II}The relation among the lepton mixing angles in the case I and case II. The best fit value $\theta_{bf}$ for $\theta$ is labelled as a star. We also indicate the points for $\theta=0, \pi/6, \pi/3, \pi/2, 2\pi/3$ and $5\pi/6$ by the sign ``+'' on the curves, and only one $\theta$ value is displayed if there are points matching together. The shown $3\sigma$ ranges for the mixing angles and their best fit values are
taken from Ref.~\cite{GonzalezGarcia:2012sz}. }
\end{figure}

\begin{description}[labelindent=-0.7em, leftmargin=0.1em]

\item[~~(I)] {$X_{\nu\mathbf{r}}=\rho_{\mathbf{r}}(1), \rho_{\mathbf{r}}(d^2)$}

In this case, the lepton mixing matrix is determined to  be
\begin{equation}
\label{eq:PMNS_I}U_{PMNS}=\frac{1}{\sqrt{3}}
\begin{pmatrix}
-\sqrt{2}\cos\left(\frac{\pi}{12}-\theta\right) &  ~1 &   ~~\sqrt{2}\sin\left(\frac{\pi}{12}-\theta\right) \\
\sqrt{2}\cos\left(\frac{\pi}{4}+\theta\right) & ~1 &  ~~\sqrt{2}\sin\left(\frac{\pi}{4}+\theta\right) \\
\sqrt{2}\sin\left(\frac{\pi}{12}+\theta\right) & ~1 &   ~~-\sqrt{2}\cos\left(\frac{\pi}{12}+\theta\right)
\end{pmatrix}P\,,
\end{equation}
where $P$ is a diagonal matrix with entry $\pm1$ or $\pm i$, and it would be neglected hereafter. In the present work, we shall work in the PDG convention~\cite{pdg}, where the PMNS matrix is cast into the form
\begin{equation}
\label{eq:pmns_pdg}U_{PMNS}=V\,\text{diag}(1,e^{i\frac{\alpha_{21}}{2}},e^{i\frac{\alpha_{31}}{2}}),
\end{equation}
with
\begin{equation}
V=\begin{pmatrix}
c_{12}c_{13}  &   s_{12}c_{13}   &   s_{13}e^{-i\delta_{CP}}  \\
-s_{12}c_{23}-c_{12}s_{23}s_{13}e^{i\delta_{CP}}   &  c_{12}c_{23}-s_{12}s_{23}s_{13}e^{i\delta_{CP}}  &  s_{23}c_{13}  \\
s_{12}s_{23}-c_{12}c_{23}s_{13}e^{i\delta_{CP}}   & -c_{12}s_{23}-s_{12}c_{23}s_{13}e^{i\delta_{CP}}  &  c_{23}c_{13}
\end{pmatrix}\,,
\end{equation}
where $c_{ij}=\cos\theta_{ij}$ and $s_{ij}=\sin\theta_{ij}$, $\delta_{CP}$ is the Dirac CP phase, $\alpha_{21}$ and $\alpha_{31}$ are the Majorana CP phases. In the following, we shall redefine the Majorana phase and introduce $\alpha^{\prime}_{31}=\alpha_{31}-2\delta_{\rm{CP}}$ for the sake of convenience. As a result, the lepton mixing parameters are predicted to be
\begin{eqnarray}
\nonumber&&\sin^2\theta_{13}=\frac{1}{3}\left[1-\cos\left(\frac{\pi}{6}-2\theta\right)\right],\qquad \sin^2\theta_{12}=\frac{1}{2+\cos\left(\frac{\pi}{6}-2\theta\right)}\,,\\
\label{eq:mixing_parameters_I}&&\sin^2\theta_{23}=\frac{1+\sin2\theta}{2+\cos\left(\frac{\pi}{6}-2\theta\right)},\qquad \tan\delta_{CP}=\tan\alpha_{21}=\tan\alpha_{31}=0\,.
\end{eqnarray}
Obviously CP is predicted to be conserved in this scenario. The mixing parameters are strongly correlated with each other:
\begin{equation}
\label{eq:correlation_I}3\sin^2\theta_{12}\cos^2\theta_{13}=1,\qquad \sin^2\theta_{23}=\frac{1}{2}\pm\frac{\sin\theta_{13}}{2\cos^2\theta_{13}}\sqrt{2-3\sin^2\theta_{13}}\,.
\end{equation}
The correlations among the mixing angles are shown in Fig.~\ref{fig:caseI_II}. Excellent agreement with the present data~\cite{GonzalezGarcia:2012sz} can be achieved. The best fitting value of $\theta$ is $\theta_{\text{bf}}\simeq0.0798$, the minimal value of $\chi^2$ is $\chi^2_{\text{min}}\simeq9.548$, and the corresponding values for the mixing angles are:
\begin{equation}
\sin^2\theta_{13}(\theta_{\text{bf}})\simeq0.0218,\quad \sin^2\theta_{12}(\theta_{\text{bf}})\simeq0.341,\quad
\sin^2\theta_{23}(\theta_{\text{bf}})\simeq0.395\,,
\end{equation}
where the atmospheric mixing angle is smaller than $\pi/4$ and therefore lies in the first octant. If we exchange the second and the third rows of the PMNS matrix in Eq.~\eqref{eq:PMNS_I}, the atmospheric mixing angle becomes
\begin{equation}
\sin^2\theta_{23}=\frac{1+\cos\left(\frac{\pi}{6}+2\theta\right)}{2+\cos\left(\frac{\pi}{6}-2\theta\right)}\,,
\end{equation}
while the predictions for the remaining mixing parameters keep intact, as shown in Eq.~\eqref{eq:mixing_parameters_I}. It is notable that the
correlations in Eq.~\eqref{eq:correlation_I} are also satisfied in this case. This pattern can also accommodate the observed lepton mixing data very well and it prefer the second octant $\theta_{23}$. The best fitting values are
\begin{eqnarray}
\nonumber& \theta_{\text{b.f.}}\simeq0.0798,\quad \chi^2_{\text{min}}\simeq9.303,\\
&\sin^2\theta_{13}(\theta_{\text{bf}})\simeq0.0218,\quad \sin^2\theta_{12}(\theta_{\text{bf}})\simeq0.341,\quad
\sin^2\theta_{23}(\theta_{\text{bf}})\simeq0.605\,.
\end{eqnarray}

\item[~~(II)] {$X_{\nu\mathbf{r}}=\rho_{\mathbf{r}}(c^2d), \rho_{\mathbf{r}}(c^2d^3)$}

The PMNS matrix is given by
\begin{equation}
U_{PMNS}=\frac{1}{\sqrt{3}}
\begin{pmatrix}
-\sqrt{2}\,e^{i\pi/4}\cos\left(\frac{\pi}{12}-\theta\right) &  ~~1 &   ~~\sqrt{2}\,e^{i\pi/4}\sin\left(\frac{\pi}{12}-\theta\right)\\
\sqrt{2}\,e^{i\pi/4}\cos\left(\frac{\pi}{4}+\theta\right) & ~~1 &  ~~\sqrt{2}\,e^{i\pi/4}\sin\left(\frac{\pi}{4}+\theta\right) \\
\sqrt{2}\,e^{i\pi/4}\sin\left(\frac{\pi}{12}+\theta\right) & ~~1 &   ~~-\sqrt{2}\,e^{i\pi/4}\cos\left(\frac{\pi}{12}+\theta\right)
\end{pmatrix}\,,
\end{equation}
where the trivial matrix $P$ has been omitted. The lepton mixing parameters are
\begin{eqnarray}
\nonumber&&\sin^2\theta_{13}=\frac{1}{3}\left[1-\cos\left(\frac{\pi}{6}-2\theta\right)\right],\qquad \sin^2\theta_{12}=\frac{1}{2+\cos\left(\frac{\pi}{6}-2\theta\right)}\,,\\
&&\sin^2\theta_{23}=\frac{1+\sin2\theta}{2+\cos\left(\frac{\pi}{6}-2\theta\right)},\qquad \tan\delta_{CP}=\cot\alpha_{21}=\tan\alpha_{31}=0\,.
\end{eqnarray}
Compared with Eq.~\eqref{eq:mixing_parameters_I}, the mixing parameters are predicted to be the same as those of case I except that now the Majorana phase $\alpha_{21}$ is $\pm\pi/2$ rather than zero.

\item[~~(III)] {$X_{\nu\mathbf{r}}=\rho_{\mathbf{r}}(a^2b), \rho_{\mathbf{r}}(a^2bd^2)$}

The lepton mixing matrix is given by
\begin{equation}
U_{PMNS}=\frac{1}{\sqrt{6}}
\begin{pmatrix}
2i\cos\theta  &   ~~\sqrt{2} &  ~~-2i\sin\theta \\
-\sqrt{3}\sin\theta-i\cos\theta  &   ~~\sqrt{2} & ~~-\sqrt{3}\cos\theta +i\sin\theta  \\
\sqrt{3}\sin\theta-i\cos\theta  & ~~\sqrt{2} &  ~~\sqrt{3}\cos\theta+i\sin\theta
\end{pmatrix}\,.
\end{equation}
The lepton mixing parameters are determined to be
\begin{eqnarray}
\nonumber&\sin^2\theta_{13}=\frac{1}{3}\left(1-\cos2\theta\right),\qquad \sin^2\theta_{12}=\frac{1}{2+\cos2\theta},\quad \sin2\theta_{23}=\frac{1}{2},\\
&\left|J_{CP}\right|=\frac{1}{6\sqrt{3}}\left|\sin2\theta\right|,\qquad \cot\delta_{CP}=\tan\alpha_{21}=\tan\alpha_{31}=0\,.
\end{eqnarray}
Good agreement with the experimental data can be achieved in this case, and the best fitting results are listed in Table~\ref{tab:caseI_II}. Note that the solar mixing angle $\theta_{12}$ is related to the reactor mixing angle $\theta_{13}$ by $3\sin^2\theta_{12}\cos^2\theta_{13}=1$, the atmospheric mixing angle is predicted to maximal and the Dirac CP is maximally broken. The correlations of $\sin^2\theta_{12}$ and $\left|J_{CP}\right|$ with respect to $\sin\theta_{13}$ are displayed in Fig.~\ref{fig:caseIII}.

\item[~~(IV)]{$X_{\nu\mathbf{r}}=\rho_{\mathbf{r}}(a^2bd), \rho_{\mathbf{r}}(a^2bd^3)$}

The lepton mixing matrix is
\begin{equation}
U_{PMNS}=\frac{1}{2\sqrt{3}}
\begin{pmatrix}
\sqrt{3}\;e^{i\theta}-e^{-i\theta} & ~~2 & ~~-i\left(\sqrt{3}\;e^{i\theta}+e^{-i\theta}\right) \\
 2 e^{-i \theta } & ~~2 & ~~2i e^{-i \theta }\\
 -\sqrt{3}\;e^{i\theta}-e^{-i\theta} & ~~2 & ~~i\left(\sqrt{3}\;e^{i\theta}-e^{-i\theta}\right)
\end{pmatrix}\,.
\end{equation}
The mixing parameters take the form
\begin{eqnarray}
\nonumber&&\hskip-0.4in \sin^2\theta_{13}=\frac{1}{3}+\frac{1}{2\sqrt{3}}\cos2\theta,\quad \sin^2\theta_{12}=\sin^2\theta_{23}=\frac{2}{4-\sqrt{3}\;\cos2\theta},\\
\nonumber&&\hskip-0.4in \left|J_{CP}\right|=\frac{1}{6\sqrt{3}}\left|\sin2\theta\right|,\quad \left|\tan\delta_{CP}\right|=\left|\frac{4-\sqrt{3}\;\cos2\theta}{1-\sqrt{3}\;\cos2\theta}\tan2\theta\right|,\\
&&\hskip-0.4in \left|\tan\alpha_{21}\right|=\left|\frac{\sin2\theta}{\sqrt{3}-2\cos2\theta}\right|,\quad \left|\tan\alpha^{\prime}_{31}\right|=\left|\frac{4\sqrt{3}\;\sin2\theta}{1-3\cos4\theta}\right|\,.
\end{eqnarray}
We see that all the mixing parameters nontrivially depend on the parameter $\theta$, and these results for the mixing parameters
are illustrated in Fig.~\ref{fig:caseIV}. However, this mixing pattern doesn't describe the the experimental data very well although not so bad. The minimum values of the $\chi^2$ functions are somewhat large: 110.741 and 111.559 for $\theta_{23}<\pi/4$ and $\theta_{23}>\pi/4$ respectively, as shown in Table~\ref{tab:caseI_II}. The best fitting values $\theta_{\text{bf}}$ is $\pi/2$, the reason is that for this value $\sin^2\theta_{13}$ is minimized as $\sin^2\theta_{13}(\theta_{\text{bf}})=\left(2-\sqrt{3}\right)/6\simeq0.0447$. In addition, all the three CP phases $\delta_{CP}$, $\alpha_{21}$ and $\alpha_{31}$
become trivial with $\sin\delta_{CP}(\theta_{\text{bf}})=\sin\alpha_{21}(\theta_{\text{bf}})=\sin\alpha_{31}(\theta_{\text{bf}})=0$ for $\theta_{\text{bf}}=\pi/2$.

\item[~~(V)]{$X_{\nu\mathbf{r}}=\rho_{\mathbf{r}}(a^2bc), \rho_{\mathbf{r}}(a^2bcd^2)$}

In this case, the lepton mixing matrix is determined to be
{\footnotesize\begin{equation}
\label{eq:PMNS_V}U_{PMNS}=\frac{1}{\sqrt{3}}
\begin{pmatrix}
e^{-\frac{i\pi}{8}}\cos\left(\frac{\pi}{24}-\theta\right)-e^{\frac{3i\pi}{8}}\cos\left(\frac{\pi}{24}+\theta\right) &  ~1 & ~-e^{-\frac{i\pi}{8}}\sin\left(\frac{\pi}{24}-\theta\right)-e^{\frac{3i\pi}{8}}\sin\left(\frac{\pi}{24} +\theta\right) \\
-e^{-\frac{i\pi}{8}}\sin\left(\frac{5\pi}{24}-\theta\right)+e^{\frac{3i\pi}{8}}\sin\left(\frac{5\pi}{24}+\theta\right) &  ~1 & ~-e^{-\frac{i\pi}{8}}\cos\left(\frac{5\pi}{24}-\theta\right)-e^{\frac{3i\pi}{8}}\cos\left(\frac{5\pi}{24}+\theta\right)\\
-e^{-\frac{i\pi}{8}}\sin\left(\frac{\pi}{8}+\theta\right)+e^{\frac{3i\pi}{8}}\sin\left(\frac{\pi}{8}-\theta\right) &   ~1 & ~e^{-\frac{i\pi}{8}}\cos\left(\frac{\pi}{8}+\theta\right)+e^{\frac{3i\pi}{8}}\cos\left(\frac{\pi}{8}-\theta\right)
\end{pmatrix}\,.
\end{equation}}
We can straightforwardly read out the lepton mixing parameters
\begin{eqnarray}
\nonumber&&\sin^2\theta_{13}=\frac{1}{3}-\frac{\sqrt{6}+\sqrt{2}}{12}\cos2\theta,\qquad \sin^2\theta_{12}=\frac{4}{8+\left(\sqrt{6}+\sqrt{2}\right)\cos2\theta},\\
\nonumber&&\sin^2\theta_{23}=\frac{4+\left(\sqrt{6}-\sqrt{2}\right)\cos2\theta}{8+\left(\sqrt{6}+\sqrt{2}\right)\cos2\theta},\qquad \left|J_{CP}\right|=\frac{1}{6\sqrt{3}}\left|\sin2\theta\right|,\\
\nonumber&&\left|\tan\delta_{CP}\right|=\left|\frac{4\sqrt{2}+\left(1+\sqrt{3}\right)\cos2\theta}{1-\sqrt{3}-\sqrt{2}\cos2\theta}\tan2\theta\right|,\\
\nonumber&&\left|\tan\alpha_{21}\right|=\left|\frac{\sqrt{6}+\sqrt{2}+4\cos2\theta+\left(\sqrt{6}-\sqrt{2}\right)\sin2\theta}{\sqrt{6}+\sqrt{2}+4\cos2\theta-\left(\sqrt{6}-\sqrt{2}\right)\sin2\theta}\right|\,,\\
\label{eq:mixing_parameters_V}&&\left|\tan\alpha^{\prime}_{31}\right|=\left|\frac{4\sin2\theta}{2-3\sqrt{3}+\left(2+\sqrt{3}\right)\cos4\theta}\right|\,.
\end{eqnarray}
Note that all the mixing parameters are nontrivial functions of $\theta$. The different mixing angles are correlated with each other as
\begin{eqnarray}
\nonumber&&3\sin^2\theta_{12}\cos^2\theta_{13}=1,\\
\nonumber&&3\sin^2\theta_{23}\cos^2\theta_{13}=3-\sqrt{3}-3\left(2-\sqrt{3}\right)\sin^2\theta_{13},\quad \theta_{23}<\pi/4,\\
\label{eq:sum_rules_V}&&3\sin^2\theta_{23}\cos^2\theta_{13}=\sqrt{3}-3\left(\sqrt{3}-1\right)\sin^2\theta_{13},\qquad~~ \theta_{23}>\pi/4\,.
\end{eqnarray}
For the measured value of $\theta_{13}$ with $\sin^2\theta_{13}=0.0227$, we obtain $\sin^2\theta_{12}\simeq0.341$ and $\sin^2\theta_{23}\simeq0.426$ ($\sin^2\theta_{23}\simeq0.574$), which are in accordance with experimental data. The best fitting values are presented in Table~\ref{tab:caseV_VI} and the mixing parameters are plotted in Fig.~\ref{fig:caseV}, we see that the CP phases $\delta_{CP}$ and $\alpha_{21}$ no longer take regular values such as 0 or $\pm\pi/2$ although $\left|\alpha^{\prime}_{31}\right|\simeq\pi/2$ is approximately fulfilled.

\item[~~(VI)]{$X_{\nu\mathbf{r}}=\rho_{\mathbf{r}}(a^2bcd), \rho_{\mathbf{r}}(a^2bcd^3)$}

The PMNS matrix takes the form
{\footnotesize\begin{equation}
\label{eq:PMNS_VI}U_{PMNS}=\frac{1}{\sqrt{3}}
\begin{pmatrix}
-e^{-\frac{i\pi}{8}}\cos\left(\frac{\pi}{24}+\theta\right)+e^{\frac{3i\pi}{8}}\cos\left(\frac{\pi}{24}-\theta\right) &  ~1 & ~-e^{-\frac{i\pi}{8}}\sin\left(\frac{\pi}{24}+\theta\right)-e^{\frac{3i\pi}{8}}\sin\left(\frac{\pi}{24}-\theta\right)\\
e^{-\frac{i\pi}{8}}\sin\left(\frac{5\pi}{24}+\theta\right)-e^{\frac{3i\pi}{8}}\sin\left(\frac{5\pi}{24}-\theta\right) & ~1 & ~-e^{-\frac{i\pi}{8}}\cos\left(\frac{5\pi}{24}+\theta\right)-e^{\frac{3i\pi}{8}}\cos\left(\frac{5\pi}{24}-\theta\right)\\
e^{-\frac{i\pi}{8}}\sin\left(\frac{\pi}{8}-\theta\right)-e^{\frac{3i\pi}{8}}\sin\left(\frac{\pi}{8}+\theta\right) &  ~1  & ~e^{-\frac{i\pi}{8}}\cos\left(\frac{\pi}{8}-\theta\right)+e^{\frac{3i\pi}{8}}\cos\left(\frac{\pi}{8}+\theta\right)
\end{pmatrix}\,.
\end{equation}}
The lepton mixing parameters read as
\begin{eqnarray}
\nonumber&&\sin^2\theta_{13}=\frac{1}{3}-\frac{\sqrt{6}+\sqrt{2}}{12}\cos2\theta,\qquad \sin^2\theta_{12}=\frac{4}{8+\left(\sqrt{6}+\sqrt{2}\right)\cos2\theta},\\
\nonumber&&\sin^2\theta_{23}=\frac{4+\left(\sqrt{6}-\sqrt{2}\right)\cos2\theta}{8+\left(\sqrt{6}+\sqrt{2}\right)\cos2\theta},\qquad \left|J_{CP}\right|=\frac{1}{6\sqrt{3}}\left|\sin2\theta\right|,\\
\nonumber&&\left|\tan\delta_{CP}\right|=\left|\frac{4\sqrt{2}+\left(1+\sqrt{3}\right)\cos2\theta}{1-\sqrt{3}-\sqrt{2}\cos2\theta}\tan2\theta\right|,\\
\nonumber&&\left|\tan\alpha_{21}\right|=\left|\frac{\sqrt{6}+\sqrt{2}+4\cos2\theta-\left(\sqrt{6}-\sqrt{2}\right)\sin2\theta}{\sqrt{6}+\sqrt{2}+4\cos2\theta+\left(\sqrt{6}-\sqrt{2}\right)\sin2\theta}\right|,\\
\label{eq:mixing_parameters_VI}&&\left|\tan\alpha^{\prime}_{31}\right|=\left|\frac{4\sin2\theta}{2-3\sqrt{3}+\left(2+\sqrt{3}\right)\cos4\theta}\right|\,.
\end{eqnarray}

Compared with Eq.~\eqref{eq:mixing_parameters_V}, we see that $\left|\tan\alpha_{21}\right|$ is predicted to be the inverse of the corresponding one of case V, and all the remaining mixing parameters coincide exactly in both cases.

\item[~~(VII)]{$X_{\nu\mathbf{r}}=\rho_{\mathbf{r}}(a^2bc^2), \rho_{\mathbf{r}}(a^2bc^2d^2)$}

The PMNS matrix is of the form
\begin{equation}
U_{PMNS}=\frac{1}{2\sqrt{3}}
\begin{pmatrix}{ccc}
e^{\frac{i\pi}{4}}\left(\sqrt{3}\;e^{i\theta}-e^{-i\theta}\right) & ~~2 &   ~~-e^{\frac{3i\pi}{4}}\left(\sqrt{3}\;e^{i\theta}+e^{-i\theta}\right)   \\
2e^{i\left(\frac{\pi}{4}-\theta\right)} & ~~2 & ~~2e^{i\left(\frac{3\pi}{4}-\theta\right)} \\
-e^{\frac{i\pi}{4}}\left(\sqrt{3}\;e^{i\theta}+e^{-i\theta}\right) & ~~2 &  ~~e^{\frac{3i\pi}{4}}\left(\sqrt{3}\;e^{i\theta}-e^{-i\theta}\right)
\end{pmatrix}\,.
\end{equation}
The lepton mixing parameters are fixed to be
\begin{eqnarray}
\nonumber&&\sin^2\theta_{13}=\frac{1}{3}+\frac{1}{2\sqrt{3}}\cos2\theta,\qquad \sin^2\theta_{12}=\sin^2\theta_{23}=\frac{2}{4-\sqrt{3}\cos2\theta},\\
\nonumber&&\left|J_{CP}\right|=\frac{1}{6\sqrt{3}}\left|\sin2\theta\right|,\qquad \left|\tan\delta_{CP}\right|=\left|\frac{4-\sqrt{3}\cos2\theta}{1-\sqrt{3}\cos2\theta}\tan2\theta\right|,\\
&&\left|\tan\alpha_{21}\right|=\left|\frac{\sqrt{3}-2\cos2\theta}{\sin2\theta}\right|,\qquad \left|\tan\alpha^{\prime}_{31}\right|=\left|\frac{4\sqrt{3}\sin2\theta}{1-3\cos4\theta}\right|
\,.
\end{eqnarray}
Obviously the lepton mixing parameters are of the same forms in case VII and case IV except the Majorana phase $\alpha$ which fulfills $\alpha^{\text{VII}}_{21}=\alpha^{\text{IV}}_{21}-\pi/2$, where the superscripts ``VII'' and ``IV'' denote the different remnant symmetries.

\item[~~(VIII)]{$X_{\nu\mathbf{r}}=\rho_{\mathbf{r}}(a^2bc^2d), \rho_{\mathbf{r}}(a^2bc^2d^3)$}

In this case, the lepton flavor mixing matrix is determined to be
\begin{equation}
U_{PMNS}=\frac{1}{\sqrt{6}}
\begin{pmatrix}{ccc}
2e^{\frac{i\pi}{4}}\cos\theta  &   ~\sqrt{2}  &  ~~-2e^{\frac{i\pi}{4}}\sin\theta  \\

-e^{\frac{i\pi}{4}}\left(\cos\theta+i\sqrt{3}\sin\theta\right) &  ~\sqrt{2}   &    ~~e^{\frac{i\pi}{4}}\left(\sin\theta-i\sqrt{3}\cos\theta\right)  \\

-e^{\frac{i\pi}{4}}\left(\cos\theta-i\sqrt{3}\sin\theta\right) &  ~\sqrt{2}   &   ~~e^{\frac{i\pi}{4}} \left(\sin\theta+i\sqrt{3} \cos\theta\right)
\end{pmatrix}\,.
\end{equation}
The lepton mixing parameters are given by
\begin{eqnarray}
\nonumber&&\sin^2\theta_{13}=\frac{1}{3}\left(1-\cos2\theta\right),\qquad \sin^2\theta_{12}=\frac{1}{2+\cos2\theta},\qquad \sin^2\theta_{23}=\frac{1}{2}\,,\\
&&\left|J_{CP}\right|=\frac{1}{6\sqrt{3}}\left|\sin2\theta\right|,\qquad
\cot\delta_{CP}=\cot\alpha_{21}=\tan\alpha_{31}=0 \,.
\end{eqnarray}
Obviously the phenomenological predictions of case VIII and case III only differ in the Majorana phase $\alpha_{21}$. It is maximally broken in case VIII while it is completely conserved in case III. Analogous to case III, the experimentally preferred values of the mixing angles can be obtained.

\item[~~(IX)]{$X_{\nu\mathbf{r}}=\rho_{\mathbf{r}}(a^2bc^3), \rho_{\mathbf{r}}(a^2bc^3d^2)$}

The lepton mixing matrix takes the form
{\footnotesize\begin{equation}
U_{PMNS}=\frac{1}{\sqrt{3}}
\begin{pmatrix}
-e^{-\frac{3i\pi}{8}}\cos\left(\frac{\pi}{24}+\theta\right)+e^{\frac{i\pi}{8}}\cos\left(\frac{\pi}{24}-\theta\right) &   ~1 & ~-e^{-\frac{3i\pi}{8}}\sin\left(\frac{\pi}{24}+\theta\right)-e^{\frac{i\pi}{8}}\sin\left(\frac{\pi}{24}-\theta\right)\\

e^{-\frac{3i\pi}{8}}\sin\left(\frac{5\pi}{24}+\theta\right)-e^{\frac{i\pi}{8}}\sin\left(\frac{5\pi}{24}-\theta\right) &  ~1 & ~-e^{-\frac{3i\pi}{8}}\cos\left(\frac{5\pi}{24}+\theta\right)-e^{\frac{i\pi}{8}}\cos\left(\frac{5\pi}{24}-\theta\right)\\

e^{-\frac{3i\pi}{8}}\sin\left(\frac{\pi}{8}-\theta\right)-e^{\frac{i\pi}{8}}\sin\left(\frac{\pi}{8}+\theta\right) &  ~1 & ~e^{-\frac{3i\pi}{8}}\cos\left(\frac{\pi}{8}-\theta\right)+e^{\frac{i\pi}{8}}\cos\left(\frac{\pi}{8}+\theta\right)
\end{pmatrix}\,,
\end{equation}}
which is the complex conjugate of the PMNS matrix of case V shown in Eq.~\eqref{eq:PMNS_V}. The lepton mixing parameters are determined to be
\begin{eqnarray}
\nonumber&&\sin^2\theta_{13}=\frac{1}{3}-\frac{\sqrt{6}+\sqrt{2}}{12}\cos2\theta,\qquad \sin^2\theta_{12}=\frac{4}{8+\left(\sqrt{6}+\sqrt{2}\right)\cos2\theta},\\
\nonumber&&\sin^2\theta_{23}=\frac{4+\left(\sqrt{6}-\sqrt{2}\right)\cos2\theta}{8+\left(\sqrt{6}+\sqrt{2}\right)\cos2\theta},\qquad \left|J_{CP}\right|=\frac{1}{6\sqrt{3}}\left|\sin2\theta\right|,\\
\nonumber&&\left|\tan\delta_{CP}\right|=\left|\frac{4\sqrt{2}+\left(1+\sqrt{3}\right)\cos2\theta}{1-\sqrt{3}-\sqrt{2}\cos2\theta}\tan2\theta\right|,\\
\nonumber&&\left|\tan\alpha_{21}\right|=\left|\frac{\sqrt{6}+\sqrt{2}+4\cos2\theta+\left(\sqrt{6}-\sqrt{2}\right)\sin2\theta}{\sqrt{6}+\sqrt{2}+4\cos2\theta-\left(\sqrt{6}-\sqrt{2}\right)\sin2\theta}\right|,\\
\label{eq:mixing_parameters_IX}&&\left|\tan\alpha^{\prime}_{31}\right|=\left|\frac{4\sin2\theta}{2-3\sqrt{3}+\left(2+\sqrt{3}\right)\cos4\theta}\right|\,.
\end{eqnarray}

The above mixing parameters are exactly the same as the ones of case V. Hence the relations in Eq.~\eqref{eq:sum_rules_V} arise as well.

\item[~~(X)]{$X_{\nu\mathbf{r}}=\rho_{\mathbf{r}}(a^2bc^3d), \rho_{\mathbf{r}}(a^2bc^3d^3)$}

The PMNS mixing matrix is given by
{\footnotesize\begin{equation}
U_{PMNS}=\frac{1}{\sqrt{3}}
\begin{pmatrix}
e^{-\frac{3i\pi}{8}}\cos\left(\frac{\pi}{24}-\theta\right)-e^{\frac{i\pi}{8}}\cos\left(\frac{\pi}{24}+\theta\right) &  ~1 & ~-e^{-\frac{3i\pi}{8}}\sin\left(\frac{\pi}{24}-\theta\right)-e^{\frac{i\pi}{8}}\sin\left(\frac{\pi}{24}+\theta\right)\\

-e^{-\frac{3i\pi}{8}}\sin\left(\frac{5\pi}{24}-\theta\right)+e^{\frac{i\pi}{8}}\sin\left(\frac{5\pi}{24}+\theta\right) & ~1 &  ~-e^{-\frac{3i\pi}{8}}\cos\left(\frac{5\pi}{24}-\theta\right)-e^{\frac{i\pi}{8}}\cos\left(\frac{5\pi}{24}+\theta\right)\\

-e^{-\frac{3i\pi}{8}}\sin\left(\frac{\pi}{8}+\theta\right)+e^{\frac{i\pi}{8}}\sin\left(\frac{\pi}{8}-\theta\right) &  ~1 &   ~e^{-\frac{3i\pi}{8}}\cos\left(\frac{\pi}{8}+\theta\right)+e^{\frac{i\pi}{8}}\cos\left(\frac{\pi}{8}-\theta\right)
\end{pmatrix}\,,
\end{equation}}
which is the complex conjugate of the predicted PMNS matrix of case VI. The lepton mixing parameters are
\begin{eqnarray}
\nonumber&&\sin^2\theta_{13}=\frac{1}{3}-\frac{\sqrt{6}+\sqrt{2}}{12}\cos2\theta,\qquad \sin^2\theta_{12}=\frac{4}{8+\left(\sqrt{6}+\sqrt{2}\right)\cos2\theta},\\
\nonumber&&\sin^2\theta_{23}=\frac{4+\left(\sqrt{6}-\sqrt{2}\right)\cos2\theta}{8+\left(\sqrt{6}+\sqrt{2}\right)\cos2\theta},\qquad \left|J_{CP}\right|=\frac{1}{6\sqrt{3}}\left|\sin2\theta\right|,\\
\nonumber&&\left|\tan\delta_{CP}\right|=\left|\frac{4\sqrt{2}+\left(1+\sqrt{3}\right)\cos2\theta}{1-\sqrt{3}-\sqrt{2}\cos2\theta}\tan2\theta\right|,\\
\nonumber&&\left|\tan\alpha_{21}\right|=\left|\frac{\sqrt{6}+\sqrt{2}+4\cos2\theta-\left(\sqrt{6}-\sqrt{2}\right)\sin2\theta}{\sqrt{6}+\sqrt{2}+4\cos2\theta+\left(\sqrt{6}-\sqrt{2}\right)\sin2\theta}\right|,\\
\label{eq:mixing_parameters_X}&&\left|\tan\alpha^{\prime}_{31}\right|=\left|\frac{4\sin2\theta}{2-3\sqrt{3}+\left(2+\sqrt{3}\right)\cos4\theta}\right|\,.
\end{eqnarray}
They are exactly the same as the phenomenological predictions of case VI, as shown in Eq.~\eqref{eq:mixing_parameters_VI}.

\end{description}
For all the cases (case I --- case X) discussed above, the second column of the mixing matrix is constrained to be $\left(1,1,1\right)^{T}/\sqrt{3}$ due to the protection of the remnant family symmetry $G_{\nu}=Z^{(2)}_2$. As a result, the relation $3\sin^2\theta_{12}\cos^2\theta_{13}=1$ is satisfied such that the solar mixing angle has a lower limit given by $\sin^2\theta_{12}\geq1/3$.

\subsection{$G_{l}=Z^{(2)}_3$, $G_{\nu}=Z^{(9)}_2$}

In this scenario, only the residual CP symmetry $H^{l}_{CP}=\left\{\rho_{\mathbf{r}}(1), \rho_{\mathbf{r}}(ac), \rho_{\mathbf{r}}(a^2cd)\right\}$ is viable, and the hermitian combination $m^{\dagger}_{l}m_{l}$ is diagonal. Hence the lepton mixing is completely determined by the neutrino sector up to  permutations of rows and columns.

\begin{figure}[hptb!]\vskip-0.3in
\includegraphics[width=0.95\textwidth]{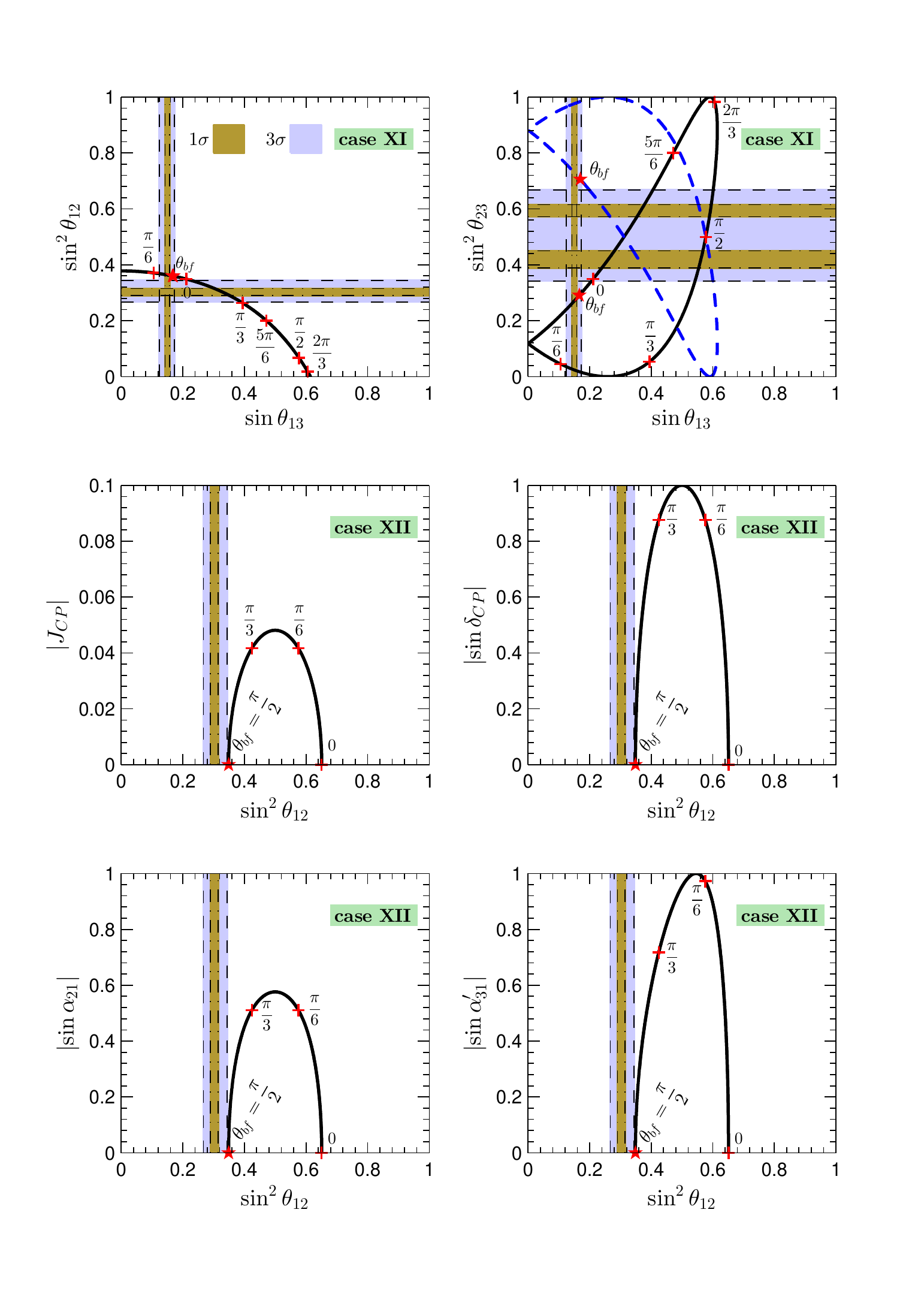} ~~~\caption{\label{fig:caseXI_XII}The relation among the lepton mixing parameters in case XI and case XII. The best fit value $\theta_{bf}$ for $\theta$ is labelled as a star. We also indicate the points for $\theta=0, \pi/6, \pi/3, \pi/2, 2\pi/3$ and $5\pi/6$ by the sign ``+'' on the curves, and only one $\theta$ value is displayed if there are points matching together. The shown $3\sigma$ ranges for the mixing angles and their best fit values are taken from Ref.~\cite{GonzalezGarcia:2012sz}. }
\end{figure}

\begin{description}[labelindent=-0.7em, leftmargin=0.1em]

\item[~~(XI)]{$X_{\nu\mathbf{r}}=\rho_{\mathbf{r}}(1), \rho_{\mathbf{r}}(a^2bd)$}

In this case, the lepton mixing matrix is determined to be
\begin{equation}
U_{PMNS}=\frac{1}{2\sqrt{3}}
\begin{pmatrix}

 -1-\sqrt{3}   &   ~\left(\sqrt{3}-1\right)\sin\theta+2\cos\theta  &  ~\left(\sqrt{3}-1\right)\cos\theta-2\sin\theta \\

2   &  ~2\sqrt{2}\cos\left(\frac{\pi}{4}-\theta\right)   &  ~2\sqrt{2}\cos\left(\frac{\pi}{4}+\theta\right)  \\

\sqrt{3}-1   &  ~-\left(1+\sqrt{3}\right)\sin\theta+2\cos\theta    &   ~-\left(1+\sqrt{3}\right)\cos\theta-2\sin\theta
\end{pmatrix}\,.
\end{equation}
The lepton mixing parameters are given by
\begin{eqnarray}
\nonumber&&\sin^2\theta_{13}=\frac{1}{12}\left[4-\sqrt{3}-\sqrt{3}\cos2\theta-2\left(\sqrt{3}-1\right)\sin2\theta\right],\\ \nonumber&&\sin^2\theta_{12}=\frac{4-\sqrt{3}+\sqrt{3}\cos2\theta+2\left(\sqrt{3}-1\right)\sin2\theta}{8+\sqrt{3}+\sqrt{3}\cos2\theta+2\left(\sqrt{3}-1\right)\sin2\theta},\\
\nonumber&&\sin^2\theta_{23}=\frac{4\left(1-\sin2\theta\right)}{8+\sqrt{3}+\sqrt{3}\cos2\theta+2\left(\sqrt{3}-1\right)\sin2\theta},\\
&&\sin\delta_{CP}=\sin\alpha_{21}=\sin\alpha_{31}=0\,.
\end{eqnarray}
Note that CP is fully conserved in this scenario. The results for the above predicted mixing angles are shown in Fig.~\ref{fig:caseXI_XII}, and the best fitting values for the mixing angles are collected in Table~\ref{tab:caseXI_XII}. We see that both $\sin^2\theta_{12}(\theta_{\text{bf}})$ and $\sin^2\theta_{23}(\theta_{\text{bf}})$ are slightly beyond the experimentally preferred $3\sigma$ range~\cite{GonzalezGarcia:2012sz}. Therefore this mixing pattern can marginally accommodate the experimental data.

\item[~~(XII)]{$X_{\nu\mathbf{r}}=\rho_{\mathbf{r}}(d^2), \rho_{\mathbf{r}}(a^2bd^3)$}

The PMNS matrix is of the form
\begin{equation}
U_{PMNS}=\frac{1}{2\sqrt{3}}
\begin{pmatrix}
-\left(1+\sqrt{3}\right)\sin\theta+2i\cos\theta   &  ~~-\left(1+\sqrt{3}\right)\cos\theta-2i\sin\theta   &  ~\sqrt{3}-1 \\

 2 i e^{-i \theta } &    2 e^{-i \theta}   &   2\\

\left(\sqrt{3}-1\right)\sin\theta+2i\cos\theta   &    ~~\left(\sqrt{3}-1\right)\cos\theta-2i\sin\theta  & ~-1-\sqrt{3}
\end{pmatrix}\,.
\end{equation}
The lepton mixing parameters are determined to be
\begin{eqnarray}
\nonumber&&\hskip-0.25in\sin^2\theta_{13}=\frac{1}{6}\left(2-\sqrt{3}\right),\qquad \sin^2\theta_{12}=\frac{1}{2}+\frac{\sqrt{3}\cos2\theta}{2\left(4+\sqrt{3}\right)},\qquad \sin^2\theta_{23}=\frac{2}{4+\sqrt{3}},\\
\nonumber&&\hskip-0.25in\left|J_{CP}\right|=\frac{1}{12\sqrt{3}}\left|\sin2\theta\right|,\qquad \left|\tan\delta_{CP}\right|=\left|\frac{\left(4+\sqrt{3}\right)\tan2\theta}{2\left(1+\sqrt{3}\right)}\right|,\\
&&\hskip-0.25in\left|\tan\alpha_{21}\right|=\left|\frac{8\left(3+\sqrt{3}\right)\sin2\theta}{29+16\sqrt{3}+3\cos4\theta}\right|,~ \left|\tan\alpha^{\prime}_{31}\right|=\left|\frac{2\left(1-\sqrt{3}\right)\sin2\theta}{3-2\sqrt{3}+\left(5-2\sqrt{3}\right)\cos2\theta}\right|\,.
\end{eqnarray}
Obviously both the reactor mixing angle $\theta_{13}$ and the atmospheric mixing angle $\theta_{23}$ are independent of the parameter $\theta$ here. The best fit value of $\theta$ is $\theta_{\text{bf}}=\pi/2$, since the minimal value of $\sin^2\theta_{12}$ is $\sin^2\theta_{12}\left(\theta_{\text{bf}}\right)=2/\left(4+\sqrt{3}\right)$. For $\theta_{\text{bf}}=\pi/2$, all the three CP phases are trivial with $\sin\alpha_{21}\left(\theta_{\text{bf}}\right)=\sin\alpha_{31}\left(\theta_{\text{bf}}\right)=\sin\delta_{CP}\left(\theta_{\text{bf}}\right)=0$ whereas the resulting $\sin^2\theta_{12}\left(\theta_{\text{bf}}\right)$ and $\sin^2\theta_{13}\left(\theta_{\text{bf}}\right)$ are slightly larger than their $3\sigma$ upper bounds~\cite{GonzalezGarcia:2012sz}, as shown in Table~\ref{tab:caseXI_XII}. The correlations of $\left|J_{CP}\right|$, $\left|\sin\delta_{CP}\right|$, $\left|\sin\alpha_{21}\right|$ and $\left|\sin\alpha^{\prime}_{31}\right|$ with respect to $\sin\theta_{13}$ are displayed in Fig.~\ref{fig:caseXI_XII}.

\item[~~(XIII)]{$X_{\nu\mathbf{r}}=\rho_{\mathbf{r}}(c^2d), \rho_{\mathbf{r}}(a^2bc^2d^2)$}

The PMNS matrix is of the form
{\small\begin{equation}
\label{eq:PMNS_XIII}U_{PMNS}=\frac{1}{2\sqrt{3}}
\begin{pmatrix}
-1-\sqrt{3}  &   ~\left(\sqrt{3}-1\right)\sin\theta+2e^{-\frac{i\pi}{4}}\cos\theta  &   ~~\left(\sqrt{3}-1\right)\cos\theta-2e^{-\frac{i\pi}{4}}\sin\theta \\

2    &   ~2\sin\theta+2e^{-\frac{i\pi}{4}}\cos\theta  &   2\cos\theta-2e^{-\frac{i\pi}{4}}\sin\theta  \\

\sqrt{3}-1   &  ~-\left(1+\sqrt{3}\right)\sin\theta+2e^{-\frac{i\pi}{4}}\cos\theta & ~~-\left(1+\sqrt{3}\right)\cos\theta-2e^{-\frac{i\pi}{4}}\sin\theta
\end{pmatrix}\,.
\end{equation}}
The lepton mixing parameters nontrivially depend on the parameter $\theta$ as follows:
\begin{eqnarray}
\nonumber&&\sin^2\theta_{13}=\frac{1}{12}\left[4-\sqrt{3}-\sqrt{3}\cos2\theta-\left(\sqrt{6}-\sqrt{2}\right)\sin2\theta\right],\\ \nonumber&&\sin^2\theta_{12}=\frac{4-\sqrt{3}+\sqrt{3}\cos2\theta+\left(\sqrt{6}-\sqrt{2}\right)\sin2\theta}{8+\sqrt{3}+\sqrt{3}\cos2\theta+\left(\sqrt{6}-\sqrt{2}\right)\sin2\theta}\,,\\
\nonumber&&\sin^2\theta_{23}=\frac{4-2\sqrt{2}\sin2\theta}{8+\sqrt{3}+\sqrt{3}\cos2\theta+\left(\sqrt{6}-\sqrt{2}\right)\sin2\theta},\qquad \left|J_{CP}\right|=\frac{1}{12\sqrt{6}}\left|\sin2\theta\right|\,,\\
\nonumber&&\left|\tan\delta_{CP}\right|=\left|\frac{\left(6-2\sqrt{3}\right)\left(1-\cos4\theta\right)+\left(6\sqrt{2}+16\sqrt{6}\right)\sin2\theta+3\sqrt{2}\sin4\theta}{24+18\sqrt{3}+\left(24-8\sqrt{3}\right)\cos2\theta-6\sqrt{2}\sin2\theta+6\sqrt{3}\cos4\theta-\left(15\sqrt{2}+4\sqrt{6}\right)\sin4\theta}\right|\,,\\
\nonumber&&\left|\tan\alpha_{21}\right|=\left|\frac{2\left(2+\sqrt{3}\right)\left(1+\cos2\theta\right)+\left(\sqrt{6}+\sqrt{2}\right)\sin2\theta}{1-\cos2\theta+\left(\sqrt{6}+\sqrt{2}\right)\sin2\theta}\right|\,,\\
\label{eq:mixing_parameters_XIII}&&\left|\tan\alpha^{\prime}_{31}\right|=\left|\frac{2\left(2+\sqrt{3}\right)\left(1-\cos2\theta\right)-\left(\sqrt{6}+\sqrt{2}\right)\sin2\theta}{1+\cos2\theta-\left(\sqrt{6}+\sqrt{2}\right)\sin2\theta}\right|\,.
\end{eqnarray}
The relation among the different mixing parameters are shown in Fig.~\ref{fig:caseXIII_XIV}. The best fitting results are listed in Table~\ref{tab:caseXIII_XIV}. The experimental data can be marginally accommodated.

\item[~~(XIV)]{$X_{\nu\mathbf{r}}=\rho_{\mathbf{r}}(c^2d^3), \rho_{\mathbf{r}}(a^2bc^2)$}

The PMNS matrix is given by
\begin{equation}
U_{PMNS}=\frac{1}{2\sqrt{3}}
\begin{pmatrix}
-1-\sqrt{3}    &  ~~i\left(1-\sqrt{3}\right)\sin\theta+2e^{-\frac{i\pi}{4}}\cos\theta & ~~i\left(1-\sqrt{3}\right)\cos\theta-2e^{-\frac{i\pi}{4}}\sin\theta   \\

2    &   ~~-2i\sin\theta+2e^{-\frac{i\pi}{4}}\cos\theta   &   -2i\cos\theta-2e^{-\frac{i\pi}{4}}\sin\theta  \\

\sqrt{3}-1   &  ~~i\left(1+\sqrt{3}\right)\sin\theta+2e^{-\frac{i\pi}{4}}\cos\theta &  ~~i\left(1+\sqrt{3}\right)\cos\theta-2e^{-\frac{i\pi}{4}}\sin\theta
\end{pmatrix}\,.
\end{equation}
It is related to the resulting PMNS matrix of case XIII via $U^{\text{XIV}}_{PMNS}=U^{\text{XIII}^{*}}_{PMNS}\;\text{diag}\left(1,-i,-i\right)$. We can straightforwardly calculate the lepton mixing parameters
\begin{eqnarray}
\nonumber&&\sin^2\theta_{13}=\frac{1}{12}\left[4-\sqrt{3}-\sqrt{3}\cos2\theta-\left(\sqrt{6}-\sqrt{2}\right)\sin2\theta\right]\,,\\
\nonumber&&\sin^2\theta_{12}=\frac{4-\sqrt{3}+\sqrt{3}\cos2\theta+\left(\sqrt{6}-\sqrt{2}\right)\sin2\theta}{8+\sqrt{3}+\sqrt{3}\cos2\theta+\left(\sqrt{6}-\sqrt{2}\right)\sin2\theta}\,,\\
\nonumber&&\sin^2\theta_{23}=\frac{4-2\sqrt{2}\sin2\theta}{8+\sqrt{3}+\sqrt{3}\cos2\theta+\left(\sqrt{6}-\sqrt{2}\right)\sin2\theta},\qquad \left|J_{CP}\right|=\frac{1}{12\sqrt{6}}\left|\sin2\theta\right|\,,\\
\nonumber&&\left|\tan\delta_{CP}\right|=\left|\frac{\left(6-2\sqrt{3}\right)(1-\cos4\theta)+\left(6\sqrt{2}+16\sqrt{6}\right)\sin2\theta+3\sqrt{2}\sin4\theta}{24+18\sqrt{3}+\left(24-8\sqrt{3}\right)\cos2\theta-6\sqrt{2}\sin2\theta+6\sqrt{3}\cos4\theta-\left(15\sqrt{2}+4\sqrt{6}\right)\sin4\theta}\right|\,,\\
\nonumber&&\left|\tan\alpha_{21}\right|=\left|\frac{2\left(2+\sqrt{3}\right)\left(1+\cos2\theta\right)+\left(\sqrt{6}+\sqrt{2}\right)\sin2\theta}{1-\cos2\theta+\left(\sqrt{6}+\sqrt{2}\right)\sin2\theta}\right|\,,\\
&&\left|\tan\alpha^{\prime}_{31}\right|=\left|\frac{2\left(2+\sqrt{3}\right)\left(1-\cos2\theta\right)-\left(\sqrt{6}+\sqrt{2}\right)\sin2\theta}{1+\cos2\theta-\left(\sqrt{6}+\sqrt{2}\right)\sin2\theta}\right|\,.
\end{eqnarray}
Compared with Eq.~\eqref{eq:mixing_parameters_XIII}, we see that the above mixing parameters are of the same form as those of case XIII.

\end{description}

\subsection{$G_{l}=Z^{(2)}_3$, $G_{\nu}=Z^{(10)}_2$}

Similar to previous cases, the lepton flavor mixing completely comes from the neutrino sector. The PMNS matrix is fixed up to permutations of rows and columns.

\begin{figure}[hptb!]\hskip-0.2in
\hskip-5mm\includegraphics[width=1.1\textwidth]{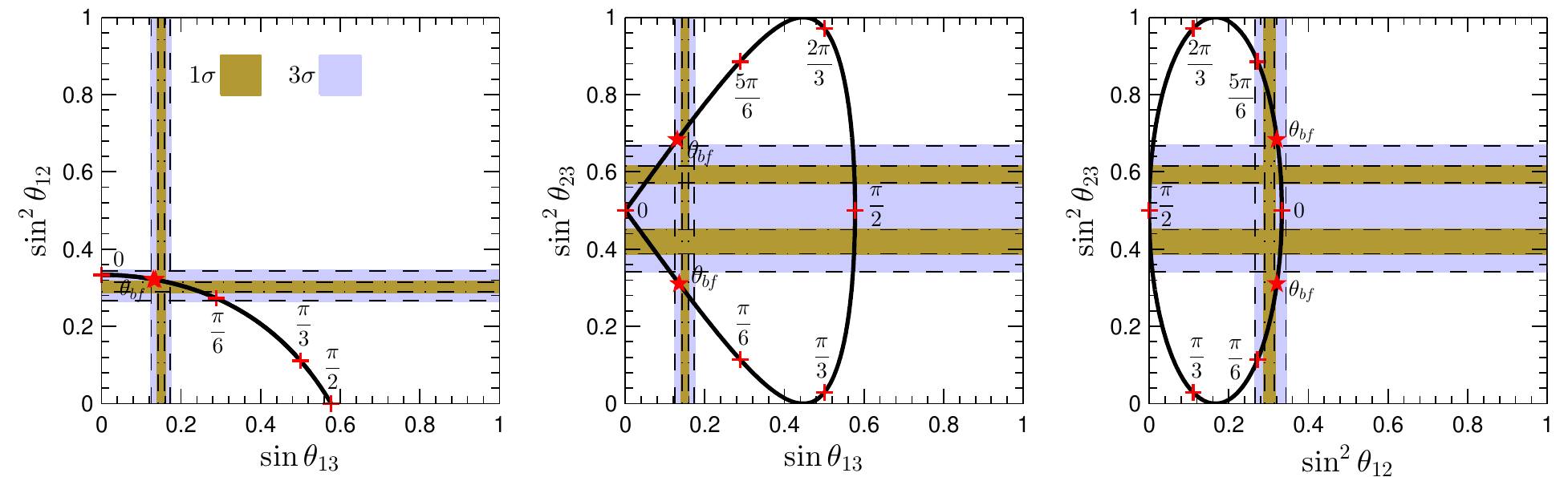} ~~~\caption{\label{fig:caseXV} The relation among the lepton mixing angles in case XV. Note that the curves for $\sin^2\theta_{12}$ with respect to $\sin\theta_{13}$ coincidence in case XV, case XVI, case XVII and case XVIII because they are predicted to be of the same form, as shown in Table~\ref{tab:caseXV_XVIII}.  The best fit value $\theta_{bf}$ for $\theta$ is labelled as a star. We also indicate the points for $\theta=0, \pi/6, \pi/3, \pi/2, 2\pi/3$ and $5\pi/6$ by the sign ``+'' on the curves, and only one $\theta$ value is displayed if there are points matching together. The shown $3\sigma$ ranges for the mixing angles and their best fit values are taken from Ref.~\cite{GonzalezGarcia:2012sz}. }
\end{figure}

\begin{figure}[hptb!]
\includegraphics[width=1.0\textwidth]{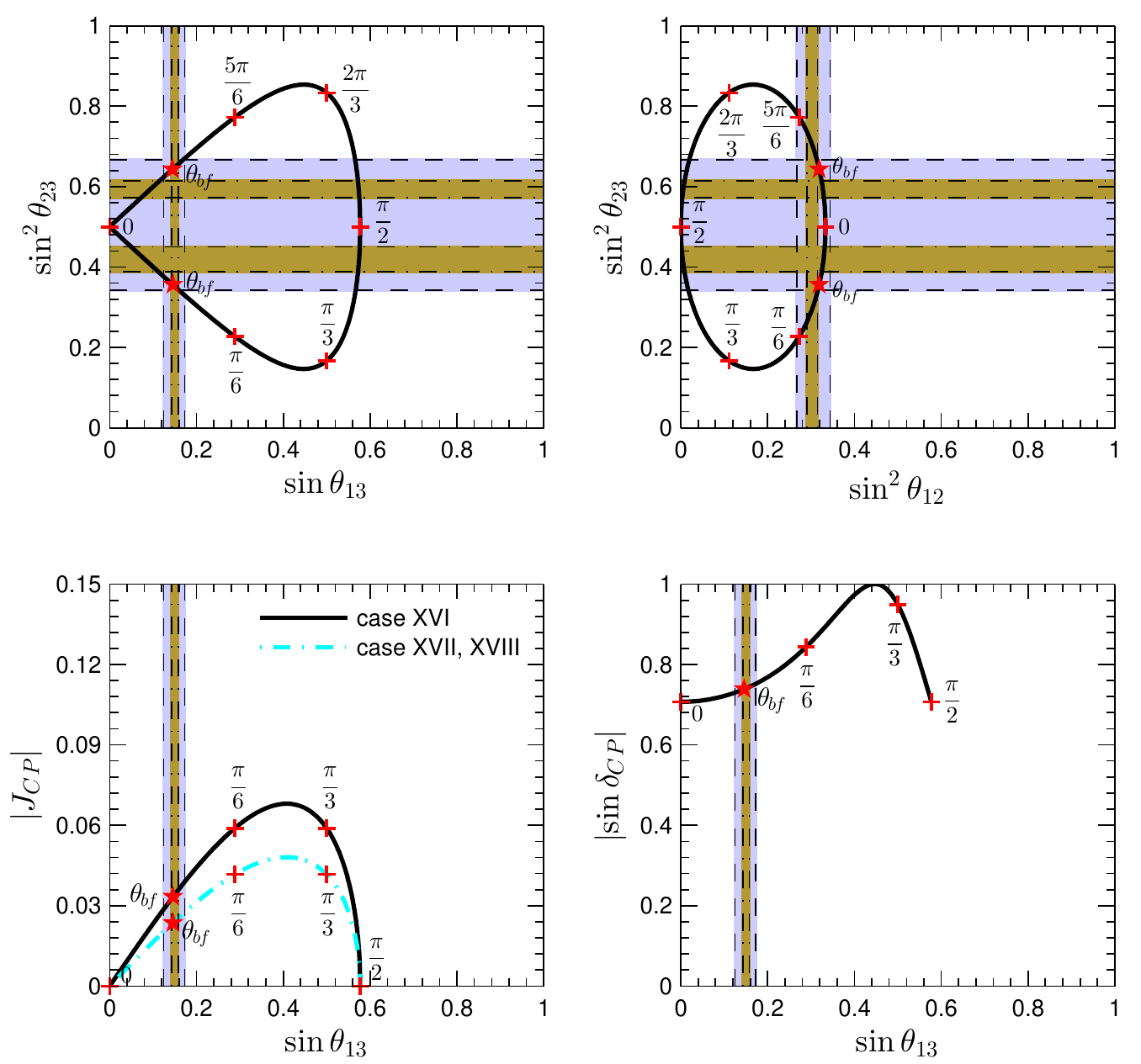} ~~~\caption{\label{fig:caseXVII}The relation among the lepton mixing parameters in case XVII and case XVIII. Notice that the curves of $\sin^2\theta_{23}$ versus $\sin\theta_{13}$ (or $\sin^2\theta_{23}$ versus $\sin\theta_{12}$) for $\theta_{23}(\theta_{\text{bf}})<\pi/4$ and $\theta_{23}(\theta_{\text{bf}})>\pi/4$ coincide with each other. The best fit value $\theta_{bf}$ for $\theta$ is labelled as a star. We also indicate the points for $\theta=0, \pi/6, \pi/3, \pi/2, 2\pi/3$ and $5\pi/6$ by the sign ``+'' on the curves, and only one $\theta$ value is displayed if there are points matching together. The shown $3\sigma$ ranges for the mixing angles and their best fit values are taken from Ref.~\cite{GonzalezGarcia:2012sz}. Note that the plot for $\sin^2\theta_{12}$ with respect to $\sin\theta_{13}$ is the same as that for case XV and can be found in Fig.~\ref{fig:caseXV}. }
\end{figure}

\begin{description}[labelindent=-0.7em, leftmargin=0.1em]

\item[~~(XV)]{$X_{\nu\mathbf{r}}=\rho_{\mathbf{r}}(1), \rho_{\mathbf{r}}(a^2bd^2)$}

In this case, the lepton mixing matrix is of the following form
\begin{equation}
U_{PMNS}=\frac{1}{\sqrt{6}}
\begin{pmatrix}
2 & \sqrt{2}\;\cos\theta & \sqrt{2}\;\sin\theta \\
-1 & ~\sqrt{2}\;\cos\theta+\sqrt{3}\;\sin\theta & ~\sqrt{2}\;\sin\theta-\sqrt{3}\;\cos\theta \\
-1 & ~\sqrt{2}\;\cos\theta-\sqrt{3}\;\sin\theta & ~\sqrt{2}\;\sin\theta+\sqrt{3}\;\cos\theta
\end{pmatrix}\,.
\end{equation}
We see that the first column of the PMNS matrix is $\left(2,-1,-1\right)^{T}/\sqrt{6}$, this mixing pattern is the so-called $\text{TM}_1$ mixing. Since the PMNS matrix is real, there is no CP violation in this scenario. The lepton mixing parameters read
\begin{eqnarray}
\nonumber&&\sin^2\theta_{13}=\frac{1}{6}\left(1-\cos2\theta\right),\qquad \sin^2\theta_{12}=\frac{1+\cos2\theta}{5+\cos2\theta},\\
\label{eq:mixing_parameters_XV}&&\sin^2\theta_{23}=\frac{1}{2}-\frac{\sqrt{6}\sin2\theta}{5+\cos2\theta},\qquad \tan\delta_{CP}=\tan\alpha_{21}=\tan\alpha_{31}=0\,.
\end{eqnarray}
As a consequence, the mixing angles are related with each other as
\begin{eqnarray}
\label{eq:relations_XV}3\cos^2\theta_{12}\cos^2\theta_{13}=2,\qquad \sin^2\theta_{23}=\frac{1}{2}\pm\frac{\sqrt{2-6\sin^2\theta_{13}}}{\cos^2\theta_{13}}\sin\theta_{13}\,.
\end{eqnarray}
The correlations among the mixing angles are presented in Fig.~\ref{fig:caseXV}. As is shown in Table~\ref{tab:caseXV_XVIII}, we can  find a value of $\theta$ for which the resulting mixing angles agree rather well with the experimental observations~\cite{GonzalezGarcia:2012sz}.

\item[~~(XVI)]{$X_{\nu\mathbf{r}}=\rho_{\mathbf{r}}(d^2), \rho_{\mathbf{r}}(a^2b)$}

In this case, the PMNS matrix is determined to be
\begin{equation}
U_{PMNS}=\frac{1}{\sqrt{6}}
\begin{pmatrix}
2 & ~~\sqrt{2}\;\cos\theta & ~~\sqrt{2}\;\sin\theta \\
-1  & ~~\sqrt{2}\;\cos\theta+i\sqrt{3}\;\sin\theta &    ~~\sqrt{2}\;\sin\theta-i\sqrt{3}\;\cos\theta \\
 -1 & ~~\sqrt{2}\;\cos\theta-i\sqrt{3}\;\sin\theta & ~~\sqrt{2}\;\sin\theta+i\sqrt{3}\;\cos\theta
\end{pmatrix}\,.
\end{equation}
The lepton mixing parameters are
\begin{eqnarray}
\nonumber&&\sin^2\theta_{13}=\frac{1}{6}\left(1-\cos2\theta\right),\qquad \sin^2\theta_{12}=\frac{1+\cos2\theta}{5+\cos2\theta},\qquad  \sin^2\theta_{23}=\frac{1}{2}\,,\\
&&\left|J_{CP}\right|=\frac{1}{6\sqrt{6}}\left|\sin2\theta\right|,\qquad \cot\delta_{CP}=\tan\alpha_{21}=\tan\alpha_{31}=0 \,.
\end{eqnarray}
Note that the atmospheric mixing angle and the Dirac CP phase are predicted to be maximal while both Majorana phases are trivial. The best fitting values are presented in Table~\ref{tab:caseXV_XVIII}, and excellent agreement with the experimental observations could be achieved.

\item[~~(XVII)]{$X_{\nu\mathbf{r}}=\rho_{\mathbf{r}}(c^2d), \rho_{\mathbf{r}}(a^2bc^2d^3)$}

The PMNS matrix is given by
\begin{equation}
\label{eq:PMNS_VII}U_{PMNS}=\frac{1}{\sqrt{6}}
\begin{pmatrix}
2 & \sqrt{2}\;e^{-\frac{i\pi}{4}}\cos\theta   & \sqrt{2}\;e^{-\frac{i\pi}{4}}\sin\theta  \\

-1   & ~\sqrt{3}\;\sin\theta+\sqrt{2}\;e^{-\frac{i\pi}{4}}\cos\theta & ~-\sqrt{3}\;\cos\theta+\sqrt{2}\;e^{-\frac{i\pi}{4}}\sin\theta \\

-1 &  ~-\sqrt{3}\;\sin\theta+\sqrt{2}\;e^{-\frac{i\pi}{4}}\cos\theta  & ~\sqrt{3}\;\cos\theta+\sqrt{2}\;e^{-\frac{i\pi}{4}}\sin\theta
\end{pmatrix}\,.
\end{equation}
The lepton mixing parameters are
\begin{eqnarray}
\nonumber&&\hskip-0.2in\sin^2\theta_{13}=\frac{1}{6}\left(1-\cos2\theta\right),~~ \sin^2\theta_{12}=\frac{1+\cos2\theta}{5+\cos2\theta},~~ \sin^2\theta_{23}=\frac{1}{2}-\frac{\sqrt{3}\sin2\theta}{5+\cos2\theta},\\
\label{eq:mixing_parameters_XVII}&&\hskip-0.2in\left|J_{CP}\right|=\frac{1}{12\sqrt{3}}\left|\sin2\theta\right|,~~ \left|\tan\delta_{CP}\right|=\left|\frac{5+\cos2\theta}{1+5\cos2\theta}\right|,~~ \cot\alpha_{21}=\cot\alpha^{\prime}_{31}=0\,.
\end{eqnarray}
We have maximal Majorana CP violation with $\left|\sin\alpha_{21}\right|=\left|\sin\alpha^{\prime}_{31}\right|=1$ in this case. There is a deviation of the atmospheric angle $\theta_{23}$ from maximal mixing. The three mixing angles are correlated with each other as
\begin{eqnarray}
\label{eq:relations_XVII}3\cos^2\theta_{12}\cos^2\theta_{13}=2,\qquad \sin^2\theta_{23}=\frac{1}{2}\pm\frac{\sqrt{1-3\sin^2\theta_{13}}}{\cos^2\theta_{13}}\sin\theta_{13}\,.
\end{eqnarray}
With the measured reactor mixing angle $\sin^2\theta_{13}=0.0227$, the other two mixing angles are determined to be $\sin^2\theta_{12}\simeq0.318$, $\sin^2\theta_{23}\simeq0.351$ or $\sin^2\theta_{23}\simeq0.649$, which are in the experimentally preferred ranges. The above results for the mixing parameters are displayed in Fig.~\ref{fig:caseXVII}. It is remarkable that the Dirac CP phase is always nontrivial in this case, and its best fitting value fulfills $\left|\sin\delta_{\text{CP}}(\theta_{\text{bf}})\right|\simeq0.738$.

\item[~~(XVIII)]{$X_{\nu\mathbf{r}}=\rho_{\mathbf{r}}(c^2d^3), \rho_{\mathbf{r}}(a^2bc^2d)$}

The lepton mixing matrix takes the form
\begin{equation}
U_{PMNS}=\frac{1}{\sqrt{6}}
\begin{pmatrix}
2  & ~~\sqrt{2}\;e^{-\frac{i\pi}{4}}\cos\theta   & ~~\sqrt{2}\;e^{-\frac{i\pi}{4}}\sin\theta  \\

-1 & ~~\sqrt{2}\;e^{-\frac{i\pi}{4}}\cos\theta-\sqrt{3}\;i\sin\theta & ~~\sqrt{2}\;e^{-\frac{i\pi}{4}}\sin\theta+\sqrt{3}\;i\cos\theta \\

-1 & ~~\sqrt{2}\;e^{-\frac{i\pi}{4}}\cos\theta+\sqrt{3}\;i\sin\theta  & ~~\sqrt{2}\;e^{-\frac{i\pi}{4}}\sin\theta-\sqrt{3}\;i\cos\theta
\end{pmatrix}\,,
\end{equation}
which is related to the PMNS matrix of case XVII via $U^{\text{XVIII}}_{PMNS}=U^{\text{XVII}^{*}}_{PMNS}\;\text{diag}\left(1,-i,-i\right)$.
The lepton mixing parameters read
\begin{eqnarray}
\nonumber&&\hskip-0.2in\sin^2\theta_{13}=\frac{1}{6}\left(1-\cos2\theta\right),~~ \sin^2\theta=\frac{1+\cos2\theta}{5+\cos2\theta},~~
\sin^2\theta_{23}=\frac{1}{2}-\frac{\sqrt{3}\sin2\theta}{5+\cos2\theta},\\
&&\hskip-0.2in\left|J_{CP}\right|=\frac{1}{12\sqrt{3}}\left|\sin2\theta\right|,~~ \left|\tan\delta_{CP}\right|=\left|\frac{5+\cos2\theta}{1+5\cos2\theta}\right|,~~ \cot\alpha_{21}=\cot\alpha^{\prime}_{31}=0\,.
\end{eqnarray}
They coincide with already predicted mixing parameters of case XVII, as shown in Eq.~\eqref{eq:mixing_parameters_XVII}. As a result, the correlations in Eq.~\eqref{eq:relations_XVII} are satisfied, and the experimental data can be accommodated very well.

\end{description}

For the above discussed case XV, case XVI, case XVII and case XVIII, the first column of the PMNS martrix is $\left(2,-1,-1\right)^{T}/\sqrt{6}$. As a consequence, the relation $3\cos^2\theta_{12}\cos^2\theta_{13}=2$ is always fulfilled such that $\theta_{12}$ has a upper bound $\sin^2\theta_{12}\leq1/3$.

\subsection{$G_{l}=K^{(3)}_4$, $G_{\nu}=Z^{(9)}_2$}

Now the contribution from the charged lepton sector to the lepton mixing is nontrivial, and it takes the form of Eq.~\eqref{eq:Ul_K43}. Combining the unitary transformation $U_{\nu}$ from the neutrino sector, which is studied in section~\ref{subsubsec:Gnu_z29}, we can straightforwardly obtain the predictions for the lepton flavor mixing matrix.

\begin{figure}[hptb!]\vskip-0.3in
\includegraphics[width=0.95\textwidth]{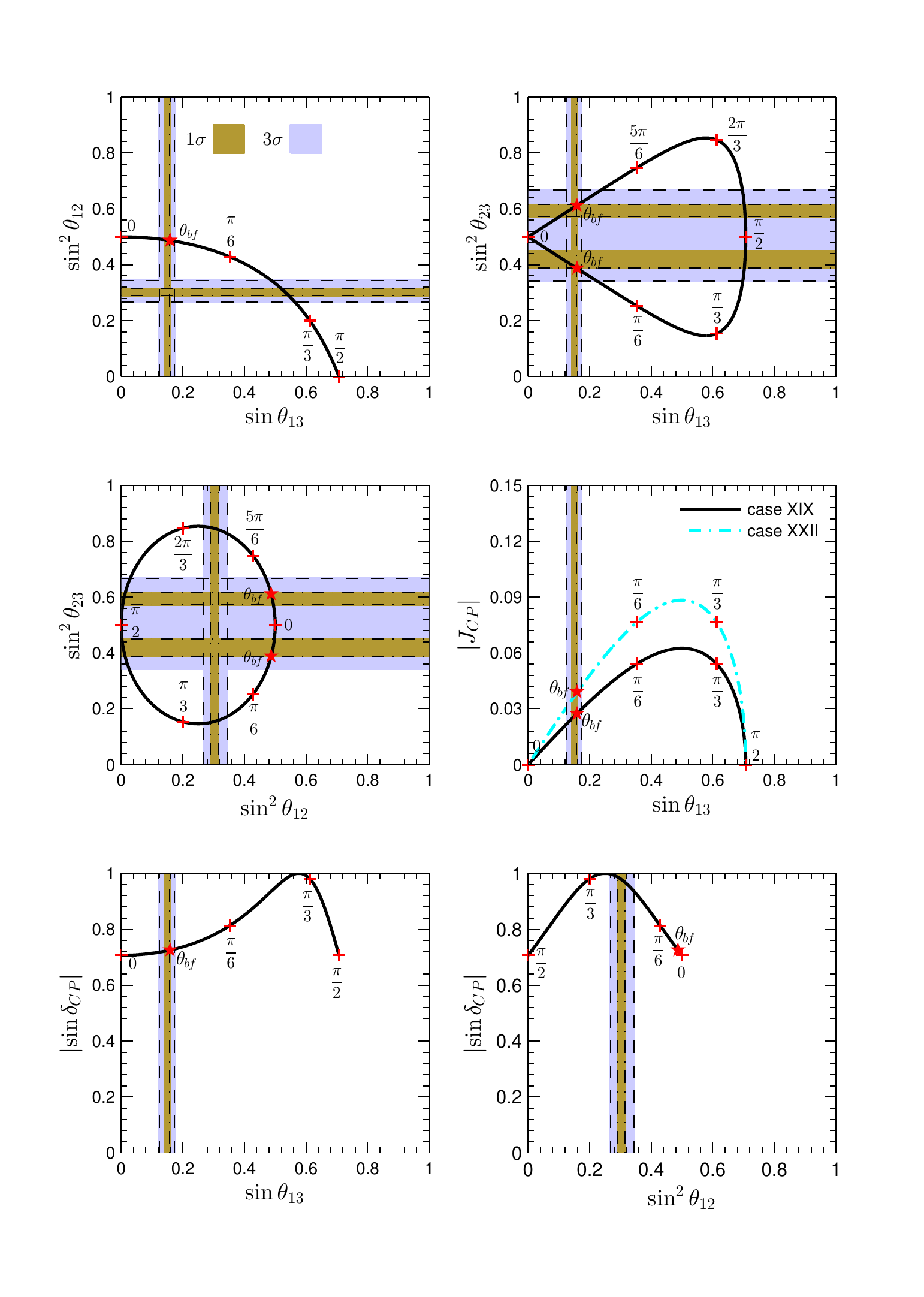} ~~~\caption{\label{fig:caseXIX_XXII}The relation among the lepton mixing parameters in cases XIX, XX, XXV, XXVI and cases XXII, XXIII. The best fit value $\theta_{bf}$ for $\theta$ is labelled as a star. We also indicate the points for $\theta=0, \pi/6, \pi/3, \pi/2, 2\pi/3$ and $5\pi/6$ by the sign ``+'' on the curves, and only one $\theta$ value is displayed if there are points matching together. The shown $3\sigma$ ranges for the mixing angles and their best fit values are taken from Ref.~\cite{GonzalezGarcia:2012sz}. Note that the figure for $\sin^2\theta_{12}$ versus $\sin\theta_{13}$ is the same in all the cases considered here. }
\end{figure}

\begin{figure}[t!]\hskip-0.2in
\hskip-5mm\includegraphics[width=1.1\textwidth]{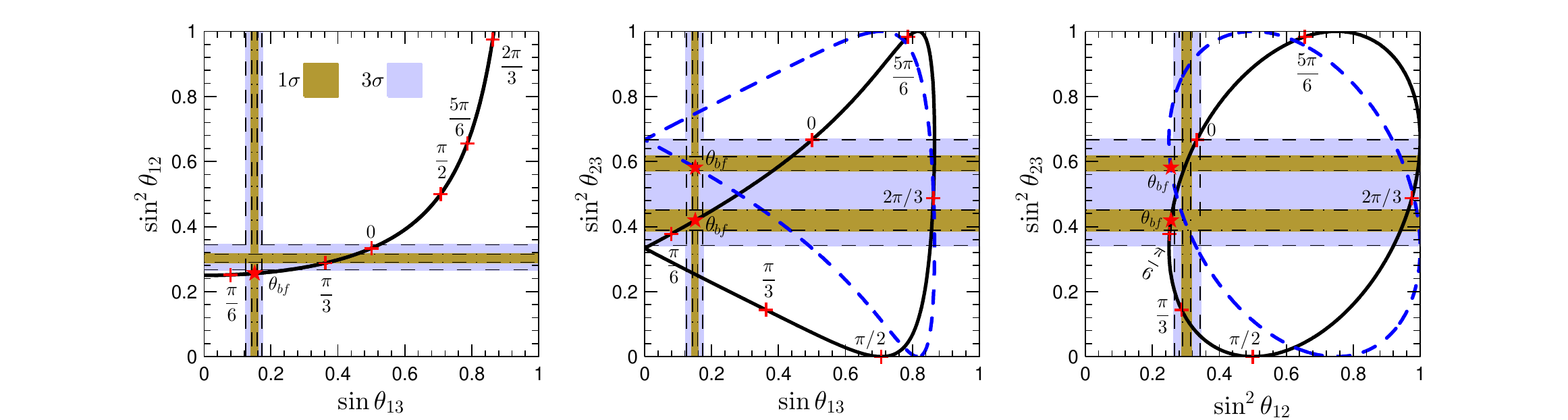} ~~~\caption{\label{fig:caseXXI} The relation among the lepton mixing angles in case XXI and case XXIV. The best fit value $\theta_{bf}$ for $\theta$ is labelled as a star. We also indicate the points for $\theta=0, \pi/6, \pi/3, \pi/2, 2\pi/3$ and $5\pi/6$ by the sign ``+'' on the curves, and only one $\theta$ value is displayed if there are points matching together. The shown $3\sigma$ ranges for the mixing angles and their best fit values are taken from Ref.~\cite{GonzalezGarcia:2012sz}. }
\end{figure}

\begin{description}[labelindent=-0.7em, leftmargin=0.1em]

\item[~~(XIX)]{$X_{\nu\mathbf{r}}=\rho_{\mathbf{r}}(1), \rho_{\mathbf{r}}(a^2bd)$}

In this case, the PMNS matrix is
\begin{equation}
\label{eq:PMNS_XIX}U_{PMNS}=\frac{1}{2}
\begin{pmatrix}
\sqrt{2} &    ~i\sqrt{2}\cos\theta & ~i\sqrt{2}\sin\theta \\
1 & ~-i\cos\theta+\sqrt{2}\;e^{-\frac{i\pi}{4}}\sin\theta
   & ~-i\sin\theta-\sqrt{2}\;e^{-\frac{i\pi}{4}}\cos\theta\\
1 & ~-i\cos\theta-\sqrt{2}\;e^{-\frac{i\pi}{4}}\sin\theta & ~-i\sin\theta+\sqrt{2}\;e^{-\frac{i\pi}{4}}\cos\theta
\end{pmatrix}\,.
\end{equation}
The lepton mixing parameters are found to be
\begin{eqnarray}
\nonumber&&\sin^2\theta_{13}=\frac{1}{4}\left(1-\cos2\theta\right),~~ \sin^2\theta_{12}=\frac{1+\cos2\theta}{3+\cos2\theta},~~ \sin^2\theta_{23}=\frac{1}{2}-\frac{\sin2\theta}{3+\cos2\theta},\\
\label{eq:mixing_parameters_XIX}&&\left|J_{CP}\right|=\frac{1}{16}\left|\sin2\theta\right|,~~ \left|\tan\delta_{CP}\right|=\left|\frac{3+\cos2\theta}{1+3\cos2\theta}\right|,~~ \tan\alpha_{21}=\tan\alpha^{\prime}_{31}=0 \,.
\end{eqnarray}
Notice that the Majorana CP is conserved, and the remaining other mixing parameters nontrivially depend on the parameter $\theta$. The mixing angles are related by
\begin{equation}
2\cos^2\theta_{12}\cos^2\theta_{13}=1,\qquad \sin^2\theta_{23}=\frac{1}{2}\pm\frac{\sqrt{2-4\sin^2\theta_{13}}}{2\cos^2\theta_{13}}\sin\theta_{13}\,.
\end{equation}
The correlations between different mixing parameters are illustrated in Fig.~\ref{fig:caseXIX_XXII}, we see that the correct values of $\theta_{13}$, $\theta_{12}$ and $\theta_{23}$ can not be reproduced in this case. For the $3\sigma$ range of the reactor mixing angle $0.0156\leq\sin^2\theta_{13}\leq0.0299$~\cite{GonzalezGarcia:2012sz}, the atmospheric mixing angle is calculated to be in the range of $[0.377,0.412]\cup [0.588,0.622]$ which is compatible with the experimental data. However, the solar angle is constrained to vary in the range of $[0.484,0.492]$ which has no overlap with the experimentally preferred $3\sigma$ region. Hence the minimal value of the $\chi^2$ function is rather large: 204.875 and 204.610 for the first octant and the second octant $\theta_{23}$ respectively.

\item[~~(XX)]{$X_{\nu\mathbf{r}}=\rho_{\mathbf{r}}(d^2), \rho_{\mathbf{r}}(a^2bd^3)$}

The PMNS matrix is
\begin{eqnarray}
\label{eq:PMNS_XX}U_{PMNS}=\frac{1}{2}
\begin{pmatrix}
 \sqrt{2} & ~\sqrt{2}\;i\cos \theta & ~\sqrt{2}\;i\sin \theta \\
 1 & ~-i\cos\theta-\sqrt{2}\; e^{\frac{i\pi}{4}} \sin \theta & ~-i\sin\theta+\sqrt{2}\;e^{\frac{i\pi}{4}}\cos\theta \\
1 & ~-i\cos\theta+\sqrt{2}\;e^{\frac{i\pi}{4}}\sin\theta & ~-i\sin\theta-\sqrt{2}\;e^{\frac{i\pi}{4}}\cos\theta
\end{pmatrix}\,.
\end{eqnarray}
From Eq.~\eqref{eq:PMNS_XIX}, we see that this PMNS matrix is closely related to case XIX's prediction as $U^{\text{XX}}_{PMNS}=U^{\text{XIX}^{*}}_{PMNS}\;\text{diag}\left(1,-1,-1\right)$. The lepton mixing parameters are
\begin{eqnarray}
\nonumber&&\sin^2\theta_{13}=\frac{1}{4}\left(1-\cos2\theta\right),~~ \sin^2\theta_{12}=\frac{1+\cos2\theta}{3+\cos2\theta},~~ \sin^2\theta_{23}=\frac{1}{2}-\frac{\sin2\theta}{3+\cos2\theta}\,,\\
\label{eq:mixing_parameters_XX}&&\left|J_{CP}\right|=\frac{1}{16}\left|\sin2\theta\right|,~~ \left|\tan\delta_{CP}\right|=\left|\frac{3+\cos2\theta}{1+3\cos2\theta}\right|,~~ \tan\alpha_{21}=\tan\alpha^{\prime}_{31}=0\,,
\end{eqnarray}
which are the same as the phenomenological predictions of case XIX. Hence the observed values of $\theta_{12}$, $\theta_{13}$ and $\theta_{23}$ can not be achieved simultaneously in this scenario.

\item[~~(XXI)]{$X_{\nu\mathbf{r}}=\rho_{\mathbf{r}}(c^2d), \rho_{\mathbf{r}}(a^2bc^2d^2)$}

In this case, the lepton mixing matrix takes the form
\begin{equation}
U_{PMNS}=\frac{1}{2}
\begin{pmatrix}
 -i \left(\sqrt{2} \cos \theta+\sin\theta\right) & ~1 & ~-i \left(\cos\theta-\sqrt{2}\sin\theta\right) \\
 i \sqrt{2} \sin \theta & ~\sqrt{2} & ~i \sqrt{2}   \cos \theta \\
 i \left(\sqrt{2} \cos \theta-\sin \theta\right)
   & ~1 & ~-i \left(\cos \theta+\sqrt{2} \sin\theta\right)
\end{pmatrix}\,.
\end{equation}
The lepton mixing parameters are
\begin{eqnarray}
\nonumber&&\sin^2\theta_{13}=\frac{1}{8}\left(3-\cos2\theta-2\sqrt{2}\sin2\theta\right),\qquad \sin^2\theta_{12}=\frac{2}{5+\cos2\theta+2\sqrt{2}\sin2\theta},\\
\label{eq:XXI_mixing_parameters}&&\sin^2\theta_{23}=\frac{2+2\cos2\theta}{5+\cos2\theta+2\sqrt{2}\sin2\theta},\qquad \tan\delta_{CP}=\tan\alpha_{21}=\tan\alpha_{31}=0\,.
\end{eqnarray}
We have CP conservation in this scenario. The three mixing angles are related with each as
\begin{eqnarray}
\nonumber&&4\sin^2\theta_{12}\cos^2\theta_{13}=1,\\
\nonumber&&9\sin^2\theta_{23}\cos^2\theta_{13}=3-2\sin^2\theta_{13}\pm2\sin\theta_{13}\sqrt{6-8\sin^2\theta_{13}}~\,,\qquad \theta_{23}<\pi/4,\\
&&9\sin^2\theta_{23}\cos^2\theta_{13}=6-7\sin^2\theta_{13}\mp2\sin\theta_{13}\sqrt{6-8\sin^2\theta_{13}}~\,,\qquad \theta_{23}<\pi/4\,.
\end{eqnarray}
As a consequence, we have $\sin^2\theta_{12}=1/\left(4\cos^2\theta_{13}\right)$ and the solar mixing angle is predicted to be close to its $3\sigma$ lower bound. The predicted mixing angles in Eq.~\eqref{eq:XXI_mixing_parameters} are shown in Fig.~\ref{fig:caseXXI}. The best fitting results are presented in Table~\ref{tab:caseXIX_XX}. The minimum value of the $\chi^2$ function is rather small: 14.811 for $\theta_{23}<\pi/4$ and 15.138 for $\theta_{23}>\pi/4$. Hence excellent agreement with the experimental data can be achieved.

\item[~~(XXII)]{$X_{\nu\mathbf{r}}=\rho_{\mathbf{r}}(c^2d^3), \rho_{\mathbf{r}}(a^2bc^2)$}

The PMNS matrix is given
\begin{equation}
U_{PMNS}=\frac{1}{2}
\begin{pmatrix}
 \sqrt{2} & ~\sqrt{2} \cos \theta & ~\sqrt{2}\sin\theta \\
 1 & ~-\cos\theta-i\sqrt{2}\sin\theta & ~-\sin\theta+i\sqrt{2}\cos\theta \\
 1 & ~-\cos\theta+i\sqrt{2}\sin\theta & ~-\sin\theta-i\sqrt{2}\cos\theta
\end{pmatrix}\,.
\end{equation}
The lepton mixing parameters are
\begin{eqnarray}
\nonumber&&\sin^2\theta_{13}=\frac{1}{4}\left(1-\cos2\theta\right),\qquad \sin^2\theta_{12}=\frac{1+\cos2\theta}{3+\cos2\theta},\qquad \sin^2\theta_{23}=\frac{1}{2}\,,\\
&&\left|J_{CP}\right|=\frac{1}{8\sqrt{2}}\left|\sin2\theta\right|,\qquad \cot\delta_{CP}=\tan\alpha_{21}=\tan\alpha_{31}=0\,.
\end{eqnarray}
We see that the atmospheric neutrino mixing is maximal, and the Dirac CP is maximally violated. In addition, the solar mixing angle and the reactor mixing angle are predicted to be of the same form as the corresponding ones of case XIX. Therefore we have the equality $2\cos^2\theta_{12}\cos^2\theta_{13}=1$ and thus the correct values of $\theta_{12}$ and $\theta_{13}$ can not be reproduced simultaneously, as shown in Fig.~\ref{fig:caseXIX_XXII}. Moderate corrections to $\theta_{12}$ and $\theta_{13}$ are necessary in order to match the experimental best fit value.

\end{description}

In short summary, for the above four cases XIX, XX, XXI and XXII, the PMNS matrix is predicted to have one column of the form $\left(\sqrt{2}, 1, 1\right)^{T}/2$ or $\left(1, \sqrt{2}, 1\right)^{T}/2$ which is in common (up to permutation) with the bimaximal mixing pattern. Only the case XXI can accommodate the observed the three lepton mixing angles, and the remaining three cases can not produce the correct values of the $\theta_{12}$ and $\theta_{13}$ simultaneously.

\subsection{$G_{l}=Z^{(1)}_{8}$,  $G_{\nu}=Z^{(9)}_2$}

In this scenario, the charged lepton matrix $m^{\dagger}_{l}m_{l}$ is diagonalized by the unitary transformation $U_{l}$ shown in Eq.~\eqref{eq:Ul_Z81}. The constraints on the light neutrino mass matrix and its diagonalization for $G_{\nu}=Z^{(9)}_2$ has been discussed in section~\ref{subsubsec:Gnu_z29}. The resulting PMNS matrix for different remnant CP symmetry in the neutrino sector can be easily obtained as follows.

\begin{description}[labelindent=-0.7em, leftmargin=0.1em]

\item[~~(XXIII)]{$X_{\nu\mathbf{r}}=\rho_{\mathbf{r}}(1), \rho_{\mathbf{r}}(a^2bd)$}

In this case, the lepton mixing matrix is
\begin{equation}
U_{PMNS}=\frac{1}{2}
\begin{pmatrix}
 \sqrt{2} & ~i \sqrt{2}\cos\theta & ~i \sqrt{2}\sin\theta \\
 1 & ~\sqrt{2}\sin\theta-i\cos\theta &
   ~-\sqrt{2}\cos\theta-i\sin\theta \\
 1 & ~-\sqrt{2}\sin\theta-i\cos\theta &
   ~\sqrt{2}\cos\theta-i\sin\theta
\end{pmatrix}\,,
\end{equation}
which is related to the PMNS matrix of the above case XXII by $U^{\text{XXIII}}_{PMNS}=U^{\text{XXII}}_{PMNS}\;\text{diag}\left(1,i,i\right)$. The lepton mixing parameters are
\begin{eqnarray}
\nonumber&&\sin^2\theta_{13}=\frac{1}{4}\left(1-\cos2\theta\right),\qquad \sin^2\theta_{12}=\frac{1+\cos2\theta}{3+\cos2\theta},\qquad \sin^2\theta_{23}=\frac{1}{2}\,\\
&&\left|J_{CP}\right|=\frac{1}{8\sqrt{2}}\left|\sin2\theta\right|,\qquad \cot\delta_{CP}=\tan\alpha_{21}=\tan\alpha_{31}=0 \,,
\end{eqnarray}
which coincide exactly with the phenomenological predictions of case XXII.

\item[~~(XXIV)]{$X_{\nu\mathbf{r}}=\rho_{\mathbf{r}}(d^2), \rho_{\mathbf{r}}(a^2bd^3)$}

In this case, the lepton mixing matrix is
\begin{equation}
U_{PMNS}=\frac{1}{2}
\begin{pmatrix}
 -i \left(\sqrt{2} \cos \theta+\sin\theta\right) & ~1 & ~-i\left(\cos \theta-\sqrt{2}\sin\theta\right) \\
 i \sqrt{2}\sin\theta & ~\sqrt{2} & ~i \sqrt{2}\cos\theta \\
 i \left(\sqrt{2} \cos \theta-\sin \theta\right)
   & ~1 & ~-i\left(\cos \theta+\sqrt{2} \sin\theta\right)
\end{pmatrix}\,.
\end{equation}
It is exactly the same as the PMNS matrix of case XXI. Hence the same lepton mixing parameters are predicted as
\begin{eqnarray}
\nonumber&&\sin^2\theta_{13}=\frac{1}{8}\left(3-\cos2\theta-2\sqrt{2}\sin2\theta\right),\qquad \sin^2\theta_{12}=\frac{2}{5+\cos2\theta+2\sqrt{2}\sin2\theta},\\
&&\sin^2\theta_{23}=\frac{2+2\cos2\theta}{5+\cos2\theta+2\sqrt{2}\sin2\theta},\qquad \tan\delta_{CP}=\tan\alpha_{21}=\tan\alpha_{31}=0\,.
\end{eqnarray}
As has been already shown, the experimental data can be accommodated very well.

\item[~~(XXV)]{$X_{\nu\mathbf{r}}=\rho_{\mathbf{r}}(c^2d), \rho_{\mathbf{r}}(a^2bc^2d^2)$}

In this case, the lepton mixing matrix is determined to be
\begin{equation}
U_{PMNS}=\frac{1}{2}
\begin{pmatrix}
 \sqrt{2} & ~i \sqrt{2} \cos \theta~ & ~i \sqrt{2}
   \sin \theta \\
 1 & ~-i\cos\theta+\sqrt{2}\;e^{-\frac{i\pi}{4}}\sin\theta~ & ~-i\sin\theta-\sqrt{2}\;e^{-\frac{i\pi}{4}}\cos\theta \\
 1 & ~-i\cos\theta-\sqrt{2}\;e^{-\frac{i\pi}{4}}\sin\theta~ & ~-i\sin\theta+\sqrt{2}\;e^{-\frac{i\pi}{4}}\cos\theta
\end{pmatrix}\,.
\end{equation}
It coincides with the PMNS matrix of case XIX. As a consequence, the same lepton mixing parameters as shown in Eq.~\eqref{eq:mixing_parameters_XIX} arise:
\begin{eqnarray}
\nonumber&&\hskip-0.3in\sin^2\theta_{13}=\frac{1}{4}\left(1-\cos2\theta\right),~~ \sin^2\theta_{12}=\frac{1+\cos2\theta}{3+\cos2\theta},~~ \sin^2\theta_{23}=\frac{1}{2}-\frac{\sin2\theta}{3+\cos2\theta},\\
&&\hskip-0.3in\left|J_{CP}\right|=\frac{1}{16}\left|\sin2\theta\right|,~~ \left|\tan\delta_{CP}\right|=\left|\frac{3+\cos2\theta}{1+3\cos2\theta}\right|,~~ \tan\alpha_{21}=\tan\alpha^{\prime}_{31}=0\,.
\end{eqnarray}

\item[~~(XXVI)]{$X_{\nu\mathbf{r}}=\rho_{\mathbf{r}}(c^2d^3), \rho_{\mathbf{r}}(a^2bc^2)$}

In this case, the PMNS matrix is given by
\begin{equation}
U_{PMNS}=\frac{1}{2}
\begin{pmatrix}
 \sqrt{2} & ~\sqrt{2} \cos \theta & ~\sqrt{2} \sin
   \theta \\
 1 & ~-\cos\theta-\sqrt{2}\;e^{-\frac{i\pi}{4}}\sin\theta & ~-\sin\theta+\sqrt{2}\;e^{-\frac{i\pi}{4}}\cos\theta \\
 1 & ~-\cos\theta+\sqrt{2}\;e^{-\frac{i\pi}{4}}\sin\theta & ~-\sin\theta-\sqrt{2}\;e^{-\frac{i\pi}{4}}\cos\theta
\end{pmatrix}\,.
\end{equation}
Compared with the case XX's PMNS matrix in Eq.~\eqref{eq:PMNS_XX}, we have $U^{\text{XXVI}}_{PMNS}=U^{\text{XX}}_{PMNS}\;\text{diag}\left(1,-i,-i\right)$. The lepton mixing parameters read as
\begin{eqnarray}
\nonumber&&\hskip-0.2in\sin^2\theta_{13}=\frac{1}{4}\left(1-\cos2\theta\right),\qquad \sin^2\theta_{12}=\frac{1+\cos2\theta}{3+\cos2\theta},\qquad \sin^2\theta_{23}=\frac{1}{2}-\frac{\sin2\theta}{3+\cos2\theta},\\
&&\hskip-0.2in\left|J_{CP}\right|=\frac{1}{16}\left|\sin2\theta\right|,\qquad \left|\tan\delta_{CP}\right|=\left|\frac{3+\cos2\theta}{1+3\cos2\theta}\right|,\qquad \tan\alpha_{21}=\tan\alpha^{\prime}_{31}=0\,.
\end{eqnarray}
They are of the same form as the corresponding ones of case XX, as shown in Eq.~\eqref{eq:mixing_parameters_XX}. Hence the relation $2\cos^2\theta_{12}\cos^2\theta_{13}=1$ is fulfilled such that the observed values of $\theta_{12}$ and $\theta_{13}$ can not be accommodated simultaneously. In summary, compared with scenario of $G_{l}=K^{(3)}_{4}$,  $G_{\nu}=Z^{(9)}_2$, no new predictions for the lepton mixing parameters are obtained. and only the case XXIV is in accordance with the present data.

\end{description}

\begin{table}[hptb]
\begin{tabular}{|c|c|c|c|c|c|c|}
\hline\hline

  &  \multicolumn{2}{|c|}{\text{\tt I}, \text{\tt II}}  &    \multicolumn{2}{|c|}{\text{\tt III}, \text{\tt VIII}}   & \multicolumn{2}{|c|}{\text{\tt IV}, \text{\tt VII}}  \\  \hline

$\sin^2\theta_{13}$   &  \multicolumn{2}{|c|}{$\frac{1}{3}\big[1-\cos\left(\frac{\pi}{6}-2\theta\right)\big]$}  &  \multicolumn{2}{|c|}{$\frac{1}{3}\left(1-\cos2\theta\right)$}  &  \multicolumn{2}{|c|}{$\frac{1}{3}+\frac{1}{2\sqrt{3}}\cos2\theta$}      \\  \hline

$\sin^2\theta_{12}$   &  \multicolumn{2}{|c|}{$\frac{1}{2+\cos\left(\frac{\pi}{6}-2\theta\right)}$}  &  \multicolumn{2}{|c|}{$\frac{1}{2+\cos2\theta}$}  &  \multicolumn{2}{|c|}{$\frac{2}{4-\sqrt{3}\;\cos2\theta}$ }  \\ \hline

$\sin^2\theta_{23}(\theta_{23}<\pi/4)$   & \multicolumn{2}{|c|}{$\frac{1+\sin2\theta}{2+\cos\left(\frac{\pi}{6}-2\theta\right)}$}   &  \multicolumn{2}{|c|}{$\frac{1}{2}$}   &  \multicolumn{2}{|c|}{$\frac{2}{4-\sqrt{3}\;\cos2\theta}$ }   \\ \cdashline{1-7}

$\sin^2\theta_{23}(\theta_{23}>\pi/4)$   &  \multicolumn{2}{|c|}{$\frac{1+\cos\left(\frac{\pi}{6}+2\theta\right)}{2+\cos\left(\frac{\pi}{6}-2\theta\right)}$}   &  \multicolumn{2}{|c|}{$\frac{1}{2}$}  &  \multicolumn{2}{|c|}{$\frac{2-\sqrt{3}\;\cos2\theta}{4-\sqrt{3}\;\cos2\theta}$}    \\  \hline

$\left|J_{\text{CP}}\right|$   &  \multicolumn{2}{|c|}{0}  &  \multicolumn{2}{|c|}{$\frac{1}{6\sqrt{3}}\left|\sin2\theta\right|$}   &  \multicolumn{2}{|c|}{$\frac{1}{6\sqrt{3}}\left|\sin2\theta\right|$}    \\ \hline

$\left|\tan\delta_{\text{CP}}\right|$ &  \multicolumn{2}{|c|}{0}  &  \multicolumn{2}{|c|}{$+\infty$}   &  \multicolumn{2}{|c|}{$\big|\frac{4-\sqrt{3}\;\cos2\theta}{1-\sqrt{3}\;\cos2\theta}\tan2\theta\big|$}   \\ \hline

\multirow{2}{*}{$\left|\tan\alpha_{21}\right|$}  &  \multicolumn{2}{|c|}{0, case I}   &  \multicolumn{2}{|c|}{0, case III}  & \multicolumn{2}{|c|}{$\big|\frac{\sin2\theta}{\sqrt{3}-2\cos2\theta}\big|$,~ case IV}      \\  \cdashline{2-7}

                  & \multicolumn{2}{|c|}{$+\infty$, case II}   &  \multicolumn{2}{|c|}{$+\infty$, case VIII}  &  \multicolumn{2}{|c|}{$\big|\frac{\sqrt{3}-2\cos2\theta}{\sin2\theta}\big|$, case VII}      \\ \hline

$\left|\tan\alpha^{\prime}_{31}\right|$  & \multicolumn{2}{|c|}{0}   &  \multicolumn{2}{|c|}{0}  &  \multicolumn{2}{|c|}{$\big|\frac{4\sqrt{3}\;\sin2\theta}{1-3\cos4\theta}\big|$}     \\ \hline\hline

\multicolumn{7}{|c|}{\tt Best Fitting}  \\\hline\hline
   &  $ \theta_{23}<\pi/4$  &  $\theta_{23}>\pi/4$ &   $ \theta_{23}<\pi/4$  &  $\theta_{23}>\pi/4$  &   $ \theta_{23}<\pi/4$  &  $\theta_{23}>\pi/4$  \\ \hline

$\chi^2_{\text{min}}$  &  9.548  &  9.303  &  14.527  &  27.254  &  110.741  &  111.559  \\\hline

$\theta_{\text{bf}}$  &   \multicolumn{2}{|c|}{0.0798}    &  \multicolumn{2}{|c|}{$\pm0.184$}    &    \multicolumn{2}{|c|}{$\pm\pi/2$}    \\\hline

$\sin^2\theta_{13}(\theta_{\text{bf}})$  &   \multicolumn{2}{|c|}{0.0218}  &  \multicolumn{2}{|c|}{0.0222}   &    \multicolumn{2}{|c|}{0.0447}    \\\hline

$\theta_{13}(\theta_{\text{bf}})/^{\circ}$  &   \multicolumn{2}{|c|}{8.498}  &  \multicolumn{2}{|c|}{8.576}   &    \multicolumn{2}{|c|}{12.200}    \\\hline

$\sin^2\theta_{12}(\theta_{\text{bf}})$  &   \multicolumn{2}{|c|}{0.341}    &   \multicolumn{2}{|c|}{0.341}    &    \multicolumn{2}{|c|}{0.349}    \\\hline

$\theta_{12}(\theta_{\text{bf}})/^{\circ}$  &   \multicolumn{2}{|c|}{35.715}    &   \multicolumn{2}{|c|}{35.724}    &    \multicolumn{2}{|c|}{36.206}    \\\hline

$\sin^2\theta_{23}(\theta_{\text{bf}})$  & 0.395   & 0.605   &   \multicolumn{2}{|c|}{0.5}    &  0.349  &  0.651  \\\hline

$\sin^2\theta_{23}(\theta_{\text{bf}})/^{\circ}$  & 38.936   & 51.072  &   \multicolumn{2}{|c|}{45}    & 36.206  &  53.794    \\\hline

$\left|\sin\delta_{\text{CP}}(\theta_{\text{bf}})\right|$  &    \multicolumn{2}{|c|}{0}    &   \multicolumn{2}{|c|}{1}    &    \multicolumn{2}{|c|}{0}     \\\hline

$\delta_{\text{CP}}(\theta_{\text{bf}})/^{\circ}$  &    \multicolumn{2}{|c|}{0}    &   \multicolumn{2}{|c|}{90}    &    \multicolumn{2}{|c|}{0}     \\\hline

\multirow{2}{*}{$\left|\sin\alpha_{21}(\theta_{\text{bf}})\right|$}  &    \multicolumn{2}{|c|}{0, case I }    &   \multicolumn{2}{|c|}{0, case III }    &    \multicolumn{2}{|c|}{0, case IV }    \\ \cdashline{2-7}

       &   \multicolumn{2}{|c|}{1, case II }    &   \multicolumn{2}{|c|}{1, case VIII }    &     \multicolumn{2}{|c|}{1, case VII }    \\\hline

\multirow{2}{*}{$\alpha_{21}(\theta_{\text{bf}})/^{\circ}$}  &    \multicolumn{2}{|c|}{0, case I }    &   \multicolumn{2}{|c|}{0, case III }    &    \multicolumn{2}{|c|}{0, case IV }    \\ \cdashline{2-7}

       &   \multicolumn{2}{|c|}{90, case II }    &   \multicolumn{2}{|c|}{90, case VIII }    &     \multicolumn{2}{|c|}{90, case VII }    \\\hline

$\left|\sin\alpha^{\prime}_{31}(\theta_{\text{bf}})\right|$  &   \multicolumn{2}{|c|}{0}     &  \multicolumn{2}{|c|}{0}    &    \multicolumn{2}{|c|}{0}      \\\hline

$\alpha^{\prime}_{31}(\theta_{\text{bf}})/^{\circ}$  &   \multicolumn{2}{|c|}{0}     &  \multicolumn{2}{|c|}{0}    &    \multicolumn{2}{|c|}{0}      \\\hline\hline

\end{tabular}
\caption{\label{tab:caseI_II}The results of the mixing parameters for the cases I, II, III, IV, VII and VIII, where $``+\infty"$ for $\left|\tan\delta_{\text{CP}}\right|$, $\left|\tan\alpha\right|$ and $\left|\tan\beta\right|$ implies that the absolute value of the corresponding CP phase is $\pi/2$. Notice that the Dirac CP phase $\delta_{\text{CP}}$ is determined up to $\delta_{\text{CP}}$, $\pi+\delta_{\text{CP}}$, $\pi-\delta_{\text{CP}}$ and $2\pi-\delta_{\text{CP}}$ in the present context, and only one representative value is displayed in this table. The same convention is taken for the Majorana CP phases $\alpha_{21}$ and $\alpha^{\prime}_{31}$. }
\end{table}

\begin{table}[hptb]
\begin{center}
\begin{tabular}{|c|c|c| }
\hline\hline
  &  \multicolumn{2}{|c|}{\text{\tt V}, \text{\tt VI}, \text{\tt IX}, \text{\tt X}}  \\ \hline

$\sin^2\theta_{13}$   &  \multicolumn{2}{|c|}{$\frac{1}{3}-\frac{\sqrt{6}+\sqrt{2}}{12}\cos2\theta$} \\  \hline

$\sin^2\theta_{12}$    &  \multicolumn{2}{|c|}{$\frac{4}{8+\left(\sqrt{6}+\sqrt{2}\right)\cos2\theta}$ } \\  \hline

$\sin^2\theta_{23}(\theta_{23}<\pi/4)$   &  \multicolumn{2}{|c|}{$\frac{4+\left(\sqrt{6}-\sqrt{2}\right)\cos2\theta}{8+\left(\sqrt{6}+\sqrt{2}\right)\cos2\theta}$} \\  \cdashline{1-3}

$\sin^2\theta_{23}(\theta_{23}>\pi/4)$   & \multicolumn{2}{|c|}{$\frac{4+2\sqrt{2}\cos2\theta}{8+\left(\sqrt{6}+\sqrt{2}\right)\cos2\theta}$}    \\  \hline

$\left|J_{\text{CP}}\right|$   &  \multicolumn{2}{|c|}{$\frac{1}{6\sqrt{3}}\left|\sin2\theta\right|$}  \\  \hline

$\left|\tan\delta_{\text{CP}}\right|$   & \multicolumn{2}{|c|}{$\big|\frac{4\sqrt{2}+\left(1+\sqrt{3}\right)\cos2\theta}{1-\sqrt{3}-\sqrt{2}\cos2\theta}\tan2\theta\big|$}  \\  \hline

\multirow{2}{*}{$\left|\tan\alpha_{21}\right|$}   &   \multicolumn{2}{|c|}{$\big|\frac{\sqrt{6}+\sqrt{2}+4\cos2\theta+\left(\sqrt{6}-\sqrt{2}\right)\sin2\theta}
{\sqrt{6}+\sqrt{2}+4\cos2\theta-\left(\sqrt{6}-\sqrt{2}\right)\sin2\theta}\big|$,~ cases V, IX} \\  \cdashline{2-3}

                  & \multicolumn{2}{|c|}{$\big|\frac{\sqrt{6}+\sqrt{2}+4\cos2\theta-\left(\sqrt{6}-\sqrt{2}\right)\sin2\theta}
                  {\sqrt{6}+\sqrt{2}+4\cos2\theta+\left(\sqrt{6}-\sqrt{2}\right)\sin2\theta}\big|$,~ cases VI, X}   \\  \hline

$\left|\tan\alpha^{\prime}_{31}\right|$   & \multicolumn{2}{|c|}{$\big|\frac{4\sin2\theta}{2-3\sqrt{3}+\left(2+\sqrt{3}\right)\cos4\theta}\big|$ } \\ \hline\hline

\multicolumn{3}{|c|}{\tt Best Fitting}  \\\hline\hline

   &   ~~~~~~~$\theta_{23}<\pi/4$~~~~~~~  &  $\theta_{23}>\pi/4$  \\ \hline

$\chi^2_{\text{min}}$  &  9.124   &  9.838 \\\hline

$\theta_{\text{bf}}$  &   \multicolumn{2}{|c|}{$\pm0.130$}  \\\hline

$\sin^2\theta_{13}(\theta_{\text{bf}})$  &    \multicolumn{2}{|c|}{0.0222}  \\\hline

$\theta_{13}(\theta_{\text{bf}})/^{\circ}$  &    \multicolumn{2}{|c|}{8.574}  \\\hline

$\sin^2\theta_{12}(\theta_{\text{bf}})$  &  \multicolumn{2}{|c|}{0.341}   \\\hline

$\theta_{12}(\theta_{\text{bf}})/^{\circ}$  &  \multicolumn{2}{|c|}{35.724}   \\\hline

$\sin^2\theta_{23}(\theta_{\text{bf}})$  &  0.426  &  0.574 \\\hline

$\theta_{23}(\theta_{\text{bf}})/^{\circ}$  &  40.754  &  49.246 \\\hline

$\left|\sin\delta_{\text{CP}}(\theta_{\text{bf}})\right|$  &   \multicolumn{2}{|c|}{0.725}   \\\hline

$\delta_{\text{CP}}(\theta_{\text{bf}})/^{\circ}$  &   \multicolumn{2}{|c|}{46.512}   \\\hline

$\left|\sin\alpha_{21}(\theta_{\text{bf}})\right|$  &   \multicolumn{2}{|c|}{0.682 ~\text{or}~ 0.731 }  \\\hline

$\alpha_{21}(\theta_{\text{bf}})/^{\circ}$  &  \multicolumn{2}{|c|}{43.023~\text{or}~46.977}  \\\hline

$\left|\sin\alpha^{\prime}_{31}(\theta_{\text{bf}})\right|$  &      \multicolumn{2}{|c|}{0.999}       \\\hline

$\alpha^{\prime}_{31}(\theta_{\text{bf}})/^{\circ}$  &      \multicolumn{2}{|c|}{87.755}       \\\hline\hline

\end{tabular}
\caption{\label{tab:caseV_VI}The results of the mixing parameters for the cases V, VI, IX and X, where $``+\infty"$ for $\left|\tan\delta_{\text{CP}}\right|$, $\left|\tan\alpha\right|$ and $\left|\tan\beta\right|$ implies that the absolute value of the corresponding CP phase is $\pi/2$. Notice that the Dirac CP phase $\delta_{\text{CP}}$ is determined up to $\delta_{\text{CP}}$, $\pi+\delta_{\text{CP}}$, $\pi-\delta_{\text{CP}}$ and $2\pi-\delta_{\text{CP}}$ in the present context, and only one representative value is displayed in this table. The same convention is taken for the Majorana CP phases $\alpha_{21}$ and $\alpha^{\prime}_{31}$.}
\end{center}
\end{table}

\begin{table}[hptb]
\begin{tabular}{|c|c|c|c|c|}
\hline\hline

  &  \multicolumn{2}{|c|}{\text{\tt XI}}  &    \multicolumn{2}{|c|}{\text{\tt XII}}     \\  \hline

$\sin^2\theta_{13}$   &  \multicolumn{2}{|c|}{$\frac{1}{12}\big[4-\sqrt{3}-\sqrt{3}\cos2\theta-2\left(\sqrt{3}-1\right)\sin2\theta\big]$}  &  \multicolumn{2}{|c|}{$\frac{1}{6}\left(2-\sqrt{3}\right)$}      \\  \hline

$\sin^2\theta_{12}$   &  \multicolumn{2}{|c|}{$\frac{4-\sqrt{3}+\sqrt{3}\cos2\theta+2\left(\sqrt{3}-1\right)\sin2\theta}{8+\sqrt{3}+\sqrt{3}\cos2\theta+2\left(\sqrt{3}-1\right)\sin2\theta}$}  &  \multicolumn{2}{|c|}{$\frac{1}{2}+\frac{\sqrt{3}\cos2\theta}{2\left(4+\sqrt{3}\right)}$}  \\ \hline

$\sin^2\theta_{23}(\theta_{23}<\pi/4)$   & \multicolumn{2}{|c|}{$\frac{4\left(1-\sin2\theta\right)}{8+\sqrt{3}+\sqrt{3}\cos2\theta+2\left(\sqrt{3}-1\right)\sin2\theta}$}   &  \multicolumn{2}{|c|}{$\frac{2}{4+\sqrt{3}}$}      \\ \cdashline{1-5}

$\sin^2\theta_{23}(\theta_{23}>\pi/4)$   &  \multicolumn{2}{|c|}{$\frac{4+\sqrt{3}+\sqrt{3}\cos2\theta+2\left(1+\sqrt{3}\right)\sin2\theta}{8+\sqrt{3}+\sqrt{3}\cos2\theta+2\left(\sqrt{3}-1\right)\sin2\theta}$}   &  \multicolumn{2}{|c|}{$\frac{2+\sqrt{3}}{4+\sqrt{3}}$}    \\  \hline

$\left|J_{\text{CP}}\right|$   &  \multicolumn{2}{|c|}{0}  &  \multicolumn{2}{|c|}{$\frac{1}{12\sqrt{3}}\left|\sin2\theta\right|$}    \\ \hline

$\left|\tan\delta_{\text{CP}}\right|$ &  \multicolumn{2}{|c|}{0}  &  \multicolumn{2}{|c|}{$\big|\frac{\left(4+\sqrt{3}\right)\tan2\theta}{2\left(1+\sqrt{3}\right)}\big|$}   \\ \hline

$\left|\tan\alpha_{21}\right|$  &  \multicolumn{2}{|c|}{0}   &  \multicolumn{2}{|c|}{$\big|\frac{8\left(3+\sqrt{3}\right)\sin2\theta}{29+16\sqrt{3}+3\cos4\theta}\big|$}  \\  \hline

$\left|\tan\alpha^{\prime}_{31}\right|$  & \multicolumn{2}{|c|}{0}   &  \multicolumn{2}{|c|}{$\big|\frac{2\left(1-\sqrt{3}\right)\sin2\theta}{3-2\sqrt{3}+\left(5-2\sqrt{3}\right)\cos2\theta}\big|$}  \\ \hline\hline

\multicolumn{5}{|c|}{\tt Best Fitting}  \\\hline\hline
   &   ~~~~~~~$\theta_{23}<\pi/4$~~~~~~~   &  $\theta_{23}>\pi/4$ &   $ \theta_{23}<\pi/4$  &  $\theta_{23}>\pi/4$  \\ \hline

$\chi^2_{\text{min}}$  &  48.862  &   54.822  &  110.741 &  111.559  \\\hline

$\theta_{\text{bf}}$  & 0.0764  &  0.0711 &  \multicolumn{2}{|c|}{$\pm\pi/2$}      \\\hline

$\sin^2\theta_{13}(\theta_{\text{bf}})$  & 0.0278  &  0.0288  &  \multicolumn{2}{|c|}{0.0447}      \\\hline

$\theta_{13}(\theta_{\text{bf}})/^{\circ}$  &  9.592 & 9.774  &  \multicolumn{2}{|c|}{12.200}      \\\hline

$\sin^2\theta_{12}(\theta_{\text{bf}})$  &   \multicolumn{2}{|c|}{0.360}    &   \multicolumn{2}{|c|}{0.349}       \\\hline

$\theta_{12}(\theta_{\text{bf}})/^{\circ}$  &   \multicolumn{2}{|c|}{36.883}    &   \multicolumn{2}{|c|}{36.206}       \\\hline

$\sin^2\theta_{23}(\theta_{\text{bf}})$  &  0.291  & 0.705   & 0.349  &  0.651  \\\hline

$\theta_{23}(\theta_{\text{bf}})/^{\circ}$  & 32.624  &   57.129  & 36.206 &  53.794 \\\hline

$\left|\sin\delta_{\text{CP}}(\theta_{\text{bf}})\right|$  &    \multicolumn{2}{|c|}{0}    &   \multicolumn{2}{|c|}{0}       \\\hline

$\delta_{\text{CP}}(\theta_{\text{bf}})/^{\circ}$  &    \multicolumn{2}{|c|}{0}    &   \multicolumn{2}{|c|}{0}       \\\hline

$\left|\sin\alpha_{21}(\theta_{\text{bf}})\right|$  &    \multicolumn{2}{|c|}{ 0}    &   \multicolumn{2}{|c|}{0}       \\ \hline

$\alpha_{21}(\theta_{\text{bf}})/^{\circ}$  &    \multicolumn{2}{|c|}{ 0}    &   \multicolumn{2}{|c|}{0}       \\ \hline

$\left|\sin\alpha^{\prime}_{31}(\theta_{\text{bf}})\right|$  &   \multicolumn{2}{|c|}{~0}     &  \multicolumn{2}{|c|}{0}     \\\hline

$\alpha^{\prime}_{31}(\theta_{\text{bf}})/^{\circ}$  &   \multicolumn{2}{|c|}{~0}     &  \multicolumn{2}{|c|}{0}     \\\hline\hline

\end{tabular}
\caption{\label{tab:caseXI_XII}The results of the mixing parameters for the case XI and case XII, where $``+\infty"$ for $\left|\tan\delta_{\text{CP}}\right|$, $\left|\tan\alpha\right|$ and $\left|\tan\beta\right|$ implies that the absolute value of the corresponding CP phase is $\pi/2$.  Notice that the Dirac CP phase $\delta_{\text{CP}}$ is determined up to $\delta_{\text{CP}}$, $\pi+\delta_{\text{CP}}$, $\pi-\delta_{\text{CP}}$ and $2\pi-\delta_{\text{CP}}$ in the present context, and only one representative value is displayed in this table. The same convention is taken for the Majorana CP phases $\alpha_{21}$ and $\alpha^{\prime}_{31}$.}
\end{table}

\begin{table}[hptb]
\begin{center}
\begin{tabular}{|c|c|c|}
\hline\hline

  & \multicolumn{2}{|c|}{\text{\tt XIII}, \text{\tt XIV}}  \\  \hline

$\sin^2\theta_{13}$   &  \multicolumn{2}{|c|}{$\frac{1}{12}\big[4-\sqrt{3}-\sqrt{3}\cos2\theta-\left(\sqrt{6}-\sqrt{2}\right)\sin2\theta\big]$}      \\  \hline

$\sin^2\theta_{12}$   &  \multicolumn{2}{|c|}{$\frac{4-\sqrt{3}+\sqrt{3}\cos2\theta+\left(\sqrt{6}-\sqrt{2}\right)\sin2\theta}{8+\sqrt{3}+\sqrt{3}\cos2\theta+\left(\sqrt{6}-\sqrt{2}\right)\sin2\theta}$ }  \\ \hline

$\sin^2\theta_{23}(\theta_{23}<\pi/4)$   &  \multicolumn{2}{|c|}{$\frac{4-2\sqrt{2}\sin2\theta}{8+\sqrt{3}+\sqrt{3}\cos2\theta+\left(\sqrt{6}-\sqrt{2}\right)\sin2\theta}$ }   \\ \cdashline{1-3}

$\sin^2\theta_{23}(\theta_{23}>\pi/4)$  &  \multicolumn{2}{|c|}{$\frac{4+\sqrt{3}+\sqrt{3}\cos2\theta+\left(\sqrt{6}+\sqrt{2}\right)\sin2\theta}{8+\sqrt{3}+\sqrt{3}\cos2\theta+\left(\sqrt{6}-\sqrt{2}\right)\sin2\theta}$}    \\  \hline

$\left|J_{\text{CP}}\right|$   &  \multicolumn{2}{|c|}{$\frac{1}{12\sqrt{6}}\left|\sin2\theta\right|$}    \\ \hline

$\left|\tan\delta_{\text{CP}}\right|$  &  \multicolumn{2}{|c|}{$\big|\frac{\left(6-2\sqrt{3}\right)\left(1-\cos4\theta\right)+\left(6\sqrt{2}+16\sqrt{6}\right)\sin2\theta+3\sqrt{2}\sin4\theta}{24+18\sqrt{3}+\left(24-8\sqrt{3}\right)\cos2\theta-6\sqrt{2}\sin2\theta+6\sqrt{3}\cos4\theta-\left(15\sqrt{2}+4\sqrt{6}\right)\sin4\theta}\big|$}   \\ \hline

$\left|\tan\alpha_{21}\right|$  & \multicolumn{2}{|c|}{$\big|\frac{2\left(2+\sqrt{3}\right)\left(1+\cos2\theta\right)+\left(\sqrt{6}+\sqrt{2}\right)\sin2\theta}{1-\cos2\theta+\left(\sqrt{6}+\sqrt{2}\right)\sin2\theta}\big|$}      \\  \hline

$\left|\tan\alpha^{\prime}_{31}\right|$  &  \multicolumn{2}{|c|}{$\big|\frac{2\left(2+\sqrt{3}\right)\left(1-\cos2\theta\right)-\left(\sqrt{6}+\sqrt{2}\right)\sin2\theta}{1+\cos2\theta-\left(\sqrt{6}+\sqrt{2}\right)\sin2\theta}\big|$}     \\ \hline\hline

\multicolumn{3}{|c|}{\tt Best Fitting}  \\\hline\hline
   &   ~~~~~~~~~~~~$\theta_{23}<\pi/4$~~~~~~~~~~~~  &  $\theta_{23}>\pi/4$  \\ \hline

$\chi^2_{\text{min}}$  &  51.645  &  57.745  \\\hline

$\theta_{\text{bf}}$  &  0.113  &   0.103    \\\hline

$\sin^2\theta_{13}(\theta_{\text{bf}})$  & 0.0290   &   0.0301 \\\hline

$\theta_{13}(\theta_{\text{bf}})/^{\circ}$  & 9.805  &  9.989 \\\hline

$\sin^2\theta_{12}(\theta_{\text{bf}})$  &  \multicolumn{2}{|c|}{0.359} \\\hline

$\theta_{12}(\theta_{\text{bf}})/^{\circ}$  &  \multicolumn{2}{|c|}{36.835 } \\\hline

$\sin^2\theta_{23}(\theta_{\text{bf}})$  &   0.289   &  0.706 \\\hline

$\theta_{23}(\theta_{\text{bf}})/^{\circ}$  &  32.514   &  57.161 \\\hline

$\left|\sin\delta_{\text{CP}}(\theta_{\text{bf}})\right|$  &  0.212 &   0.189 \\\hline

$\delta_{\text{CP}}(\theta_{\text{bf}})/^{\circ}$  & 12.235  &   10.886\\\hline

$\left|\sin\alpha_{21}(\theta_{\text{bf}})\right|$  &    \multicolumn{2}{|c|}{0.998}      \\ \hline

$\alpha_{21}(\theta_{\text{bf}})/^{\circ}$  &    \multicolumn{2}{|c|}{86.732}      \\ \hline

$\left|\sin\alpha^{\prime}_{31}(\theta_{\text{bf}})\right|$  & 0.520  &  0.469 \\ \hline

$\alpha^{\prime}_{31}(\theta_{\text{bf}})/^{\circ}$  & 31.358  & 27.944  \\ \hline
\hline

\end{tabular}
\caption{\label{tab:caseXIII_XIV}The results of the mixing parameters for the case XIII and case XIV, where $``+\infty"$ for $\left|\tan\delta_{\text{CP}}\right|$, $\left|\tan\alpha\right|$ and $\left|\tan\beta\right|$ implies that the absolute value of the corresponding CP phase is $\pi/2$.  Notice that the Dirac CP phase $\delta_{\text{CP}}$ is determined up to $\delta_{\text{CP}}$, $\pi+\delta_{\text{CP}}$, $\pi-\delta_{\text{CP}}$ and $2\pi-\delta_{\text{CP}}$ in the present context, and only one representative value is displayed in this table. The same convention is taken for the Majorana CP phases $\alpha_{21}$ and $\alpha^{\prime}_{31}$.}
\end{center}
\end{table}

\begin{table}[hptb]
\begin{tabular}{|c|c|c|c|c|c|c|}
\hline\hline

  &  \multicolumn{2}{|c|}{\text{\tt XV}}  &    \multicolumn{2}{|c|}{\text{\tt XVI}}   & \multicolumn{2}{|c|}{\text{\tt XVII}, \text{\tt XVIII}}  \\  \hline

$\sin^2\theta_{13}$   &  \multicolumn{6}{|c|}{$\frac{1}{6}\left(1-\cos2\theta\right)$}    \\  \hline

$\sin^2\theta_{12}$   &  \multicolumn{6}{|c|}{$\frac{1+\cos2\theta}{5+\cos2\theta}$}   \\ \hline

$\sin^2\theta_{23}(\theta_{23}<\pi/4)$   & \multicolumn{2}{|c|}{$\frac{1}{2}-\frac{\sqrt{6}\sin2\theta}{5+\cos2\theta}$}   &  \multicolumn{2}{|c|}{$\frac{1}{2}$}   &  \multicolumn{2}{|c|}{$\frac{1}{2}-\frac{\sqrt{3}\sin2\theta}{5+\cos2\theta}$ }   \\ \cdashline{1-7}

$\sin^2\theta_{23}(\theta_{23}>\pi/4)$   &  \multicolumn{2}{|c|}{$\frac{1}{2}+\frac{\sqrt{6}\sin2\theta}{5+\cos2\theta}$}   &  \multicolumn{2}{|c|}{$\frac{1}{2}$}  &  \multicolumn{2}{|c|}{$\frac{1}{2}+\frac{\sqrt{3}\sin2\theta}{5+\cos2\theta}$}    \\  \hline

$\left|J_{\text{CP}}\right|$   &  \multicolumn{2}{|c|}{0}  &  \multicolumn{2}{|c|}{$\frac{1}{6\sqrt{6}}\left|\sin2\theta\right|$}   &  \multicolumn{2}{|c|}{$\frac{1}{12\sqrt{3}}\left|\sin2\theta\right|$}    \\ \hline

$\left|\tan\delta_{\text{CP}}\right|$ &  \multicolumn{2}{|c|}{0}  &  \multicolumn{2}{|c|}{$+\infty$}   &  \multicolumn{2}{|c|}{$\left|\frac{5+\cos2\theta}{1+5\cos2\theta}\right|$}   \\ \hline

$\left|\tan\alpha_{21}\right|$  &  \multicolumn{4}{|c|}{0}    & \multicolumn{2}{|c|}{$+\infty$}      \\  \hline

$\left|\tan\alpha^{\prime}_{31}\right|$  & \multicolumn{4}{|c|}{0}    &  \multicolumn{2}{|c|}{$+\infty$}     \\ \hline\hline

\multicolumn{7}{|c|}{\tt Best Fitting}  \\\hline\hline
   &   $ \theta_{23}<\pi/4$  &  $\theta_{23}>\pi/4$ &   $ \theta_{23}<\pi/4$  &  $\theta_{23}>\pi/4$  &   $ \theta_{23}<\pi/4$  &  $\theta_{23}>\pi/4$  \\ \hline

$\chi^2_{\text{min}}$  &  22.270  &  25.815  & 6.993  & 19.720   &  7.264  &  7.726 \\\hline

$\theta_{\text{bf}}$  &  0.237  &  0.228    &  \multicolumn{2}{|c|}{$\pm0.266$}    &  0.256  & 0.253 \\\hline

$\sin^2\theta_{13}(\theta_{\text{bf}})$  & 0.0184  &  0.0170 &  \multicolumn{2}{|c|}{0.0230}   & 0.0214 &  0.0210  \\\hline

$\theta_{13}(\theta_{\text{bf}})/^{\circ}$  & 7.786  &  7.502 &  \multicolumn{2}{|c|}{8.731}   &  8.402 &  8.325 \\\hline

$\sin^2\theta_{12}(\theta_{\text{bf}})$  &  \multicolumn{2}{|c|}{0.321}    &   \multicolumn{2}{|c|}{0.318}    &  \multicolumn{2}{|c|}{0.319}   \\\hline

$\theta_{12}(\theta_{\text{bf}})/^{\circ}$  &  \multicolumn{2}{|c|}{34.503}    &   \multicolumn{2}{|c|}{34.303}    &  \multicolumn{2}{|c|}{34.376}   \\\hline

$\sin^2\theta_{23}(\theta_{\text{bf}})$  &  0.310 & 0.683   &   \multicolumn{2}{|c|}{0.5}    &  0.356  &  0.643  \\\hline

$\theta_{23}(\theta_{\text{bf}})/^{\circ}$  & 33.850  &  55.734  &   \multicolumn{2}{|c|}{45}    & 36.604   & 53.319   \\\hline

$\left|\sin\delta_{\text{CP}}(\theta_{\text{bf}})\right|$  &    \multicolumn{2}{|c|}{0}    &   \multicolumn{2}{|c|}{1}    &    \multicolumn{2}{|c|}{0.738}     \\\hline

$\delta_{\text{CP}}(\theta_{\text{bf}})/^{\circ}$  &    \multicolumn{2}{|c|}{0}    &   \multicolumn{2}{|c|}{90}    &    \multicolumn{2}{|c|}{47.612 }     \\\hline

$\left|\sin\alpha_{21}(\theta_{\text{bf}})\right|$  &    \multicolumn{4}{|c|}{0}       &    \multicolumn{2}{|c|}{1}    \\ \hline

$\alpha_{21}(\theta_{\text{bf}})_{21}/^{\circ}$  &    \multicolumn{4}{|c|}{0}       &    \multicolumn{2}{|c|}{90}    \\ \hline

$\left|\sin\alpha^{\prime}_{31}(\theta_{\text{bf}})\right|$  &   \multicolumn{4}{|c|}{0}     &    \multicolumn{2}{|c|}{1}      \\ \hline

$\alpha^{\prime}_{31}(\theta_{\text{bf}})/^{\circ}$  &   \multicolumn{4}{|c|}{0}     &    \multicolumn{2}{|c|}{90}      \\ \hline\hline

\end{tabular}
\caption{\label{tab:caseXV_XVIII}The results of the mixing parameters for the cases XV, XVI, XVII and XVIII, where $``+\infty"$ for $\left|\tan\delta_{\text{CP}}\right|$, $\left|\tan\alpha\right|$ and $\left|\tan\beta\right|$ implies that the absolute value of the corresponding CP phase is $\pi/2$.  Notice that the Dirac CP phase $\delta_{\text{CP}}$ is determined up to $\delta_{\text{CP}}$, $\pi+\delta_{\text{CP}}$, $\pi-\delta_{\text{CP}}$ and $2\pi-\delta_{\text{CP}}$ in the present context, and only one representative value is displayed in this table. The same convention is taken for the Majorana CP phases $\alpha_{21}$ and $\alpha^{\prime}_{31}$.}
\end{table}

\begin{table}[hptb]
\begin{tabular}{|c|c|c|c|c|c|c|}
\hline\hline

  &  \multicolumn{2}{|c|}{\text{\tt XIX}, \text{\tt XX}, \text{\tt XXV}, \text{\tt XXVI}} & \multicolumn{2}{|c|}{\text{\tt XXII}, \text{\tt XXIII}}   &    \multicolumn{2}{|c|}{\text{\tt XXI}, \text{\tt XXIV}}    \\  \hline

$\sin^2\theta_{13}$   &  \multicolumn{4}{|c|}{$\frac{1}{4}\left(1-\cos2\theta\right)$}  &  \multicolumn{2}{|c|}{$\frac{1}{8}\left(3-\cos2\theta-2\sqrt{2}\sin2\theta\right)$}  \\  \hline

$\sin^2\theta_{12}$   &  \multicolumn{4}{|c|}{$\frac{1+\cos2\theta}{3+\cos2\theta}$}   &  \multicolumn{2}{|c|}{$\frac{2}{5+\cos2\theta+2\sqrt{2}\sin2\theta}$}  \\ \hline

$\sin^2\theta_{23}(\theta_{23}<\pi/4)$   & \multicolumn{2}{|c|}{$\frac{1}{2}-\frac{\sin2\theta}{3+\cos2\theta}$}   &  \multicolumn{2}{|c|}{$\frac{1}{2}$ }  &  \multicolumn{2}{|c|}{$\frac{2+2\cos2\theta}{5+\cos2\theta+2\sqrt{2}\sin2\theta}$}   \\ \cdashline{1-7}

$\sin^2\theta_{23}(\theta_{23}>\pi/4)$   &  \multicolumn{2}{|c|}{$\frac{1}{2}+\frac{\sin2\theta}{3+\cos2\theta}$}   &  \multicolumn{2}{|c|}{$\frac{1}{2}$}  &  \multicolumn{2}{|c|}{$\frac{3-\cos2\theta+2\sqrt{2}\sin2\theta}{5+\cos2\theta+2\sqrt{2}\sin2\theta}$}  \\  \hline

$\left|J_{\text{CP}}\right|$   &  \multicolumn{2}{|c|}{$\frac{1}{16}\left|\sin2\theta\right|$}  &  \multicolumn{2}{|c|}{$\frac{1}{8\sqrt{2}}\left|\sin2\theta\right|$}  &  \multicolumn{2}{|c|}{0}      \\ \hline

$\left|\tan\delta_{\text{CP}}\right|$ &  \multicolumn{2}{|c|}{$\left|\frac{3+\cos2\theta}{1+3\cos2\theta}\right|$}  &  \multicolumn{2}{|c|}{$+\infty$}  &  \multicolumn{2}{|c|}{0}    \\ \hline

$\left|\tan\alpha_{21}\right|$  &  \multicolumn{6}{|c|}{0}       \\  \hline

$\left|\tan\alpha^{\prime}_{31}\right|$  & \multicolumn{6}{|c|}{0}     \\ \hline\hline

\multicolumn{7}{|c|}{\tt Best Fitting}  \\\hline\hline
   &   $ \theta_{23}<\pi/4$  &  $\theta_{23}>\pi/4$ &   $ \theta_{23}<\pi/4$  &  $\theta_{23}>\pi/4$  &   $ \theta_{23}<\pi/4$  &  $\theta_{23}>\pi/4$  \\ \hline

$\chi^2_{\text{min}}$  &  204.875 &  204.610  & 209.331  &  222.058  & 14.811   &  15.138 \\\hline

$\theta_{\text{bf}}$  &  \multicolumn{2}{|c|}{0.227}  & \multicolumn{2}{|c|}{$\pm0.229$}  &  \multicolumn{2}{|c|}{0.439}  \\\hline

$\sin^2\theta_{13}(\theta_{\text{bf}})$  &   \multicolumn{2}{|c|}{0.0253}  &  \multicolumn{2}{|c|}{0.0257}   &    \multicolumn{2}{|c|}{0.0230}    \\\hline

$\theta_{13}(\theta_{\text{bf}})/^{\circ}$  &   \multicolumn{2}{|c|}{9.147}  &  \multicolumn{2}{|c|}{9.233}   &    \multicolumn{2}{|c|}{8.741}    \\\hline

$\sin^2\theta_{12}(\theta_{\text{bf}})$  &   \multicolumn{2}{|c|}{0.487}    &   \multicolumn{2}{|c|}{0.487}    &    \multicolumn{2}{|c|}{0.256}    \\\hline

$\theta_{12}(\theta_{\text{bf}})/^{\circ}$  &   \multicolumn{2}{|c|}{44.257}    &   \multicolumn{2}{|c|}{44.243}    &    \multicolumn{2}{|c|}{30.390}    \\\hline

$\sin^2\theta_{23}(\theta_{\text{bf}})$  &   0.388 &  0.612  &   \multicolumn{2}{|c|}{0.5}    &  0.419  &  0.581  \\\hline

$\theta_{23}(\theta_{\text{bf}})/^{\circ}$  &  38.506 &  51.492 &   \multicolumn{2}{|c|}{45}    & 40.358  &  49.668  \\\hline

$\left|\sin\delta_{\text{CP}}(\theta_{\text{bf}})\right|$  &    \multicolumn{2}{|c|}{0.726}    &   \multicolumn{2}{|c|}{1}    &    \multicolumn{2}{|c|}{0}     \\\hline

$\delta_{\text{CP}}(\theta_{\text{bf}})/^{\circ}$  &    \multicolumn{2}{|c|}{46.525}    &   \multicolumn{2}{|c|}{90}    &    \multicolumn{2}{|c|}{0}     \\\hline

$\left|\sin\alpha_{21}(\theta_{\text{bf}})\right|$  &    \multicolumn{6}{|c|}{0}    \\ \hline

$\alpha_{21}(\theta_{\text{bf}})/^{\circ}$  &    \multicolumn{6}{|c|}{0}    \\ \hline

$\left|\sin\alpha^{\prime}_{31}(\theta_{\text{bf}})\right|$  &   \multicolumn{6}{|c|}{0}    \\\hline

$\alpha^{\prime}_{31}(\theta_{\text{bf}})/^{\circ}$  &   \multicolumn{6}{|c|}{0}    \\\hline\hline

\end{tabular}
\caption{\label{tab:caseXIX_XX}The results of the mixing parameters for the cases XIX, XX, XXI, XXII, XXIII, XXIV, XXV and XXVI, where $``+\infty"$ for $\left|\tan\delta_{\text{CP}}\right|$, $\left|\tan\alpha\right|$ and $\left|\tan\beta\right|$ implies that the absolute value of the corresponding CP phase is $\pi/2$.  Notice that the Dirac CP phase $\delta_{\text{CP}}$ is determined up to $\delta_{\text{CP}}$, $\pi+\delta_{\text{CP}}$, $\pi-\delta_{\text{CP}}$ and $2\pi-\delta_{\text{CP}}$ in the present context, and only one representative value is displayed in this table. The same convention is taken for the Majorana CP phases $\alpha_{21}$ and $\alpha^{\prime}_{31}$.}
\end{table}


\begin{figure}[hptb!]
\begin{center}
\includegraphics[width=1.0\textwidth]{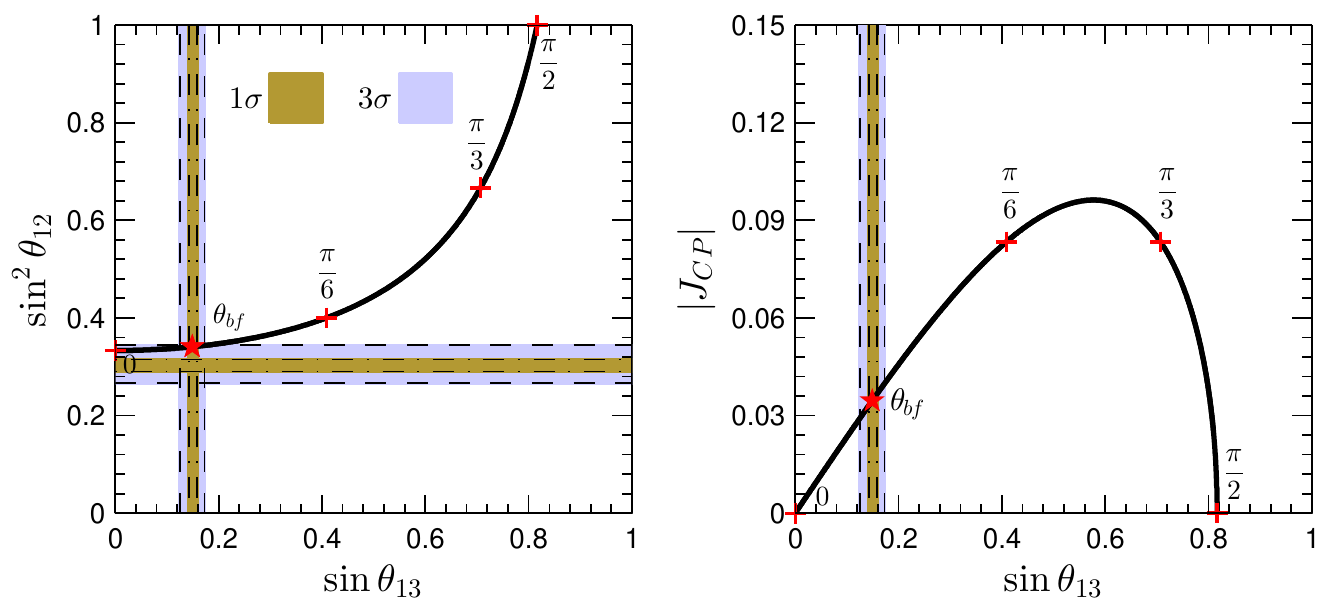} \caption{\label{fig:caseIII}The correlations of $\sin\theta_{13}$, $\sin^2\theta_{12}$ and $\left|J_{CP}\right|$ in case III and case VIII. The best fit value $\theta_{bf}$ for $\theta$ is labelled as a star. We also indicate the points for $\theta=0, \pi/6, \pi/3, \pi/2, 2\pi/3$ and $5\pi/6$ by the sign ``+'' on the curves, and only one $\theta$ value is displayed if there are points matching together. The shown $3\sigma$ ranges for the mixing angles and their best fit values are
taken from Ref.~\cite{GonzalezGarcia:2012sz}. }
\end{center}
\end{figure}

\begin{figure}[hptb!]
\includegraphics[width=0.95\textwidth]{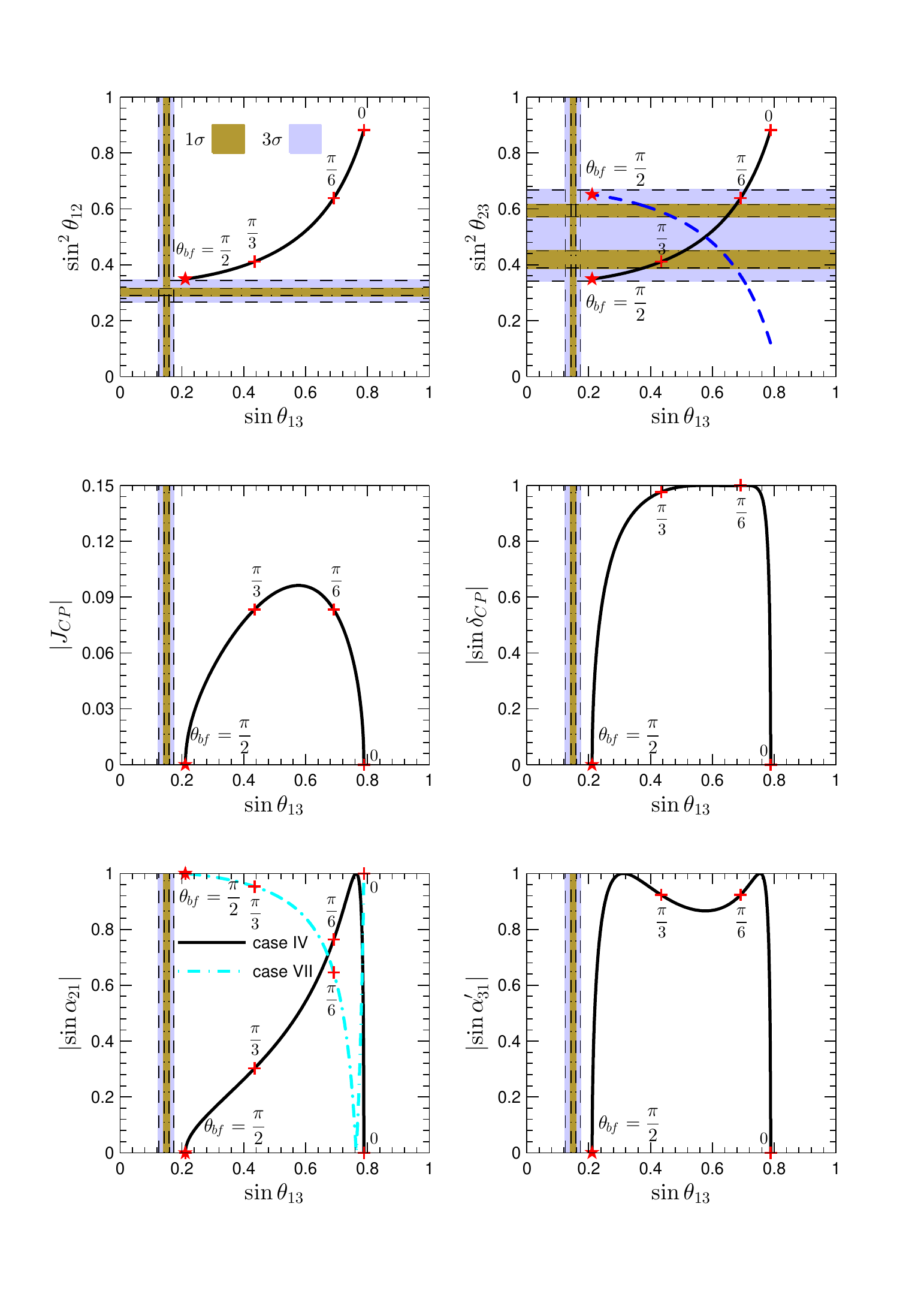} ~~~\caption{\label{fig:caseIV}The relation among the lepton mixing parameters in case IV and case VII. The best fit value $\theta_{bf}$ for $\theta$ is labelled as a star. We also indicate the points for $\theta=0, \pi/6, \pi/3, \pi/2, 2\pi/3$ and $5\pi/6$ by the sign ``+'' on the curves, and only one $\theta$ value is displayed if there are points matching together. The shown $3\sigma$ ranges for the mixing angles and their best fit values are taken from Ref.~\cite{GonzalezGarcia:2012sz}. }
\end{figure}

\begin{figure}[hptb!]\vskip-0.3in
\includegraphics[width=0.95\textwidth]{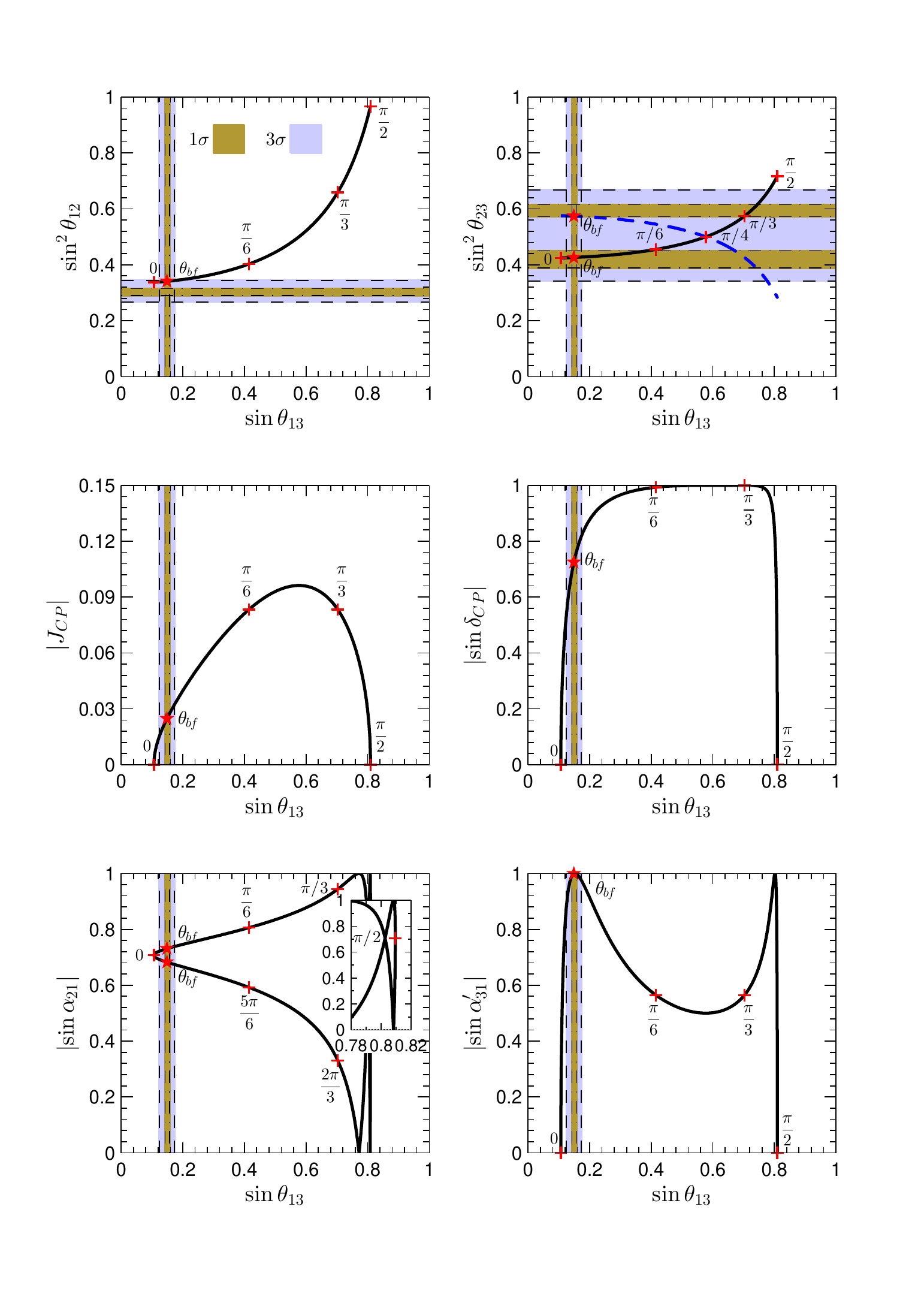} ~~~\caption{\label{fig:caseV}The correlation between the different lepton mixing parameters in case V, case VI, case IX and case X. The best fit value $\theta_{bf}$ for $\theta$ is labelled as a star. We also indicate the points for $\theta=0, \pi/6, \pi/3, \pi/2, 2\pi/3$ and $5\pi/6$ by the sign ``+'' on the curves, and only one $\theta$ value is displayed if there are points matching together. The shown $3\sigma$ ranges for the mixing angles and their best fit values are taken from Ref.~\cite{GonzalezGarcia:2012sz}. Note that the plots for $\left|\sin\alpha_{21}\right|$ with respect to $\sin\theta_{13}$ in cases V, IX and case VI, X coincide with each other since they are related by the transformation $\theta\rightarrow-\theta$, as shown in Table~\ref{tab:caseV_VI}.}
\end{figure}

\begin{figure}[hptb!]\vskip-0.3in
\includegraphics[width=0.95\textwidth]{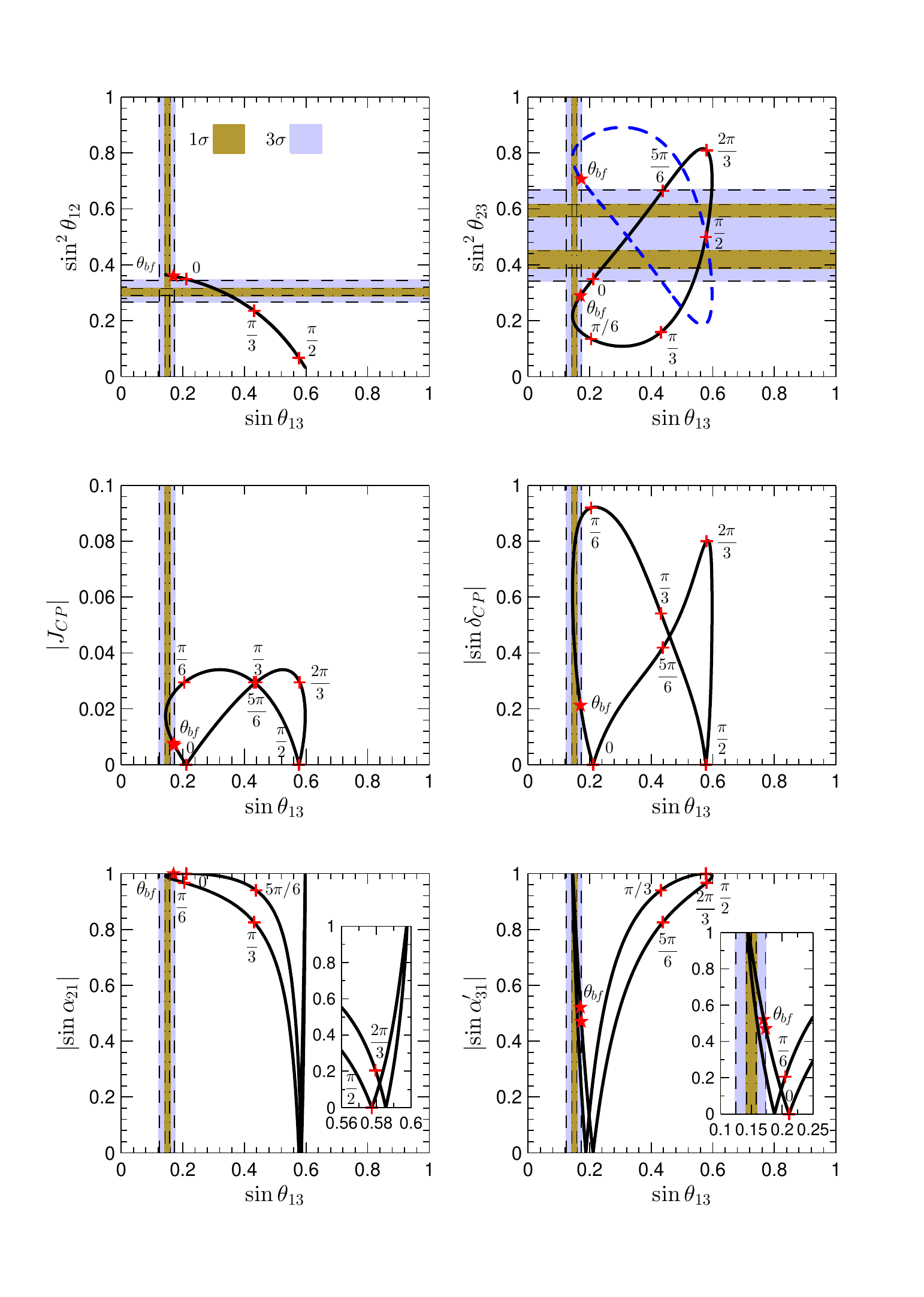} ~~~\caption{\label{fig:caseXIII_XIV}The relation among the lepton mixing parameters in case XIII and case XIV. The best fit value $\theta_{bf}$ for $\theta$ is labelled as a star. We also indicate the points for $\theta=0, \pi/6, \pi/3, \pi/2, 2\pi/3$ and $5\pi/6$ by the sign ``+'' on the curves, and only one $\theta$ value is displayed if there are points matching together. The shown $3\sigma$ ranges for the mixing angles and their best fit values are taken from Ref.~\cite{GonzalezGarcia:2012sz}. }
\end{figure}

\section{Conclusions}
\label{5}
\cleqn

We have performed a comprehensive study of the $\Delta (96)$ family symmetry combined with the generalised CP symmetry $H_{\rm{CP}}$. We have investigated the lepton mixing parameters which can be obtained from the original symmetry $\Delta (96)\rtimes H_{\rm{CP}}$ breaking to different remnant symmetries in the neutrino and charged lepton sectors, namely $G_{\nu}$ and $G_l$ subgroups in the neutrino and the charged lepton sector respectively, and the remnant CP symmetries from the breaking of $H_{\rm{CP}}$ are $H^{\nu}_{\rm{CP}}$ and $H^{l}_{\rm{CP}}$, respectively, where all cases correspond to a preserved symmetry smaller than the full Klein symmetry, as in the semi-direct approach, leading to predictions which depend on a single undetermined parameter, which may be fitted to the reactor angle for example.

The semi-direct approach to $\Delta (96)\rtimes H_{\rm{CP}}$, in which a smaller symmetry than the full Klein symmetry is preserved, clearly leads to a very rich set of possible cases which we have systematically studied here. We have discussed 26 possible cases, including a global $\chi^2$ determination of the best fit parameters and the correlations between mixing parameters, in each case. Excellent agreement with the presently observed lepton mixing angles can be achieved in some cases. It is remarkable that the CP phases are predicted to take irregular values rather than 0, $\pi$ or $\pm\pi/2$ in cases V, VI, IX and X, as shown in Table~\ref{tab:caseV_VI}. It remains to be seen if any of these possibilities will closely correspond to the observed future precise determination of leptonic mixing angles and CP violating parameters in the future.

\section*{Acknowledgements}

We are grateful to Alexander J. Stuart for his participation in the early stage of the work. The research was partially supported by the National Natural Science Foundation of China under Grant Nos. 11275188 and 11179007. SK acknowledges support from the STFC Consolidated ST/J000396/1 grant and the EU ITN grant INVISIBLES 289442

\newpage

\appendix

\section{\label{sec:appendix_A}Group theory of $\Delta(96)$}

\cleqn

$\Delta(96)$ belongs to the group series $\Delta(6n^2)$ with $n=4$, and it is a non-abelian finite subgroup of SU(3) of order 96. $\Delta(96)$ is isomorphic to $(Z_4\times Z_4)\rtimes S_3$, where $S_3$ is isomorphic to $Z_3\rtimes Z_2$, and it can be conveniently defined by four generators $a$, $b$, $c$ and $d$ obeying the relations~\cite{Escobar:2008vc}:
\begin{eqnarray}
\nonumber&a^3~=~b^2~=~(ab)^2~=~c^4~=~d^4~=~1\\
\nonumber&cd~=~dc\\
\nonumber&aca^{-1}=c^{-1}d^{-1},\quad ada^{-1}=c\\
\label{eq:abcd_generators}& bcb^{-1}=d^{-1}, \quad bdb^{-1}=c^{-1}\,,
\end{eqnarray}
The elements $a$ and $b$ are the generators of $S_3$ while
$c$ and $d$ generate $Z_4\times Z_4$, and the last two lines define the semidirect product $``\rtimes"$. Note that the generator $d=bc^3b$ is not independent. In order to see clearly the connection between the lepton flavor mixing and $\Delta(96)$ family symmetry, it is useful to express $\Delta(96)$ in terms of the ``canonical'' $S$, $T$ and $U$ generators~\cite{King:2013eh}, where $S$ and $U$ usually generate the remnant Klein group $Z^{S}_2\times Z^{U}_2$ in the neutrino sector while $T$ is the generator of the residual symmetry group $Z^{T}_3$ in the charged lepton sector. They satisfy the multiplication rules
\begin{eqnarray}
\nonumber&& S^2=T^3=U^2=(ST)^3=1,\qquad SU=US,\\
&&(TU)^8=1,\qquad (TUT^2U)^3=1,\qquad (UTSUT^2UT)^2=1\,.
\end{eqnarray}
Note that the generators $S$ and $T$ alone generate the well-known group $A_4$. The identities relating the two sets of generators are as follows,
\begin{eqnarray}
\nonumber&S=d^2,\qquad T=ac,\qquad U=a^2bd,\\
\label{eq:generators}&a=SUT^2U,\qquad b=T^2UT,\qquad c=UTSUT,\qquad d=TUT^2SUT\,.
\end{eqnarray}
The $\Delta(96)$ group has ten conjugacy classes:
\begin{eqnarray}
\nonumber&&1C_1: 1\\
\nonumber&&3C_4: cd^2=(T^2U)^2, cd^3=(UT^2)^2, c^2d^3=TUT^2UT\,,\\
\nonumber&&3C_2: c^2=TST^2, d^2=S, c^2d^2=T^2ST\,,\\
\nonumber&&3C^{\prime}_4: c^2d=T^2UTUT^2,c^3d=(TU)^2, c^3d^2=(UT)^2\,,\\ 
\nonumber&&6C_4: c=UTSUT, d=TUT^2SUT, cd=S(UT^2)^2, c^3=S(UT)^2, d^3=TSUT^2UT, \\
\nonumber&&~~\qquad c^3d^3=UT^2SUT^2\,,\\
\nonumber&&32C_3: a=SUT^2U, ac=T, ac^2=UT^2SU, ac^3=T^2ST^2, ad=T(TU)^2, ad^2=SUT^2US, \\
\nonumber&&~~\qquad ad^3=T^2UTSU, acd=UTSUT^2, acd^2=TS, acd^3=SUTUT^2, ac^2d=T^2SUTU, \\
\nonumber&&~~\qquad ac^2d^2=UT^2U, ac^2d^3=T(UT^2)^2, ac^3d=(UT)^2T, ac^3d^2=ST, ac^3d^3=(T^2U)^2T, \\
\nonumber&&~~\qquad a^2=UTSU, a^2c=SUT^2UT, a^2c^2=SUTUS, a^2c^3=UT^2UT, a^2d=TSUT^2U, \\
\nonumber&&~~\qquad a^2d^2=UTU, a^2d^3=T^2(UT)^2, a^2cd=T^2, a^2cd^2=UT^2SUT, a^2cd^3=T^2S,\\
\nonumber&&~~\qquad a^2c^2d=TUT^2SU, a^2c^2d^2=SUTU, a^2c^2d^3=TUT^2U, a^2c^3d=ST^2,\\
\nonumber&&~~\qquad a^2c^3d^2=(TU)^2T^2, a^2c^3d^3=TST\,,\\
\nonumber&&12C_2: ab=TSUT^2, abc=UT^2SUTU, abc^2=TUT^2, abc^3=UT^2UTU, a^2b=UTSUT^2UT,\\
\nonumber&&~~\qquad a^2bd=U, a^2bd^2=UTUT^2UST, a^2bd^3=SU, b=T^2UT, bcd=UTSUT^2U, \\
\nonumber&&~~\qquad bc^2d^2=T^2SUT, bc^3d^3=UTUT^2U\,,\\
\nonumber&&12C_8: abd=UTS, abcd=T^2U, abc^2d=SUT, abc^3d=ST^2US, a^2bc^3=T^2SUT^2,\\
\nonumber&&~~\qquad a^2bc^3d=STUT, a^2bc^3d^2=ST^2UST^2, a^2bc^3d^3=TUT, bc=TSU, bc^2d=UT^2,\\
\nonumber&&~~\qquad bc^3d^2=STU, bd^3=UTST\,,\\
\nonumber&&12C_4: abd^2=STUT^2, abcd^2=UT(UT^2)^2, abc^2d^2=TUT^2S, abc^3d^2=UT^2UTUS, \\
\nonumber&&~~\qquad a^2bc^2=UTUT^2UT, a^2bc^2d=UTST^2, a^2bc^2d^2=UT^2UTUT^2, a^2bc^2d^3=UT^2ST,\\
\nonumber&&~~\qquad bc^2=ST^2UT, bc^3d=UTUT^2US, bd^2=T^2UTS, bcd^3=UT^2(UT)^2\,,\\
\nonumber&&12C^{\prime}_{8}: abd^3=UT, abcd^3=T^2SU, abc^2d^3=UT^2ST^2, abc^3d^3=ST^2U, a^2bc=T^2UT^2,\\
\nonumber&&~~\qquad a^2bcd=TSUT, a^2bcd^2=ST^2UT^2, a^2bcd^3=STUST, bc^3=STUS, bd=SUT^2,\\
\label{eq:conjugacy_calss}&&~~\qquad bcd^2=TU, bc^2d^3=UT^2S\,.
\end{eqnarray}
Note that the conjugacy class is denoted in the notation of $kC_{n}$, where $k$ stands for the number of elements in the class and the subscript $n$ indicates the order of the elements. $\Delta(96)$ has two singlet irreducible representations $\mathbf{1}$ and $\mathbf{1^{\prime}}$, one doublet irreducible representation $\mathbf{2}$, six triplet irreducible representations $\mathbf{3}$, $\mathbf{3^{\prime}}$, $\mathbf{\overline{3}}$, $\mathbf{\overline{3}^{\prime}}$, $\mathbf{\widetilde{3}}$, $\mathbf{\widetilde{3}^{\prime}}$, and one sextet ${\bf 6}$. We note that $\mathbf{\overline{3}}$ and $\mathbf{\overline{3}^{\prime}}$ are the complex conjugate representations of $\mathbf{3}$ and $\mathbf{3^{\prime}}$ respectively, and the representations $\mathbf{3}$, $\mathbf{3^{\prime}}$, $\mathbf{\overline{3}}$, $\mathbf{\overline{3}^{\prime}}$ and $\mathbf{6}$ are the faithful representations of $\Delta(96)$, while $\mathbf{\widetilde{3}}$ and $\mathbf{\widetilde{3}^{\prime}}$ are not. Our choice of the basis for the representation matrices of $S$, $T$
and $U$ is listed in Table~\ref{tab:representation}, where we have defined
\begin{table}[t!]
\begin{center}
\begin{tabular}{|c|c|c|c|}\hline\hline
 ~~  &  $S$  &   $T$      &    $U$  \\ \hline
$\mathbf{1}$, $\mathbf{1}^{\prime}$ &  1  & 1  &  $\pm1$ \\ \hline

$\mathbf{2}$  &   $\left(\begin{array}{cc}1  &  0  \\
0  & 1
\end{array}\right)$   &   $\left(\begin{array}{cc}\omega^2 & 0 \\
0  &  \omega \end{array}\right)$  &   $\left(\begin{array}{cc}0&1 \\
1& 0\end{array}\right)$  \\\hline

$\mathbf{3}$,  $\mathbf{3^{\prime}}$ &   $S_{\mathbf{3}}$  &  $T_{\mathbf{3}}$   &   $\pm U_{\mathbf{3}}$  \\  \hline

$\mathbf{\overline{3}}$,  $\mathbf{\overline{3}^{\prime}}$ &   $S_{\mathbf{3}}$  &  $T^{*}_{\mathbf{3}}$   &   $\pm U_{\mathbf{3}}$  \\  \hline

$\mathbf{\widetilde{3}}$, $\mathbf{\widetilde{3}^{\prime}}$  &   $\mathds{1}_{3\times3}$  &  $T_{\mathbf{3}}$   &   $\mp P_{12}S_{\mathbf{3}}$  \\  \hline

$\mathbf{6}$  &   $\left(\begin{array}{cc}
S_{\mathbf{3}}  &  0  \\
0  &   S_{\mathbf{3}}
\end{array}\right)$   &  $\left(\begin{array}{cc}
T_{\mathbf{3}}  &  0 \\
0 &  T_{\mathbf{3}}
\end{array}\right)$  &  $\left(\begin{array}{cc}
\widetilde{U}  &  \widetilde{U}-P_{12}S_{\mathbf{3}}  \\
\widetilde{U}-P_{12}S_{\mathbf{3}}  & -\widetilde{U}
\end{array}\right)$ \\ \hline\hline
\end{tabular}
\caption{\label{tab:representation}The representation matrices for the $\Delta(96)$ generators $S$, $T$ and $U$ in different irreducible representations, where $\omega=e^{2\pi i/3}$ is the cube root of unit, $\mathds{1}_{3\times3}$ denotes the $3\times3$ unity matrix, and the matrices $S_{\mathbf{3}}$, $T_{\mathbf{3}}$, $U_{\mathbf{3}}$, $P_{12}$ and $\widetilde{U}$ are given in Eq.~\eqref{eq:representation_matrix}.}
\end{center}
\end{table}
\begin{eqnarray}
\nonumber& P_{12}=\begin{pmatrix}
0  &  1  &  0 \\
1  &  0   &  0  \\
0  &   0  & 1
\end{pmatrix},\quad \widetilde{U}=\frac{1}{3}\begin{pmatrix}
1  &  1  &  1    \\
1  &  1 &  1  \\
1  &  1   &  1
\end{pmatrix},\quad S_{\mathbf{3}}=\frac{1}{3}\begin{pmatrix}
-1  &  2  &  2 \\
2  &  -1  &  2  \\
2  &  2   &  -1
\end{pmatrix}, \\
\label{eq:representation_matrix}&
T_{\mathbf{3}}=\begin{pmatrix}
\omega  &  0  &  0  \\
0  &  \omega^2  &  0  \\
0  &  0  &  1
\end{pmatrix},~~ U_{\mathbf{3}}=\frac{1}{3}\begin{pmatrix}
-1-\sqrt{3}   &   ~-1    &   ~-1+\sqrt{3}  \\
 -1   & ~-1+\sqrt{3}   &   ~-1-\sqrt{3}  \\
-1+\sqrt{3}    &   ~-1-\sqrt{3}   &  ~-1
\end{pmatrix}\,.
\end{eqnarray}
Then we can straightforwardly obtain the character table of $\Delta(96)$ as shown in Table~\ref{tab:character_table}. Furthermore, the Kronecker products between various irreducible representations are as follows:
\begin{eqnarray}
\nonumber&&\mathbf{1}^{\prime}\otimes\mathbf{2}=\mathbf{2},~\mathbf{1^{\prime}}\otimes\mathbf{r}=\mathbf{r^{\prime}}, ~\mathbf{1^{\prime}\otimes\mathbf{r^{\prime}}}=\mathbf{r},~\mathbf{1}^{\prime}\otimes\mathbf{6}=\mathbf{6},~ \mathbf{2}\otimes\mathbf{2}=\mathbf{1}\oplus\mathbf{1}^{\prime}\oplus\mathbf{2},\\ \nonumber&&\mathbf{2}\otimes\mathbf{r}=\mathbf{2}\otimes\mathbf{r^{\prime}}=\mathbf{r}\oplus\mathbf{r^{\prime}},~\mathbf{2}\otimes\mathbf{6}=\mathbf{6}\oplus\mathbf{6},~
\mathbf{3}\otimes\mathbf{3}=\mathbf{3^{\prime}}\otimes\mathbf{3^{\prime}}=\mathbf{\overline{3}}\oplus\mathbf{\overline{3}^{\prime}}\oplus\mathbf{\widetilde{3}^{\prime}},\\
\nonumber&&\mathbf{3}\otimes\mathbf{3^{\prime}}=\mathbf{\overline{3}}\oplus\mathbf{\overline{3}^{\prime}}\oplus\mathbf{\widetilde{3}},
~\mathbf{3}\otimes\mathbf{\overline{3}}=\mathbf{3^{\prime}}\otimes\mathbf{\overline{3}^{\prime}}=\mathbf{1}\oplus\mathbf{2}\oplus\mathbf{6},
~\mathbf{3}\otimes\mathbf{\overline{3}^{\prime}}=\mathbf{3^{\prime}}\otimes\mathbf{\overline{3}}=\mathbf{1^{\prime}}\oplus\mathbf{2}\oplus\mathbf{6},\\
\nonumber&&\mathbf{3}\otimes\mathbf{\widetilde{3}}=\mathbf{3^{\prime}}\otimes\mathbf{\widetilde{3}^{\prime}}=\mathbf{\overline{3}^{\prime}}\oplus\mathbf{6},
~\mathbf{3}\otimes\mathbf{\widetilde{3}^{\prime}}=\mathbf{3^{\prime}}\otimes\mathbf{\widetilde{3}}=\mathbf{\overline{3}}\oplus\mathbf{6},\\
\nonumber&&\mathbf{3}\otimes\mathbf{6}=\mathbf{3^{\prime}}\otimes\mathbf{6}=\mathbf{3}\oplus\mathbf{3^{\prime}}\oplus\mathbf{\widetilde{3}}\oplus\mathbf{\widetilde{3}^{\prime}}\oplus\mathbf{6},
~\mathbf{\overline{3}}\otimes\mathbf{\overline{3}}=\mathbf{\overline{3}^{\prime}}\otimes\mathbf{\overline{3}^{\prime}}=\mathbf{3}\oplus\mathbf{3^{\prime}}\oplus\mathbf{\widetilde{3}^{\prime}},\\
\nonumber&&\mathbf{\overline{3}}\otimes\mathbf{\overline{3}^{\prime}}=\mathbf{3}\oplus\mathbf{3^{\prime}}\oplus\mathbf{\widetilde{3}},~\mathbf{\overline{3}}\otimes\mathbf{\widetilde{3}}=\mathbf{\overline{3}^{\prime}}\otimes\mathbf{\widetilde{3}^{\prime}}=\mathbf{3^{\prime}}\oplus\mathbf{6},
~\mathbf{\overline{3}}\otimes\mathbf{\widetilde{3}^{\prime}}=\mathbf{\overline{3}^{\prime}}\otimes\mathbf{\widetilde{3}}=\mathbf{3}\oplus\mathbf{6},\\
\nonumber&&\mathbf{\overline{3}}\otimes\mathbf{6}=\mathbf{\overline{3}^{\prime}}\otimes\mathbf{6}=\mathbf{\overline{3}}\oplus\mathbf{\overline{3}^{\prime}}\oplus\mathbf{\widetilde{3}}\oplus\mathbf{\widetilde{3}^{\prime}}\oplus\mathbf{6},
~\mathbf{\widetilde{3}}\otimes\mathbf{\widetilde{3}}=\mathbf{\widetilde{3}^{\prime}}\otimes\mathbf{\widetilde{3}^{\prime}}=\mathbf{1}\oplus\mathbf{2}\oplus\mathbf{\widetilde{3}}\oplus\mathbf{\widetilde{3}^{\prime}},\\
\nonumber&&\mathbf{\widetilde{3}}\otimes\mathbf{\widetilde{3}^{\prime}}=\mathbf{1^{\prime}}\oplus\mathbf{2}\oplus\mathbf{\widetilde{3}}\oplus\mathbf{\widetilde{3}^{\prime}},~\mathbf{\widetilde{3}}\otimes\mathbf{6}=\mathbf{\widetilde{3}^{\prime}}\otimes\mathbf{6}=\mathbf{3}\otimes\mathbf{3^{\prime}}\oplus\mathbf{\overline{3}}\oplus\mathbf{\overline{3}^{\prime}}\oplus\mathbf{6},\\
&&\mathbf{6}\otimes\mathbf{6}=\mathbf{1}\oplus\mathbf{1^{\prime}}\oplus\mathbf{2}_S\oplus\mathbf{2}_A\oplus\mathbf{3}\oplus\mathbf{3^{\prime}}\oplus\mathbf{\overline{3}}\oplus\mathbf{\overline{3}^{\prime}}\oplus\mathbf{\widetilde{3}}\oplus\mathbf{\widetilde{3}^{\prime}}\oplus\mathbf{6}_S\oplus\mathbf{6}_A
\end{eqnarray}
where $\mathbf{r}$ denotes the triplet representations $\mathbf{3}$, $\mathbf{\overline{3}}$ and $\mathbf{\widetilde{3}}$, and the subscript $S(A)$ denotes symmetric (antisymmetric) combinations. Given the explicit forms of the generators in Table~\ref{tab:representation}, one can calculate the corresponding Clebsch-Gordan (CG) coefficients. In the following, we report the CG coefficients of $\Delta(96)$ in the form $\alpha\otimes\beta$, where the $\alpha_i$ denote the elements of the representation on the left of the product, and $\beta_i$ indicate those of the representation on the right of the product.

\begin{table}[t!]
\begin{center}
\begin{tabular}{|c|c|c|c|c|c|c|c|c|c|c|}\hline\hline
   &\multicolumn{10}{c|}{\tt Conjugate Classes}\\\cline{2-11}
   & $1C_1$ & $3C_4$ & $3C_2$  & $3C^{\prime}_4$ & $6C_4$ & $32C_3$ & $12C_2$  & $12C_8$  &  $12C_4$  &  $12C^{\prime}_{8}$  \\\hline

$G$  & 1  &  $cd^2$  &  $c^2$  &  $c^2d$  &  $c$   &  $a$  &  $b$   &  $bc$ &  $bc^2$   &  $bd$ \\\hline

$\mathbf{1}$&1&1&1&1&1 &1&1&1&1&1\\\hline

$\mathbf{1}'$ &1&1&1&1&1 &1&$-1$&$-1$&$-1$&$-1$ \\\hline

$\mathbf{2}$ & 2 & 2 & 2 & 2 & 2 & $-1$  & 0  & 0  &   0  & 0  \\\hline

$\mathbf{3}$& 3 & $-1 + 2 i$ & $-1$ & $-1 - 2 i$ & 1 & 0 & $-1$ & $i$  & 1 & $-i$
\\\hline

$\mathbf{3^{\prime}}$ & 3&$-1 + 2i$ & $-1$ & $-1- 2i$ & 1 & 0 & 1 & $-i$ & $-1$ & $i$ \\\hline

$\mathbf{\overline{3}}$& 3 & $-1 - 2 i$  & $-1$  & $-1 + 2 i$  & 1  & 0  & $-1$  & $-i$  & 1  & $i$   \\\hline

$\mathbf{\overline{3}^{\prime}}$& 3 & $-1 - 2 i$  & $-1$ & $-1 + 2 i$ & 1  & 0  & 1  & $i$  & $-1$  & $-i$ \\\hline

$\mathbf{\widetilde{3}}$& 3  & $-1$  & 3  & $-1$  & $-1$   & 0  & $-1$  & 1  & $-1$  & 1  \\\hline

$\mathbf{\widetilde{3}^{\prime}}$& 3 &  $-1$  & 3  & $-1$  & $-1$  & 0  & 1  & $-1$  & 1  & $-1$  \\\hline

$\mathbf{6}$& 6  & 2  & $-2$  & 2  & $-2$  & 0  & 0  & 0  & 0  & 0  \\\hline\hline

\end{tabular}
\caption{\label{tab:character_table}Character table of the $\Delta(96)$ group, where $kC_{n}$ denotes the classes with $k$ elements which have order $n$, $G$ is a representative of the class $kC_{n}$ in terms of the generators $a$, $b$, $c$ and $d$.}
\end{center}
\end{table}

\begin{itemize}[leftmargin=1.5em]

\item{$\mathbf{1^{\prime}}\otimes\mathbf{2}=\mathbf{2}$}

\begin{equation}
\nonumber\mathbf{2}\sim
\begin{pmatrix}
 \alpha_1\beta_1 \\
 -\alpha_1\beta_2
\end{pmatrix}\,.
\end{equation}


\item{$\mathbf{1^{\prime}}\otimes\mathbf{r}=\mathbf{r^{\prime}}$ with $\mathbf{r}=\mathbf{3}, \mathbf{\overline{3}}, \mathbf{\widetilde{3}}$}

\begin{equation}
\nonumber\mathbf{r^{\prime}}\sim
\begin{pmatrix}
 \alpha_1\beta_1 \\
 \alpha_1\beta_2 \\
 \alpha_1\beta_3
\end{pmatrix}\,.
\end{equation}


\item{$\mathbf{1^{\prime}\otimes\mathbf{r^{\prime}}}=\mathbf{r}$ with $\mathbf{r}=\mathbf{3}, \mathbf{\overline{3}}, \mathbf{\widetilde{3}}$}

\begin{equation}
\nonumber\mathbf{r}\sim
\begin{pmatrix}
 \alpha_1\beta_1 \\
 \alpha_1\beta_2 \\
 \alpha_1\beta_3
\end{pmatrix}\,.
\end{equation}


\item{$\mathbf{1^{\prime}}\otimes\mathbf{6}=\mathbf{6}$}

\begin{equation}
\nonumber\mathbf{6}\sim
\begin{pmatrix}
 \alpha _1 \beta _4 \\
 \alpha _1 \beta _5 \\
 \alpha _1 \beta _6 \\
 -\alpha _1 \beta _1 \\
 -\alpha _1 \beta _2 \\
 -\alpha _1 \beta _3
\end{pmatrix}\,.
\end{equation}


\item{$\mathbf{2}\otimes\mathbf{2}=\mathbf{1}\oplus\mathbf{1^{\prime}}\oplus\mathbf{2}$}

\begin{equation}
\nonumber\mathbf{1}\sim\alpha_1\beta_2+\alpha_2\beta_1,\quad \mathbf{1^{\prime}}\sim\alpha_1\beta_2-\alpha_2\beta_1,\quad \mathbf{2}\sim
\begin{pmatrix}
 \alpha_2\beta_2 \\
 \alpha_1\beta_1
\end{pmatrix}\,.
\end{equation}


\item{$\mathbf{2}\otimes\mathbf{r}=\mathbf{r}\oplus\mathbf{r^{\prime}}$ with $\mathbf{r}=\mathbf{3}, \mathbf{\widetilde{3}}$}

\begin{equation}
\nonumber\mathbf{r}\sim
\begin{pmatrix}{c}
 \alpha_1\beta_2+\alpha_2\beta_3 \\
 \alpha_1\beta_3+\alpha_2\beta_1 \\
 \alpha_1\beta_1+\alpha_2\beta_2
\end{pmatrix},\quad \mathbf{r^{\prime}}\sim
\begin{pmatrix}
 \alpha_1\beta_2-\alpha_2\beta_3 \\
 \alpha_1\beta_3-\alpha_2\beta_1 \\
 \alpha_1\beta_1-\alpha_2\beta_2
\end{pmatrix}\,.
\end{equation}


\item{$\mathbf{2}\otimes\mathbf{r^{\prime}}=\mathbf{r}\oplus\mathbf{r^{\prime}}$ with $\mathbf{r}=\mathbf{3}, \mathbf{\widetilde{3}}$}

\begin{equation}
\nonumber\mathbf{r}\sim
\begin{pmatrix}
 \alpha_1\beta_2-\alpha_2\beta_3 \\
 \alpha_1\beta_3-\alpha_2\beta_1 \\
 \alpha_1\beta_1-\alpha_2\beta_2
\end{pmatrix},\quad \mathbf{r^{\prime}}\sim
\begin{pmatrix}
 \alpha_1\beta_2+\alpha_2\beta_3 \\
 \alpha_1\beta_3+\alpha_2\beta_1 \\
 \alpha_1\beta_1+\alpha_2\beta_2
\end{pmatrix}\,.
\end{equation}


\item{$\mathbf{2}\otimes\mathbf{\overline{3}}=\mathbf{\overline{3}}\oplus\mathbf{\overline{3}^{\prime}}$}

\begin{equation}
\nonumber\mathbf{\overline{3}}\sim
\begin{pmatrix}
 \alpha_1\beta_3+\alpha_2\beta_2\\
 \alpha_1\beta_1+\alpha_2\beta_3 \\
 \alpha_1\beta_2+\alpha_2\beta_1
\end{pmatrix},\quad \mathbf{\overline{3}^{\prime}}\sim
\begin{pmatrix}
 \alpha_1\beta_3-\alpha_2\beta_2 \\
 \alpha_1\beta_1-\alpha_2\beta_3 \\
 \alpha_1\beta_2-\alpha_2\beta_1
\end{pmatrix}\,.
\end{equation}


\item{$\mathbf{2}\otimes\mathbf{\overline{3}^{\prime}}=\mathbf{\overline{3}}\oplus\mathbf{\overline{3}^{\prime}}$}

\begin{equation}
\nonumber\mathbf{\overline{3}}\sim
\begin{pmatrix}
 \alpha_1\beta_3-\alpha_2\beta_2 \\
 \alpha_1\beta_1-\alpha_2\beta_3 \\
 \alpha_1\beta_2-\alpha_2\beta_1
\end{pmatrix},\quad \mathbf{\overline{3}^{\prime}}\sim
\begin{pmatrix}
 \alpha_1\beta_3+\alpha_2\beta_2 \\
 \alpha_1\beta_1+\alpha_2\beta_3 \\
 \alpha_1\beta_2+\alpha_2\beta_1
\end{pmatrix}\,.
\end{equation}


\item{$\mathbf{2}\otimes\mathbf{6}=\mathbf{6}\oplus\mathbf{6}$}

\begin{equation}
\nonumber\mathbf{6}\sim
\begin{pmatrix}
 \alpha _1 \beta _2+\alpha _2 \beta _3 \\
 \alpha _1 \beta _3+\alpha _2 \beta _1 \\
 \alpha _1 \beta _1+\alpha _2 \beta _2 \\
 \alpha _1 \beta _5+\alpha _2 \beta _6 \\
 \alpha _1 \beta _6+\alpha _2 \beta _4 \\
 \alpha _1 \beta _4+\alpha _2 \beta _5
\end{pmatrix},\qquad \mathbf{6}\sim
\begin{pmatrix}
 \alpha _1 \beta _5-\alpha _2 \beta _6 \\
 \alpha _1 \beta _6-\alpha _2 \beta _4 \\
 \alpha _1 \beta _4-\alpha _2 \beta _5 \\
 \alpha _2 \beta _3 -\alpha _1 \beta _2\\
 \alpha _2 \beta _1-\alpha _1 \beta _3 \\
 \alpha _2 \beta _2-\alpha _1 \beta _1
\end{pmatrix}\,.
\end{equation}


\item{$\mathbf{3}\otimes\mathbf{3}=\mathbf{3^{\prime}}\otimes\mathbf{3^{\prime}}=\mathbf{\overline{3}}\oplus\mathbf{\overline{3}^{\prime}}\oplus\mathbf{\widetilde{3}^{\prime}}$}

\begin{equation}
\nonumber\mathbf{\overline{3}}\sim
\begin{pmatrix}
 \alpha_2\beta_3-\alpha_3\beta_2 \\
 \alpha_3\beta_1-\alpha_1\beta_3 \\
 \alpha_1\beta_2-\alpha_2\beta_1
\end{pmatrix},~~\mathbf{\overline{3}^{\prime}}\sim
\begin{pmatrix}
 2\alpha_1\beta_1-\alpha_2\beta_3-\alpha_3\beta_2 \\
 2\alpha_2\beta_2-\alpha_1\beta_3-\alpha_3\beta_1 \\
 2\alpha_3\beta_3-\alpha_1\beta_2-\alpha_2\beta_1
\end{pmatrix},~~\mathbf{\widetilde{3}^{\prime}}\sim
\begin{pmatrix}
 \alpha_1\beta_3+\alpha_2\beta_2+\alpha_3\beta_1 \\
 \alpha_1\beta_1+\alpha_2\beta_3+\alpha_3\beta_2 \\
 \alpha_1\beta_2+\alpha_2\beta_1+\alpha_3\beta_3
\end{pmatrix}\,.
\end{equation}


\item{$\mathbf{3}\otimes\mathbf{3^{\prime}}=\mathbf{\overline{3}}\oplus\mathbf{\overline{3}^{\prime}}\oplus\mathbf{\widetilde{3}}$}

\begin{equation}
\nonumber\mathbf{\overline{3}}\sim
\begin{pmatrix}
 2\alpha_1\beta_1-\alpha_2\beta_3-\alpha_3\beta_2 \\
 2\alpha_2\beta_2-\alpha_1\beta_3-\alpha_3\beta_1 \\
 2\alpha_3\beta_3-\alpha_1\beta_2-\alpha_2\beta_1
\end{pmatrix},~~\mathbf{\overline{3}^{\prime}}\sim
\begin{pmatrix}
 \alpha_2\beta_3-\alpha_3\beta_2 \\
 \alpha_3\beta_1-\alpha_1\beta_3 \\
 \alpha_1\beta_2-\alpha_2\beta_1
\end{pmatrix},~~\mathbf{\widetilde{3}}\sim
\begin{pmatrix}
 \alpha_1\beta_3+\alpha_2\beta_2+\alpha_3\beta_1\\
 \alpha_1\beta_1+\alpha_2\beta_3+\alpha_3\beta_2 \\
 \alpha_1\beta_2+\alpha_2\beta_1+\alpha_3\beta_3
\end{pmatrix}\,.
\end{equation}


\item{$\mathbf{3}\otimes\mathbf{\overline{3}}=\mathbf{3^{\prime}}\otimes\mathbf{\overline{3}^{\prime}}=\mathbf{1}\oplus\mathbf{2}\oplus\mathbf{6}$}

\begin{equation}
\nonumber\mathbf{1}\sim\alpha_1\beta_1+\alpha_2\beta_2+\alpha_3\beta_3 ,~~\mathbf{2}\sim
\begin{pmatrix}
 \alpha_1\beta_2+\alpha_2\beta_3+\alpha_3\beta_1 \\
 \alpha_1\beta_3+\alpha_2\beta_1+\alpha_3\beta_2
\end{pmatrix},~~\mathbf{6}\sim
\begin{pmatrix}
 \sqrt{3} \left(\alpha_1\beta_3-\alpha_3\beta_2\right) \\
 \sqrt{3} \left(\alpha_3\beta_1-\alpha_2\beta_3\right) \\
 \sqrt{3} \left(\alpha_2\beta_2-\alpha_1\beta_1\right) \\
 2\alpha_2\beta_1-\alpha_1\beta_3-\alpha_3\beta_2 \\
2\alpha_1\beta_2-\alpha_2 \beta_3 -\alpha_3 \beta_1 \\
2\alpha_3\beta_3-\alpha_1 \beta_1-\alpha_2 \beta_2
\end{pmatrix}\,.
\end{equation}


\item{$\mathbf{3}\otimes\mathbf{\overline{3}^{\prime}}=\mathbf{3^{\prime}}\otimes\mathbf{\overline{3}}=\mathbf{1^{\prime}}\oplus\mathbf{2}\oplus\mathbf{6}$}

\begin{equation}
\nonumber\mathbf{1^{\prime}}\sim \alpha_1\beta_1+\alpha_2\beta_2+\alpha_3\beta_3,~~ \mathbf{2}\sim
\begin{pmatrix}
 \alpha_1\beta_2+\alpha_2\beta_3+\alpha_3\beta_1 \\
 -\alpha_1\beta_3-\alpha_2\beta_1-\alpha_3\beta_2
\end{pmatrix},~~ \mathbf{6}\sim
\begin{pmatrix}
2\alpha_2\beta_1-\alpha_1\beta_3-\alpha_3\beta_2 \\
2\alpha_1\beta_2-\alpha_2\beta_3-\alpha_3\beta_1 \\
2\alpha_3\beta _3-\alpha_1\beta_1-\alpha_2\beta_2 \\
\sqrt{3}\left(\alpha_3\beta_2-\alpha_1\beta_3\right) \\
\sqrt{3}\left(\alpha_2\beta_3-\alpha_3\beta_1\right) \\
\sqrt{3}\left(\alpha_1\beta_1-\alpha_2\beta_2\right)
\end{pmatrix}\,.
\end{equation}


\item{$\mathbf{3}\otimes\mathbf{\widetilde{3}}=\mathbf{3^{\prime}}\otimes\mathbf{\widetilde{3}^{\prime}}=\mathbf{\overline{3}^{\prime}}\oplus\mathbf{6}$}

\begin{equation}
\nonumber\mathbf{\overline{3}^{\prime}}\sim
\begin{pmatrix}
 \alpha_1\beta_1+\alpha_2\beta_3 +\alpha_3\beta_2\\
 \alpha_1\beta_3+\alpha_2\beta_2+\alpha_3\beta_1 \\
 \alpha_1\beta_2+\alpha_2\beta_1+\alpha_3\beta_3
\end{pmatrix},~\quad \mathbf{6}\sim
\begin{pmatrix}
\sqrt{3}\left(\alpha_3\beta_1-\alpha_2\beta_2\right) \\
\sqrt{3}\left(\alpha_1\beta_1-\alpha_3\beta_2\right) \\
\sqrt{3}\left(\alpha_2\beta_1-\alpha_1\beta_2\right) \\
2\alpha_1\beta_3-\alpha_2\beta_2-\alpha_3\beta_1 \\
2\alpha_2\beta_3-\alpha_1\beta_1-\alpha_3\beta_2 \\
2\alpha_3\beta_3-\alpha_1\beta_2-\alpha_2\beta _1
\end{pmatrix}\,.
\end{equation}


\item{$\mathbf{3}\otimes\mathbf{\widetilde{3}^{\prime}}=\mathbf{3^{\prime}}\otimes\mathbf{\widetilde{3}}=\mathbf{\overline{3}}\oplus\mathbf{6}$}

\begin{equation}
\nonumber\mathbf{\overline{3}}\sim
\begin{pmatrix}
 \alpha_1\beta_1+\alpha_2\beta_3+\alpha_3\beta_2 \\
 \alpha_1\beta_3+\alpha_2\beta_2+\alpha_3\beta_1 \\
 \alpha_1\beta_2+\alpha_2\beta_1+\alpha_3\beta_3
\end{pmatrix},\quad \mathbf{6}\sim
\begin{pmatrix}
2\alpha_1\beta_3-\alpha_2\beta _2-\alpha_3\beta_1 \\
2\alpha_2\beta_3-\alpha_1\beta _1-\alpha_3\beta_2 \\
2\alpha_3\beta_3-\alpha_1\beta_2-\alpha_2\beta_1 \\
\sqrt{3}\left(\alpha_2\beta_2-\alpha_3\beta_1\right) \\
\sqrt{3}\left(\alpha_3\beta_2-\alpha_1\beta_1\right) \\
\sqrt{3}\left(\alpha_1\beta_2-\alpha_2\beta_1\right)
\end{pmatrix}\,.
\end{equation}


\item{$\mathbf{3}\otimes\mathbf{6}=\mathbf{3}\oplus\mathbf{3^{\prime}}\oplus\mathbf{\widetilde{3}}\oplus\mathbf{\widetilde{3}^{\prime}}\oplus\mathbf{6}$}

\begin{equation}
\nonumber\mathbf{3}\sim
\begin{pmatrix}
2\alpha_2\beta_5-\alpha_1\beta_6-\alpha_3\beta_4+\sqrt{3}\left(\alpha_3\beta_1-\alpha_1\beta_3\right) \\
2\alpha_1\beta_4-\alpha_2\beta_6-\alpha_3\beta_5+\sqrt{3}\left(\alpha_2\beta_3-\alpha_3\beta_2\right) \\
2\alpha_3\beta_6-\alpha_1\beta_5-\alpha_2\beta_4+\sqrt{3}\left(\alpha_1\beta_2-\alpha_2\beta_1\right)
\end{pmatrix}\,,
\end{equation}

\begin{equation}
\nonumber\mathbf{3^{\prime}}\sim
\begin{pmatrix}
2\alpha_2\beta_2-\alpha_1\beta_3-\alpha_3\beta_1+\sqrt{3}\left(\alpha_1\beta_6-\alpha_3\beta_4\right) \\
2\alpha_1\beta_1-\alpha_2\beta_3-\alpha_3\beta_2+\sqrt{3}\left(\alpha_3\beta_5-\alpha_2\beta_6\right) \\
2\alpha_3\beta_3-\alpha_1\beta_2-\alpha_2\beta_1+\sqrt{3}\left(\alpha_2\beta_4-\alpha_1\beta_5\right)
\end{pmatrix}\,,
\end{equation}

\begin{equation}
\nonumber\mathbf{\widetilde{3}}\sim
\begin{pmatrix}
\alpha_1\beta_6+\alpha_2\beta_5+\alpha_3\beta_4+\sqrt{3}\left(\alpha_1\beta_3+\alpha_2\beta_2+\alpha_3\beta_1\right)\\
\alpha_1\beta_4+\alpha_2\beta_6+\alpha_3\beta_5-\sqrt{3}\left(\alpha_1\beta_1+\alpha_2\beta_3+\alpha_3\beta_2\right)\\
-2\left(\alpha_1\beta_5+\alpha_2\beta_4+\alpha_3\beta_6\right)
\end{pmatrix}\,,
\end{equation}

\begin{equation}
\nonumber\mathbf{\widetilde{3}^{\prime}}\sim
\begin{pmatrix}
\alpha_1\beta_3+\alpha_2\beta_2+\alpha_3\beta_1-\sqrt{3}\left(\alpha_1\beta_6+\alpha_2\beta_5+\alpha_3\beta_4\right)\\
\alpha_1\beta_1+\alpha_2\beta_3+\alpha_3\beta_2+\sqrt{3}\left(\alpha_1\beta_4+\alpha_2\beta_6+\alpha_3\beta_5\right)\\
-2\left(\alpha_1\beta_2+\alpha_2\beta_1+\alpha_3\beta_3\right)
\end{pmatrix}\,,
\end{equation}

\begin{equation}
\nonumber\mathbf{6}\sim
\begin{pmatrix}
2\alpha_2\beta_5-\alpha_1\beta_6-\alpha_3\beta_4+\sqrt{3}\left(\alpha_1\beta_3-\alpha_3\beta_1\right)\\
2\alpha_1\beta_4-\alpha_2\beta_6-\alpha_3\beta_5+\sqrt{3}\left(\alpha_3\beta_2-\alpha_2\beta_3\right)\\
2\alpha_3\beta_6-\alpha_1\beta_5-\alpha_2\beta_4+\sqrt{3}\left(\alpha_2\beta_1-\alpha_1\beta_2\right)\\
-2\alpha_2\beta_2+\alpha_1\beta_3+\alpha_3\beta_1+\sqrt{3}\left(\alpha_1\beta_6-\alpha_3\beta_4\right)\\
-2\alpha_1\beta_1+\alpha_2\beta_3+\alpha_3\beta_2+\sqrt{3}\left(\alpha_3\beta_5-\alpha_2\beta_6\right)\\
-2\alpha_3\beta_3+\alpha_1\beta_2+\alpha_2\beta_1+\sqrt{3}\left(\alpha_2\beta_4-\alpha_1\beta_5\right)
\end{pmatrix}\,.
\end{equation}


\item{$\mathbf{3^{\prime}}\otimes\mathbf{6}=\mathbf{3}\oplus\mathbf{3^{\prime}}\oplus\mathbf{\widetilde{3}}\oplus\mathbf{\widetilde{3}^{\prime}}\oplus\mathbf{6}$}

\begin{equation}
\nonumber\mathbf{3}\sim
\begin{pmatrix}
2\alpha_2\beta_2-\alpha_1\beta_3-\alpha_3\beta_1+\sqrt{3}\left(\alpha_1\beta_6-\alpha_3\beta_4\right)\\
2\alpha_1\beta_1-\alpha_2\beta_3-\alpha_3\beta_2+\sqrt{3}\left(\alpha_3\beta_5-\alpha_2\beta_6\right)\\
2\alpha_3\beta_3-\alpha_1\beta_2-\alpha_2\beta_1+\sqrt{3}\left(\alpha_2\beta_4-\alpha_1\beta_5\right)
\end{pmatrix}\,,
\end{equation}

\begin{equation}
\nonumber\mathbf{3^{\prime}}\sim
\begin{pmatrix}
2\alpha_2\beta_5-\alpha_1\beta_6-\alpha_3\beta_4+\sqrt{3}\left(\alpha_3\beta_1-\alpha_1\beta_3\right)\\
2\alpha_1\beta_4-\alpha_2\beta_6-\alpha_3\beta_5+\sqrt{3}\left(\alpha_2\beta_3-\alpha_3\beta_2\right)\\
2\alpha_3\beta_6-\alpha_1\beta_5-\alpha_2\beta_4+\sqrt{3}\left(\alpha_1\beta_2-\alpha_2\beta_1\right)
\end{pmatrix}\,,
\end{equation}

\begin{equation}
\nonumber\mathbf{\widetilde{3}}\sim
\begin{pmatrix}
\alpha_1\beta_3+\alpha_2\beta_2+\alpha_3\beta_1-\sqrt{3}\left(\alpha_1\beta_6+\alpha_2\beta_5+\alpha_3\beta_4\right)\\
\alpha_1\beta_1+\alpha_2\beta_3+\alpha_3\beta_2+\sqrt{3}\left(\alpha_1\beta_4+\alpha_2\beta_6+\alpha_3\beta_5\right)\\
-2\left(\alpha_1\beta_2+\alpha_2\beta_1+\alpha_3\beta _3\right)
\end{pmatrix}\,,
\end{equation}

\begin{equation}
\nonumber\mathbf{\widetilde{3}^{\prime}}\sim
\begin{pmatrix}
\alpha_1\beta_6+\alpha_2\beta_5+\alpha_3\beta_4+\sqrt{3}\left(\alpha_1\beta_3+\alpha_2\beta_2+\alpha_3\beta_1\right)\\
\alpha_1\beta_4+\alpha_2\beta_6+\alpha_3\beta_5-\sqrt{3}\left(\alpha_1\beta_1+\alpha_2\beta_3+\alpha_3\beta_2\right)\\
-2\left(\alpha_1\beta_5+\alpha_2\beta_4+\alpha_3\beta _6\right)
\end{pmatrix}\,,
\end{equation}

\begin{equation}
\nonumber\mathbf{6}\sim
\begin{pmatrix}
2\alpha_2\beta_2-\alpha_1\beta_3-\alpha_3\beta_1+\sqrt{3}\left(\alpha_3\beta_4-\alpha_1\beta_6\right)\\
2\alpha_1\beta_1-\alpha_2\beta_3-\alpha_3\beta_2+\sqrt{3}\left(\alpha_2\beta_6-\alpha_3\beta_5\right)\\
2\alpha_3\beta_3-\alpha_1\beta_2-\alpha_2\beta_1+\sqrt{3}\left(\alpha_1\beta_5-\alpha_2\beta_4\right)\\
2\alpha_2\beta_5-\alpha_1\beta_6-\alpha_3\beta_4+\sqrt{3}\left(\alpha_1\beta_3-\alpha_3\beta_1\right)\\
2\alpha_1\beta_4-\alpha_2\beta_6-\alpha_3\beta_5+\sqrt{3}\left(\alpha_3\beta_2-\alpha_2\beta_3\right)\\
2\alpha_3\beta_6-\alpha_1\beta_5-\alpha_2\beta_4+\sqrt{3}\left(\alpha_2\beta_1-\alpha_1\beta_2\right)
\end{pmatrix}\,.
\end{equation}


\item{$\mathbf{\overline{3}}\otimes\mathbf{\overline{3}}=\mathbf{\overline{3}^{\prime}}\otimes\mathbf{\overline{3}^{\prime}}=\mathbf{3}\oplus\mathbf{3^{\prime}}\oplus\mathbf{\widetilde{3}^{\prime}}$}

\begin{equation}
\nonumber\mathbf{3}\sim
\begin{pmatrix}
 \alpha_2\beta_3-\alpha_3\beta_2 \\
 \alpha_3\beta_1-\alpha_1\beta_3 \\
 \alpha_1\beta_2-\alpha_2\beta_1
\end{pmatrix},~~\mathbf{3^{\prime}}\sim
\begin{pmatrix}
 2\alpha_1\beta_1-\alpha_2\beta_3-\alpha_3\beta_2 \\
 2\alpha_2\beta_2-\alpha_1\beta_3-\alpha_3\beta_1 \\
 2\alpha_3\beta_3-\alpha_1\beta_2-\alpha_2\beta_1
\end{pmatrix},~~\mathbf{\widetilde{3}^{\prime}}\sim
\begin{pmatrix}
 \alpha_1\beta_1+\alpha_2\beta_3+\alpha_3\beta_2 \\
 \alpha_1\beta_3+\alpha_2\beta_2+\alpha_3\beta_1 \\
 \alpha_1\beta_2+\alpha_2\beta_1+\alpha_3\beta_3
\end{pmatrix}\,.
\end{equation}


\item{$\mathbf{\overline{3}}\otimes\mathbf{\overline{3}^{\prime}}=\mathbf{3}\oplus\mathbf{3^{\prime}}\oplus\mathbf{\widetilde{3}}$}

\begin{equation}
\nonumber\mathbf{3}\sim
\begin{pmatrix}
 2\alpha_1\beta_1-\alpha_2\beta_3-\alpha_3\beta_2 \\
 2\alpha_2\beta_2-\alpha_1\beta_3-\alpha_3\beta_1 \\
 2\alpha_3\beta_3-\alpha_1\beta_2-\alpha_2\beta_1
\end{pmatrix},~~\mathbf{3^{\prime}}\sim
\begin{pmatrix}
 \alpha_2\beta_3-\alpha_3\beta_2 \\
 \alpha_3\beta_1-\alpha_1\beta_3 \\
 \alpha_1\beta_2-\alpha_2\beta_1
\end{pmatrix},~~\mathbf{\widetilde{3}}\sim
\begin{pmatrix}
 \alpha_1\beta_1+\alpha_2\beta_3+\alpha_3\beta_2 \\
 \alpha_1\beta_3+\alpha_2\beta_2+\alpha_3\beta_1 \\
 \alpha_1\beta_2+\alpha_2\beta_1+\alpha_3\beta_3
\end{pmatrix}\,.
\end{equation}


\item{$\mathbf{\overline{3}}\otimes\mathbf{\widetilde{3}}=\mathbf{\overline{3}^{\prime}}\otimes\mathbf{\widetilde{3}^{\prime}}=\mathbf{3^{\prime}}\oplus\mathbf{6}$}

\begin{equation}
\nonumber\mathbf{3^{\prime}}\sim
\begin{pmatrix}
 \alpha_1\beta_2+\alpha_2\beta_3+\alpha_3\beta_1 \\
 \alpha_1\beta_3+\alpha_2\beta_1+\alpha_3\beta_2 \\
 \alpha_1\beta_1+\alpha_2\beta_2+\alpha_3\beta_3
\end{pmatrix},\quad \mathbf{6}\sim
\begin{pmatrix}
\sqrt{3}\left(\alpha_1\beta_2-\alpha_3\beta_1\right) \\
\sqrt{3}\left(\alpha_3\beta_2-\alpha_2\beta_1\right) \\
\sqrt{3}\left(\alpha_2\beta_2-\alpha_1\beta_1\right) \\
2\alpha_2\beta_3-\alpha_1\beta_2-\alpha_3\beta_1 \\
2\alpha_1\beta_3-\alpha_2\beta_1-\alpha_3\beta_2 \\
2\alpha_3\beta_3-\alpha_1\beta_1-\alpha_2\beta_2
\end{pmatrix}\,.
\end{equation}


\item{$\mathbf{\overline{3}}\otimes\mathbf{\widetilde{3}^{\prime}}=\mathbf{\overline{3}^{\prime}}\otimes\mathbf{\widetilde{3}}=\mathbf{3}\oplus\mathbf{6}$}

\begin{equation}
\nonumber\mathbf{3}\sim
\begin{pmatrix}
 \alpha_1\beta_2+\alpha_2\beta_3+\alpha_3\beta_1 \\
 \alpha_1\beta_3+\alpha_2\beta_1+\alpha_3\beta_2 \\
 \alpha_1\beta_1+\alpha_2\beta_2+\alpha_3\beta_3
\end{pmatrix},\quad \mathbf{6}\sim
\begin{pmatrix}
2\alpha_2\beta_3-\alpha_1\beta_2-\alpha _3\beta_1 \\
2\alpha_1\beta_3-\alpha_2\beta_1-\alpha_3\beta_2 \\
2\alpha_3\beta_3-\alpha_1\beta_1-\alpha_2\beta_2 \\
\sqrt{3}\left(\alpha_3\beta_1-\alpha_1\beta_2\right) \\
\sqrt{3}\left(\alpha_2\beta_1-\alpha_3\beta_2\right) \\
\sqrt{3}\left(\alpha_1\beta_1-\alpha_2\beta_2\right)
\end{pmatrix}\,.
\end{equation}


\item{$\mathbf{\overline{3}}\otimes\mathbf{6}=\mathbf{\overline{3}}\oplus\mathbf{\overline{3}^{\prime}}\oplus\mathbf{\widetilde{3}}\oplus\mathbf{\widetilde{3}^{\prime}}\oplus\mathbf{6}$}

\begin{equation}
\nonumber\mathbf{\overline{3}}\sim
\begin{pmatrix}
2\alpha_2\beta_4-\alpha_1\beta_6-\alpha_3\beta_5+\sqrt{3}\left(\alpha_3\beta_2-\alpha_1\beta_3\right)\\
2\alpha_1\beta_5-\alpha_2\beta_6-\alpha_3\beta_4+\sqrt{3}\left(\alpha_2\beta_3-\alpha_3\beta_1\right)\\
2\alpha_3\beta_6-\alpha_1\beta_4-\alpha_2\beta_5+\sqrt{3}\left(\alpha_1\beta_1-\alpha_2\beta_2\right)
\end{pmatrix}\,,
\end{equation}

\begin{equation}
\nonumber\mathbf{\overline{3}^{\prime}}\sim
\begin{pmatrix}
2\alpha_2\beta_1-\alpha_1\beta_3-\alpha_3\beta_2+\sqrt{3}\left(\alpha_1\beta_6-\alpha_3\beta_5\right)\\
2\alpha_1\beta_2-\alpha_2\beta_3-\alpha_3\beta_1+\sqrt{3}\left(\alpha_3\beta_4-\alpha_2\beta_6\right)\\
2\alpha_3\beta_3-\alpha_1\beta_1-\alpha_2\beta_2+\sqrt{3}\left(\alpha_2\beta_5-\alpha_1\beta_4\right)
\end{pmatrix}\,,
\end{equation}

\begin{equation}
\nonumber\mathbf{\widetilde{3}}\sim
\begin{pmatrix}
\alpha_1\beta_5+\alpha_2\beta_6+\alpha_3\beta_4-\sqrt{3}\left(\alpha_1\beta_2+\alpha_2\beta_3+\alpha_3\beta_1\right)\\
\alpha_1\beta_6+\alpha_2\beta_4+\alpha_3\beta_5+\sqrt{3}\left(\alpha_1\beta_3+\alpha_2\beta_1+\alpha_3\beta_2\right)\\
-2\left(\alpha_1\beta_4+\alpha_2\beta_5+\alpha_3\beta_6\right)
\end{pmatrix}\,,
\end{equation}

\begin{equation}
\nonumber\mathbf{\widetilde{3}^{\prime}}\sim
\begin{pmatrix}
\alpha_1\beta_2+\alpha_2\beta_3+\alpha_3\beta_1+\sqrt{3}\left(\alpha_1\beta_5+\alpha_2\beta_6+\alpha_3\beta_4\right)\\
\alpha_1\beta_3+\alpha_2\beta_1+\alpha_3\beta_2-\sqrt{3}\left(\alpha_1\beta_6+\alpha_2\beta_4+\alpha_3\beta_5\right)\\
-2\left(\alpha_1\beta_1+\alpha_2\beta_2+\alpha_3\beta_3\right)
\end{pmatrix}\,,
\end{equation}

\begin{equation}
\nonumber\mathbf{6}\sim
\begin{pmatrix}
2\alpha_1\beta_5-\alpha_2\beta_6-\alpha_3\beta_4+\sqrt{3}\left(\alpha_3\beta_1-\alpha_2\beta_3\right)\\
2\alpha_2\beta_4-\alpha_1\beta_6-\alpha_3\beta_5+\sqrt{3}\left(\alpha_1\beta_3-\alpha_3\beta_2\right)\\
2\alpha_3\beta_6-\alpha_1\beta_4-\alpha_2\beta_5+\sqrt{3}\left(\alpha_2\beta_2-\alpha_1\beta_1\right)\\
-2\alpha_1\beta_2+\alpha_2\beta_3+\alpha_3\beta_1+\sqrt{3}\left(\alpha_3\beta_4-\alpha_2\beta_6\right)\\
-2\alpha_2\beta_1+\alpha_1\beta_3+\alpha_3\beta_2+\sqrt{3}\left(\alpha_1\beta_6-\alpha_3\beta_5\right)\\
-2\alpha_3\beta_3+\alpha_1\beta_1+\alpha_2\beta_2+\sqrt{3}\left(\alpha_2\beta_5-\alpha_1\beta_4\right)
\end{pmatrix}\,.
\end{equation}


\item{$\mathbf{\overline{3}^{\prime}}\otimes\mathbf{6}=\mathbf{\overline{3}}\oplus\mathbf{\overline{3}^{\prime}}\oplus\mathbf{\widetilde{3}}\oplus\mathbf{\widetilde{3}^{\prime}}\oplus\mathbf{6}$}

\begin{equation}
\nonumber\mathbf{\overline{3}}\sim
\begin{pmatrix}
2\alpha_2\beta_1-\alpha_1\beta_3-\alpha_3\beta_2+\sqrt{3}\left(\alpha_1\beta_6-\alpha_3\beta_5\right)\\
2\alpha_1\beta_2-\alpha_2\beta_3-\alpha_3\beta_1+\sqrt{3}\left(\alpha_3\beta_4-\alpha_2\beta_6\right)\\
2\alpha_3\beta_3-\alpha_1\beta_1-\alpha_2\beta_2+\sqrt{3}\left(\alpha_2\beta_5-\alpha_1\beta_4\right)
\end{pmatrix}\,,
\end{equation}

\begin{equation}
\nonumber\mathbf{\overline{3}^{\prime}}\sim
\begin{pmatrix}
2\alpha_2\beta_4-\alpha_1\beta_6-\alpha_3\beta_5+\sqrt{3}\left(\alpha_3\beta_2-\alpha_1\beta_3\right)\\
2\alpha_1\beta_5-\alpha_2\beta_6-\alpha_3\beta_4+\sqrt{3}\left(\alpha_2\beta_3-\alpha_3\beta_1\right)\\
2\alpha_3\beta_6-\alpha_1\beta_4-\alpha_2\beta_5+\sqrt{3}\left(\alpha_1\beta_1-\alpha_2\beta_2\right)
\end{pmatrix}\,,
\end{equation}

\begin{equation}
\nonumber\mathbf{\widetilde{3}}\sim
\begin{pmatrix}
\alpha_1\beta_2+\alpha_2\beta_3+\alpha_3\beta_1+\sqrt{3}\left(\alpha_1\beta_5+\alpha_2\beta_6+\alpha_3\beta_4\right)\\
\alpha_1\beta_3+\alpha_2\beta_1+\alpha_3\beta_2-\sqrt{3}\left(\alpha_1\beta_6+\alpha_2\beta_4+\alpha_3\beta_5\right)\\
-2\left(\alpha_1\beta_1+\alpha_2\beta_2+\alpha_3\beta _3\right)
\end{pmatrix}\,,
\end{equation}

\begin{equation}
\nonumber\mathbf{\widetilde{3}^{\prime}}\sim
\begin{pmatrix}
\alpha_1\beta_5+\alpha_2\beta_6+\alpha_3\beta_4-\sqrt{3}\left(\alpha_1\beta_2+\alpha_2\beta_3+\alpha_3\beta_1\right)\\
\alpha_1\beta_6+\alpha_2\beta_4+\alpha_3\beta_5+\sqrt{3}\left(\alpha_1\beta_3+\alpha_2\beta_1+\alpha_3\beta_2\right)\\
-2\left(\alpha_1\beta_4+\alpha_2\beta_5+\alpha_3\beta_6\right)
\end{pmatrix}\,,
\end{equation}

\begin{equation}
\nonumber\mathbf{6}\sim
\begin{pmatrix}
2\alpha_1\beta_2-\alpha_2\beta_3-\alpha_3\beta_1+\sqrt{3}\left(\alpha_2\beta_6-\alpha_3\beta_4\right)\\
2\alpha_2\beta_1-\alpha_1\beta_3-\alpha_3\beta_2+\sqrt{3}\left(\alpha_3\beta_5-\alpha_1\beta_6\right)\\
2\alpha_3\beta_3-\alpha_1\beta_1-\alpha_2\beta_2+\sqrt{3}\left(\alpha_1\beta_4-\alpha_2\beta_5\right)\\
2\alpha_1\beta_5-\alpha_2\beta_6-\alpha_3\beta_4+\sqrt{3}\left(\alpha_3\beta_1-\alpha_2\beta_3\right)\\
2\alpha_2\beta_4-\alpha_1\beta_6-\alpha_3\beta_5+\sqrt{3}\left(\alpha_1\beta_3-\alpha_3\beta_2\right)\\
2\alpha_3\beta_6-\alpha_1\beta_4-\alpha_2\beta_5+\sqrt{3}\left(\alpha_2\beta_2-\alpha_1\beta_1\right)
\end{pmatrix}\,.
\end{equation}


\item{$\mathbf{\widetilde{3}}\otimes\mathbf{\widetilde{3}}=\mathbf{\widetilde{3}^{\prime}}\otimes\mathbf{\widetilde{3}^{\prime}}=\mathbf{1}\oplus\mathbf{2}\oplus\mathbf{\widetilde{3}}\oplus\mathbf{\widetilde{3}^{\prime}}$}

\begin{eqnarray}
\nonumber&&\mathbf{1}\sim\alpha_1\beta_2+\alpha_2\beta_1+\alpha_3\beta_3,\qquad \mathbf{2}\sim
\begin{pmatrix}
 \alpha_1\beta_1+\alpha_2\beta_3+\alpha_3\beta_2 \\
 \alpha_1\beta_3+\alpha_2\beta_2+\alpha_3\beta_1
\end{pmatrix},  \\
\nonumber&&\mathbf{\widetilde{3}}\sim
\begin{pmatrix}
 \alpha_1\beta_3-\alpha_3\beta_1 \\
 \alpha_3\beta_2-\alpha_2\beta_3 \\
 \alpha_2\beta_1-\alpha_1\beta_2
\end{pmatrix},\qquad  \mathbf{\widetilde{3}^{\prime}}\sim
\begin{pmatrix}
2\alpha_2\beta_2-\alpha_1\beta_3-\alpha_3\beta_1\\
2\alpha_1\beta_1-\alpha_2\beta_3-\alpha_3\beta_2 \\
2\alpha_3\beta_3-\alpha_1\beta_2-\alpha_2\beta_1
\end{pmatrix}\,.
\end{eqnarray}


\item{$\mathbf{\widetilde{3}}\otimes\mathbf{\widetilde{3}^{\prime}}=\mathbf{1^{\prime}}\oplus\mathbf{2}\oplus\mathbf{\widetilde{3}}\oplus\mathbf{\widetilde{3}^{\prime}}$}

\begin{eqnarray}
\nonumber&&\mathbf{1^{\prime}}\sim\alpha_1\beta_2+\alpha_2\beta_1+\alpha_3\beta_3,\qquad \mathbf{2}\sim
\begin{pmatrix}
 \alpha_1\beta_1+\alpha_2\beta_3+\alpha_3\beta_2 \\
 -\alpha_1\beta_3-\alpha_2\beta_2-\alpha_3\beta_1
\end{pmatrix},\\
\nonumber&& \mathbf{\widetilde{3}}\sim
\begin{pmatrix}
2\alpha_2\beta_2-\alpha_1\beta_3-\alpha_3\beta_1 \\
2\alpha_1\beta_1-\alpha_2\beta_3-\alpha_3\beta_2 \\
2\alpha_3\beta_3-\alpha_1\beta_2-\alpha_2\beta_1
\end{pmatrix},\qquad \mathbf{\widetilde{3}^{\prime}}\sim
\begin{pmatrix}
 \alpha_1\beta_3-\alpha_3\beta_1 \\
 \alpha_3\beta_2-\alpha_2\beta_3 \\
 \alpha_2\beta_1-\alpha_1\beta_2
\end{pmatrix}\,.
\end{eqnarray}


\item{$\mathbf{\widetilde{3}}\otimes\mathbf{6}=\mathbf{3}\otimes\mathbf{3^{\prime}}\oplus\mathbf{\overline{3}}\oplus\mathbf{\overline{3}^{\prime}}\oplus\mathbf{6}$}

\begin{equation}
\nonumber\mathbf{3}\sim
\begin{pmatrix}
2\alpha_3\beta_4-\alpha_1\beta_6-\alpha_2\beta_5+\sqrt{3}\left(\alpha_2\beta_2-\alpha_1\beta_3\right)\\
2\alpha_3\beta_5-\alpha_1\beta_4-\alpha_2\beta_6+\sqrt{3}\left(\alpha_2\beta_3-\alpha_1\beta_1\right)\\
2\alpha_3\beta_6-\alpha_1\beta_5-\alpha_2\beta_4+\sqrt{3}\left(\alpha_2\beta_1-\alpha_1\beta_2\right)
\end{pmatrix}\,,
\end{equation}

\begin{equation}
\nonumber\mathbf{3^{\prime}}\sim
\begin{pmatrix}
2\alpha_3\beta_1-\alpha_1\beta_3-\alpha_2\beta_2+\sqrt{3}\left(\alpha_1\beta_6-\alpha_2\beta_5\right)\\
2\alpha_3\beta_2-\alpha_1\beta_1-\alpha_2\beta_3+\sqrt{3}\left(\alpha_1\beta_4-\alpha_2\beta_6\right)\\
2\alpha_3\beta_3-\alpha_1\beta_2-\alpha_2\beta_1+\sqrt{3}\left(\alpha_1\beta_5-\alpha_2\beta_4\right)
\end{pmatrix}\,,
\end{equation}

\begin{equation}
\nonumber\mathbf{\overline{3}}\sim
\begin{pmatrix}
2\alpha_3\beta_5-\alpha_1\beta_4-\alpha_2\beta_6+\sqrt{3}\left(\alpha_1\beta_1-\alpha_2\beta_3\right)\\
2\alpha_3\beta_4-\alpha_1\beta_6-\alpha_2\beta_5+\sqrt{3}\left(\alpha_1\beta_3-\alpha_2\beta_2\right)\\
2\alpha_3\beta_6-\alpha_1\beta_5-\alpha_2\beta_4+\sqrt{3}\left(\alpha_1\beta_2-\alpha_2\beta_1\right)
\end{pmatrix}\,,
\end{equation}

\begin{equation}
\nonumber\mathbf{\overline{3}^{\prime}}\sim
\begin{pmatrix}
2\alpha_3\beta_2-\alpha_1\beta_1-\alpha_2\beta_3+\sqrt{3}\left(\alpha_2\beta_6-\alpha_1\beta_4\right)\\
2\alpha_3\beta_1-\alpha_1\beta_3-\alpha_2\beta_2+\sqrt{3}\left(\alpha_2\beta_5-\alpha_1\beta_6\right)\\
2\alpha_3\beta_3-\alpha_1\beta_2-\alpha_2\beta_1+\sqrt{3}\left(\alpha_2\beta_4-\alpha_1\beta_5\right)
\end{pmatrix}\,,
\end{equation}

\begin{equation}
\nonumber\mathbf{6}\sim
\begin{pmatrix}
\alpha_1\beta _6+\alpha_2\beta_5+\alpha_3\beta_4 \\
\alpha_1\beta_4+\alpha_2\beta_6+\alpha_3\beta_5 \\
\alpha_1\beta_5+\alpha_2\beta_4+\alpha_3\beta_6 \\
\alpha_1\beta_3+\alpha_2\beta_2+\alpha_3\beta_1 \\
\alpha_1\beta_1+\alpha_2\beta_3+\alpha_3\beta_2 \\
\alpha_1\beta_2+\alpha_2\beta_1+\alpha_3\beta_3
\end{pmatrix}\,.
\end{equation}


\item{$\mathbf{\widetilde{3}^{\prime}}\otimes\mathbf{6}=\mathbf{3}\oplus\mathbf{3^{\prime}}\oplus\mathbf{\overline{3}}\oplus\mathbf{\overline{3}^{\prime}}\oplus\mathbf{6}$}

\begin{equation}
\nonumber\mathbf{3}\sim
\begin{pmatrix}
2\alpha_3\beta_1-\alpha_1\beta_3-\alpha_2\beta_2+\sqrt{3}\left(\alpha_1\beta_6-\alpha_2\beta_5\right)\\
2\alpha_3\beta_2-\alpha_1\beta_1-\alpha_2\beta_3+\sqrt{3}\left(\alpha_1\beta_4-\alpha_2\beta_6\right)\\
2\alpha_3\beta_3-\alpha_1\beta_2-\alpha_2\beta_1+\sqrt{3}\left(\alpha_1\beta_5-\alpha_2\beta_4\right)
\end{pmatrix}\,,
\end{equation}

\begin{equation}
\nonumber\mathbf{3^{\prime}}\sim
\begin{pmatrix}
2\alpha_3\beta_4-\alpha_1\beta_6-\alpha_2\beta_5+\sqrt{3}\left(\alpha_2\beta_2-\alpha_1\beta_3\right)\\
2\alpha_3\beta_5-\alpha_1\beta_4-\alpha_2\beta_6+\sqrt{3}\left(\alpha_2\beta_3-\alpha_1\beta_1\right)\\
2\alpha_3\beta_6-\alpha_1\beta_5-\alpha_2\beta_4+\sqrt{3}\left(\alpha_2\beta_1-\alpha_1\beta_2\right)
\end{pmatrix}\,,
\end{equation}

\begin{equation}
\nonumber\mathbf{\overline{3}}\sim
\begin{pmatrix}
2\alpha_3\beta_2-\alpha_1\beta_1-\alpha_2\beta_3+\sqrt{3}\left(\alpha_2\beta_6-\alpha_1\beta_4\right)\\
2\alpha_3\beta_1-\alpha_1\beta_3-\alpha_2\beta_2+\sqrt{3}\left(\alpha_2\beta_5-\alpha_1\beta_6\right)\\
2\alpha_3\beta_3-\alpha_1\beta_2-\alpha_2\beta_1+\sqrt{3}\left(\alpha_2\beta_4-\alpha_1\beta_5\right)
\end{pmatrix}\,,
\end{equation}

\begin{equation}
\nonumber\mathbf{\overline{3}^{\prime}}\sim
\begin{pmatrix}
2\alpha_3\beta_5-\alpha_1\beta_4-\alpha_2\beta_6+\sqrt{3}\left(\alpha_1\beta_1-\alpha_2\beta_3\right)\\
2\alpha_3\beta_4-\alpha_1\beta_6-\alpha_2\beta_5+\sqrt{3}\left(\alpha_1\beta_3-\alpha_2\beta_2\right)\\
2\alpha_3\beta_6-\alpha_1\beta_5-\alpha_2\beta_4+\sqrt{3}\left(\alpha_1\beta_2-\alpha_2\beta_1\right)
\end{pmatrix}\,,
\end{equation}

\begin{equation}
\nonumber\mathbf{6}\sim
\begin{pmatrix}
\alpha_1\beta_3+\alpha_2\beta_2+\alpha_3\beta_1 \\
\alpha_1\beta_1+\alpha_2\beta_3 +\alpha_3\beta_2\\
\alpha_1\beta_2+\alpha_2\beta_1+\alpha_3\beta_3 \\
-\alpha_1\beta_6-\alpha_2\beta_5-\alpha_3\beta_4 \\
-\alpha_1\beta_4-\alpha_2\beta_6-\alpha_3\beta_5 \\
-\alpha_1\beta_5-\alpha_2\beta_4-\alpha_3\beta_6
\end{pmatrix}\,.
\end{equation}


\item{$\mathbf{6}\otimes\mathbf{6}=\mathbf{1}\oplus\mathbf{1^{\prime}}\oplus\mathbf{2}_S\oplus\mathbf{2}_A\oplus\mathbf{3}\oplus\mathbf{3^{\prime}}\oplus\mathbf{\overline{3}}\oplus\mathbf{\overline{3}^{\prime}}\oplus\mathbf{\widetilde{3}}\oplus\mathbf{\widetilde{3}^{\prime}}\oplus\mathbf{6}_S\oplus\mathbf{6}_A$}

\begin{equation}
\nonumber\mathbf{1}\sim\alpha_1\beta_2+\alpha_2\beta_1+\alpha_3\beta_3+\alpha_4\beta_5+\alpha_5\beta_4+\alpha_6\beta_6\,,
\end{equation}

\begin{equation}
\nonumber\mathbf{1^{\prime}}\sim\alpha_1\beta_5+\alpha_2\beta_4+\alpha_3\beta_6-\alpha_4\beta_2-\alpha_5\beta_1-\alpha_6\beta_3\,,
\end{equation}

\begin{equation}
\nonumber\mathbf{2}_S\sim
\begin{pmatrix}
\alpha_1\beta_1+\alpha_2\beta_3+\alpha_3\beta_2+\alpha_4\beta_4+\alpha_5\beta_6+\alpha_6\beta_5\\
\alpha_1\beta_3+\alpha_2\beta_2+\alpha_3\beta_1+\alpha_4\beta_6+\alpha_5\beta_5+\alpha_6\beta_4
\end{pmatrix}\,,
\end{equation}

\begin{equation}
\nonumber\mathbf{2}_A\sim
\begin{pmatrix}
\alpha_1\beta_4+\alpha_2\beta_6+\alpha_3\beta_5-\alpha_4\beta_1-\alpha_5\beta_3-\alpha_6\beta_2\\
-\alpha_1\beta_6-\alpha_2\beta_5-\alpha_3\beta_4+\alpha_4\beta_3+\alpha_5\beta_2+\alpha_6\beta_1
\end{pmatrix}\,,
\end{equation}

\begin{equation}
\nonumber\mathbf{3}\sim
\begin{pmatrix}
2\alpha_2\beta_5-\alpha_1\beta_6-\alpha_3\beta_4-2\alpha_5\beta_2+\alpha_4\beta_3+\alpha_6\beta_1+\sqrt{3}\left(\alpha_1\beta_3-\alpha_3\beta_1+\alpha_4\beta_6-\alpha_6\beta_4\right)\\
2\alpha_1\beta_4-\alpha_2\beta_6-\alpha_3\beta_5-2\alpha_4\beta_1+\alpha_5\beta_3+\alpha_6\beta_2+\sqrt{3}\left(\alpha_3\beta_2-\alpha_2\beta_3+\alpha_6\beta_5-\alpha_5\beta_6\right)\\
2\alpha_3\beta_6-\alpha_1\beta_5-\alpha_2\beta_4-2\alpha_6\beta_3+\alpha_4\beta_2+\alpha_5\beta_1+\sqrt{3}\left(\alpha_2\beta_1-\alpha_1\beta_2+\alpha_5\beta_4-\alpha_4\beta_5\right)
\end{pmatrix}\,,
\end{equation}

\begin{equation}
\nonumber\mathbf{3^{\prime}}\sim
\begin{pmatrix}
2\alpha_2\beta_2-\alpha_1\beta_3-\alpha_3\beta_1+2\alpha_5\beta_5-\alpha_4\beta_6-\alpha_6\beta_4+\sqrt{3}\left(\alpha_3\beta_4-\alpha_1\beta_6+\alpha_4\beta_3-\alpha_6\beta_1\right)\\
2\alpha_1\beta_1-\alpha_2\beta_3-\alpha_3\beta_2+2\alpha_4\beta_4-\alpha_5\beta_6-\alpha_6\beta_5+\sqrt{3}\left(\alpha_2\beta_6-\alpha_3\beta_5+\alpha_6\beta_2-\alpha_5\beta_3\right)\\
2\alpha_3\beta_3-\alpha_1\beta_2-\alpha_2\beta_1+2\alpha_6\beta_6-\alpha_4\beta_5-\alpha_5\beta_4+\sqrt{3}\left(\alpha_1\beta_5-\alpha_2\beta_4+\alpha_5\beta_1-\alpha_4\beta_2\right)
\end{pmatrix}\,,
\end{equation}

\begin{equation}
\nonumber\mathbf{\overline{3}}\sim
\begin{pmatrix}
2\alpha_1\beta_4-\alpha_2\beta_6-\alpha_3\beta_5-2\alpha_4\beta_1+\alpha_5\beta_3+\alpha_6\beta_2+\sqrt{3}\left(\alpha_2\beta_3-\alpha_3\beta_2+\alpha_5\beta_6-\alpha_6\beta_5\right)\\
2\alpha_2\beta_5-\alpha_1\beta_6-\alpha_3\beta_4-2\alpha_5\beta_2+\alpha_4\beta_3+\alpha_6\beta_1+\sqrt{3}\left(\alpha_3\beta_1-\alpha_1\beta_3+\alpha_6\beta_4-\alpha_4\beta_6\right)\\
2\alpha_3\beta_6-\alpha_1\beta_5-\alpha_2\beta_4-2\alpha_6\beta_3+\alpha_4\beta_2+\alpha_5\beta_1+\sqrt{3}\left(\alpha_1\beta_2-\alpha_2\beta_1+\alpha_4\beta_5-\alpha_5\beta_4\right)
\end{pmatrix}\,,
\end{equation}

\begin{equation}
\nonumber\mathbf{\overline{3}^{\prime}}\sim
\begin{pmatrix}
2\alpha_1\beta_1-\alpha_2\beta_3-\alpha_3\beta_2+2\alpha_4\beta_4-\alpha_5\beta_6-\alpha_6\beta_5+\sqrt{3}\left(\alpha_3\beta_5-\alpha_2\beta_6+\alpha_5\beta_3-\alpha_6\beta_2\right)\\
2\alpha_2\beta_2-\alpha_1\beta_3-\alpha_3\beta_1+2\alpha_5\beta_5-\alpha_4\beta_6-\alpha_6\beta_4+\sqrt{3}\left(\alpha_1\beta_6-\alpha_3\beta_4+\alpha_6\beta_1-\alpha_4\beta_3\right)\\
2\alpha_3\beta_3-\alpha_1\beta_2-\alpha_2\beta_1+2\alpha_6\beta_6-\alpha_4\beta_5-\alpha_5\beta_4+\sqrt{3}\left(\alpha_2\beta_4-\alpha_1\beta_5+\alpha_4\beta_2-\alpha_5\beta_1\right)
\end{pmatrix}\,,
\end{equation}

\begin{equation}
\nonumber\mathbf{\widetilde{3}}\sim
\begin{pmatrix}
\alpha_1\beta_6+\alpha_2\beta_5+\alpha_3\beta_4+\alpha_4\beta_3+\alpha_5\beta_2+\alpha_6\beta_1\\
\alpha_1\beta_4+\alpha_2\beta_6+\alpha_3\beta_5+\alpha_4\beta_1+\alpha_5\beta_3+\alpha_6\beta_2\\
\alpha_1\beta_5+\alpha_2\beta_4+\alpha_3\beta_6+\alpha_4\beta_2+\alpha_5\beta_1+\alpha_6\beta_3
\end{pmatrix}\,,
\end{equation}

\begin{equation}
\nonumber\mathbf{\widetilde{3}^{\prime}}\sim
\begin{pmatrix}
\alpha_1\beta_3+\alpha_2\beta_2+\alpha_3\beta_1-\alpha_4\beta_6-\alpha_5\beta_5-\alpha_6\beta_4\\
\alpha_1\beta_1+\alpha_2\beta_3+\alpha_3\beta_2-\alpha_4\beta_4-\alpha_5\beta_6-\alpha_6\beta_5\\
\alpha_1\beta_2+\alpha_2\beta_1+\alpha_3\beta_3-\alpha_4\beta_5-\alpha_5\beta_4-\alpha_6\beta_6
\end{pmatrix}\,,
\end{equation}

\begin{equation}
\nonumber\mathbf{6}_S\sim
\begin{pmatrix}
2\alpha_2\beta_5-\alpha_1\beta_6-\alpha_3\beta_4+2\alpha_5\beta_2-\alpha_4\beta_3-\alpha_6\beta_1\\
2\alpha_1\beta_4-\alpha_2\beta_6-\alpha_3\beta_5+2\alpha_4\beta_1-\alpha_5\beta_3-\alpha_6\beta_2\\
2\alpha_3\beta_6-\alpha_1\beta_5-\alpha_2\beta_4+2\alpha_6\beta_3-\alpha_4\beta_2-\alpha_5\beta_1\\
2\alpha_2\beta_2-\alpha_1\beta_3-\alpha_3\beta_1-2\alpha_5\beta_5+\alpha_4\beta_6+\alpha_6\beta_4\\
2\alpha_1\beta_1-\alpha_2\beta_3-\alpha_3\beta_2-2\alpha_4\beta_4+\alpha_5\beta_6+\alpha_6\beta_5\\
2\alpha_3\beta_3-\alpha_1\beta_2-\alpha_2\beta_1-2\alpha_6\beta_6+\alpha_4\beta_5+\alpha_5\beta_4
\end{pmatrix}\,,
\end{equation}

\begin{equation}
\nonumber\mathbf{6}_A\sim
\begin{pmatrix}
\alpha_1\beta_3-\alpha_3\beta_1+\alpha_6\beta_4-\alpha_4\beta_6 \\
\alpha_3\beta_2-\alpha_2\beta_3+\alpha_5\beta_6-\alpha_6\beta_5 \\
\alpha_2\beta_1-\alpha_1\beta_2+\alpha_4\beta_5-\alpha_5\beta_4 \\
\alpha_3\beta_4-\alpha_4\beta_3+\alpha_6\beta_1-\alpha_1\beta _6 \\
\alpha_2\beta_6-\alpha_6\beta_2+\alpha_5\beta_3-\alpha_3\beta_5 \\
\alpha_1\beta_5-\alpha_2\beta_4+\alpha_4\beta_2-\alpha_5\beta_1
\end{pmatrix}\,.
\end{equation}

\end{itemize}


\begin{thebibliography}{}


\bibitem{An:2012eh}
  F.~P.~An {\it et al.}  [DAYA-BAY Collaboration],
   Phys.\ Rev.\ Lett.\  {\bf 108}, 171803 (2012)  [arXiv:1203.1669
   [hep-ex]];  
  Chin.\  Phys.\ C {\bf 37}, 011001 (2013)  [arXiv:1210.6327 [hep-ex]].


\bibitem{Ahn:2012nd}
  J.~K.~Ahn {\it et al.}  [RENO Collaboration],
  Phys.\ Rev.\ Lett.\  {\bf 108}, 191802 (2012)  [arXiv:1204.0626 [hep-ex]].


\bibitem{Abe:2011fz}
  Y.~Abe {\it et al.}  [DOUBLE-CHOOZ Collaboration],
  Phys.\ Rev.\ Lett.\  {\bf 108}, 131801 (2012)  [arXiv:1112.6353 [hep-ex]];
  Phys.\ Rev.\ D {\bf 86}, 052008 (2012)  [arXiv:1207.6632 [hep-ex]].


\bibitem{Tortola:2012te}
 D.~V.~Forero, M.~Tortola and J.~W.~F.~Valle,
  Phys.\ Rev.\ D {\bf 86}, 073012 (2012)
  [arXiv:1205.4018 [hep-ph]].


\bibitem{GonzalezGarcia:2012sz}
  M.~C.~Gonzalez-Garcia, M.~Maltoni, J.~Salvado and T.~Schwetz,
  JHEP {\bf 1212}, 123 (2012)  [arXiv:1209.3023 [hep-ph]].


\bibitem{Capozzi:2013csa}
  F.~Capozzi, G.~L.~Fogli, E.~Lisi, A.~Marrone, D.~Montanino and A.~Palazzo,
  arXiv:1312.2878 [hep-ph].


\bibitem{Altarelli:2010gt}
  G.~Altarelli and F.~Feruglio,
  Rev.\ Mod.\ Phys.\  {\bf 82}, 2701 (2010)
  [arXiv:1002.0211 [hep-ph]];
  S.~F.~King and C.~Luhn,
  Rept.\ Prog.\ Phys.\  {\bf 76} (2013) 056201
  [arXiv:1301.1340 [hep-ph]];
  S.~F.~King, A.~Merle, S.~Morisi, Y.~Shimizu and M.~Tanimoto,
  arXiv:1402.4271 [hep-ph];
  H.~Ishimori, T.~Kobayashi, H.~Ohki, Y.~Shimizu, H.~Okada and M.~Tanimoto,
  Prog.\ Theor.\ Phys.\ Suppl.\  {\bf 183}, 1 (2010)
  [arXiv:1003.3552 [hep-th]];
  W.~Grimus and P.~O.~Ludl,
  J.\ Phys.\ A {\bf 45}, 233001 (2012)
  [arXiv:1110.6376 [hep-ph]].


\bibitem{Ecker:1981wv}
  G.~Ecker, W.~Grimus and W.~Konetschny,
  Nucl.\ Phys.\ B {\bf 191} (1981) 465;
  G.~Ecker, W.~Grimus and H.~Neufeld,
  Nucl.\ Phys.\ B {\bf 247} (1984) 70;
  G.~Ecker, W.~Grimus and H.~Neufeld,
  J.\ Phys.\ A {\bf 20} (1987) L807;
  H.~Neufeld, W.~Grimus and G.~Ecker,
  Int.\ J.\ Mod.\ Phys.\ A {\bf 3}, 603 (1988).

\bibitem{Grimus:1995zi}
  W.~Grimus and M.~N.~Rebelo,
  Phys.\ Rept.\  {\bf 281}, 239 (1997)
  [arXiv:9506272[hep-ph]].


\bibitem{Harrison:2002kp}
  P.~F.~Harrison and W.~G.~Scott,
  Phys.\ Lett.\ B {\bf 535}, 163 (2002)
  [hep-ph/0203209];
  P.~F.~Harrison and W.~G.~Scott,
  Phys.\ Lett.\ B {\bf 547}, 219 (2002)
  [hep-ph/0210197];
  P.~F.~Harrison and W.~G.~Scott,
  Phys.\ Lett.\ B {\bf 594}, 324 (2004)
  [hep-ph/0403278].


\bibitem{Grimus:2003yn}
  W.~Grimus and L.~Lavoura,
  Phys.\ Lett.\ B {\bf 579}, 113 (2004)
  [hep-ph/0305309];
 W.~Grimus and L.~Lavoura,
  arXiv:1207.1678;  
  P.~M.~Ferreira, W.~Grimus, L.~Lavoura and P.~O.~Ludl,
  JHEP {\bf 1209}, 128 (2012)
  [arXiv:1206.7072].


\bibitem{Farzan:2006vj}
  Y.~Farzan and A.~Y.~.Smirnov,
  JHEP {\bf 0701}, 059 (2007)
  [hep-ph/0610337].



\bibitem{Feruglio:2012cw}
  F.~Feruglio, C.~Hagedorn and R.~Ziegler,
  JHEP {\bf 1307}, 027 (2013)
  [arXiv:1211.5560 [hep-ph]].


\bibitem{Ding:2013hpa}
  G.~-J.~Ding, S.~F.~King, C.~Luhn and A.~J.~Stuart,
  JHEP {\bf 1305}, 084 (2013)
  [arXiv:1303.6180 [hep-ph]].


\bibitem{Li:2013jya}
  C.~-C.~Li and G.~-J.~Ding,
  Nucl.\ Phys.\ B {\bf 881}, 206 (2014)
  [arXiv:1312.4401 [hep-ph]].


\bibitem{Feruglio:2013hia}
  F.~Feruglio, C.~Hagedorn and R.~Ziegler,
  arXiv:1303.7178 [hep-ph].

  
\bibitem{Luhn:2013lkn}
  C.~Luhn,
  Nucl.\ Phys.\ B {\bf 875}, 80 (2013)
  [arXiv:1306.2358 [hep-ph]].


\bibitem{Ding:2013bpa}
  G.~-J.~Ding, S.~F.~King and A.~J.~Stuart,
  JHEP {\bf 1312} (2013) 006
  [arXiv:1307.4212].

\bibitem{Krishnan:2012me}
  R.~Krishnan, P.~F.~Harrison and W.~G.~Scott,
  JHEP {\bf 1304}, 087 (2013)
  [arXiv:1211.2000 [hep-ph]].


\bibitem{Mohapatra:2012tb}
  R.~N.~Mohapatra and C.~C.~Nishi,
  Phys.\ Rev.\ D {\bf 86}, 073007 (2012)  [arXiv:1208.2875 [hep-ph]].


\bibitem{Nishi:2013jqa}
  C.~C.~Nishi,
  Phys.\ Rev.\ D {\bf 88}, 033010 (2013)
  [arXiv:1306.0877 [hep-ph]].


\bibitem{Ding:2013nsa}
  G.~-J.~Ding and Y.~-L.~Zhou,
  arXiv:1312.5222 [hep-ph].



\bibitem{Holthausen:2012dk}
  M.~Holthausen, M.~Lindner and M.~A.~Schmidt,
  JHEP {\bf 1304}, 122 (2013)
  [arXiv:1211.6953 [hep-ph]].


\bibitem{Chen:2014tpa}
  M.~-C.~Chen, M.~Fallbacher, K.~T.~Mahanthappa, M.~Ratz and A.~Trautner,
  arXiv:1402.0507 [hep-ph].



\bibitem{Branco:1983tn}
  G.~C.~Branco, J.~M.~Gerard and W.~Grimus,
  Phys.\ Lett.\ B {\bf 136}, 383 (1984);  
  I.~de Medeiros Varzielas and D.~Emmanuel-Costa,
  Phys.\ Rev.\ D {\bf 84}, 117901 (2011)  [arXiv:1106.5477 [hep-ph]]; %
  I.~de Medeiros Varzielas, D.~Emmanuel-Costa and P.~Leser,
  Phys.\ Lett.\ B {\bf 716}, 193 (2012)  [arXiv:1204.3633 [hep-ph]];
  I.~de Medeiros Varzielas,
  JHEP {\bf 1208}, 055 (2012)  [arXiv:1205.3780 [hep-ph]];
  G.~Bhattacharyya, I.~de Medeiros Varzielas and P.~Leser,
  Phys.\ Rev.\ Lett.\  {\bf 109}, 241603 (2012)  [arXiv:1210.0545 [hep-ph]]; 
  I.~d.~M.~Varzielas and D.~Pidt,
  arXiv:1307.0711 [hep-ph].


\bibitem{Chen:2009gf}
  M.~-C.~Chen and K.~T.~Mahanthappa,
  Phys.\ Lett.\ B {\bf 681}, 444 (2009)  [arXiv:0904.1721 [hep-ph]];
  A.~Meroni, S.~T.~Petcov and M.~Spinrath,
Phys.\ Rev.\ D {\bf 86}, 113003 (2012)  [arXiv:1205.5241 [hep-ph]];
  S.~Antusch, S.~F.~King and M.~Spinrath,
  Phys.\ Rev.\ D {\bf 87}, 096018 (2013)  [arXiv:1301.6764 [hep-ph]].




\bibitem{Antusch:2011sx}
  S.~Antusch, S.~F.~King, C.~Luhn and M.~Spinrath,
   Nucl.\ Phys.\ B {\bf 850}, 477 (2011)  [arXiv:1103.5930 [hep-ph]];
  S.~Antusch, M.~Holthausen, M.~A.~Schmidt and M.~Spinrath,
   arXiv:1307.0710 [hep-ph].



\bibitem{Girardi:2013sza}
  I.~Girardi, A.~Meroni, S.~T.~Petcov and M.~Spinrath,
  JHEP {\bf 1402}, 050 (2014)
  [arXiv:1312.1966 [hep-ph]].



\bibitem{King:2014rwa}
  S.~F.~King and T.~Neder,
  arXiv:1403.1758 [hep-ph];
see also
  S.~F.~King, T.~Neder and A.~J.~Stuart,
  Phys.\ Lett.\ B {\bf 726} (2013) 312
  [arXiv:1305.3200 [hep-ph]].



\bibitem{Ding:2012xx}
  G.~-J.~Ding,
  Nucl.\ Phys.\ B {\bf 862}, 1 (2012)
  [arXiv:1201.3279 [hep-ph]].


\bibitem{Toorop:2011jn}
  R.~d.~A.~Toorop, F.~Feruglio and C.~Hagedorn,
  Phys.\ Lett.\ B {\bf 703}, 447 (2011)
  [arXiv:1107.3486 [hep-ph]];
  R.~de Adelhart Toorop, F.~Feruglio and C.~Hagedorn,
  Nucl.\ Phys.\ B {\bf 858}, 437 (2012)
  [arXiv:1112.1340 [hep-ph]].



\bibitem{King:2012in}
  S.~F.~King, C.~Luhn and A.~J.~Stuart,
  Nucl.\ Phys.\ B {\bf 867}, 203 (2013)
  [arXiv:1207.5741 [hep-ph]].



\bibitem{pdg}
   J. Beringer {\it et al.} [Particle Data Group Collaboration],
  Phys.\ Rev.\ D {\bf 86} (2012) 010001.



\bibitem{Escobar:2008vc}
  J.~A.~Escobar and C.~Luhn,
  J.\ Math.\ Phys.\  {\bf 50}, 013524 (2009)  [arXiv:0809.0639 [hep-th]].  



\bibitem{King:2013eh}
  S.~F.~King and C.~Luhn,
  Rept.\ Prog.\ Phys.\  {\bf 76}, 056201 (2013)
  [arXiv:1301.1340 [hep-ph]].


\end{thebibliography}
\end{document}